\documentclass[preprint,12pt,authoryear]{elsarticle}
\usepackage[round]{natbib} 

\usepackage[margin=2.5cm]{geometry}
\usepackage{enumitem}
\newlist{inlinelist}{enumerate*}{1}
\setlist*[inlinelist,1]{%
  label=(\roman*),
}
\usepackage{soul}
\usepackage{hhline}
\usepackage{subfig}
\usepackage{multirow}
\usepackage{xcolor}
\usepackage{graphicx}
\usepackage{upgreek}
\usepackage{amssymb}
\usepackage{amsfonts,amsthm,bm,amsmath} 
\usepackage{appendix}
\usepackage{times}
\usepackage{caption}
\usepackage{subfig}
\newcommand{\psubref}[1]{\protect\subref{#1}}
\usepackage{multirow}
\usepackage{xcolor}
\usepackage[linesnumbered,ruled,vlined]{algorithm2e}
\usepackage{tablefootnote}

\SetCommentSty{mycommfont}

\SetKwInput{KwInput}{Input}                
\SetKwInput{KwOutput}{Output}              

\usepackage[colorlinks]{hyperref} 
\hypersetup{ 
    colorlinks=true,       
    linkcolor=red,          
    citecolor=blue,        
    filecolor=magenta,      
    urlcolor=cyan           
}

\newcommand{\fref}[1]{Fig.~\ref{#1}}

\newcommand{\sref}[1]{Section~\ref{#1}}
\newcommand{\tref}[1]{Table~\ref{#1}}

\begin{document}

\begin{frontmatter}

\title{Geom-DeepONet: A Point-cloud-based Deep Operator Network for Field Predictions on 3D Parameterized Geometries}
\author[]{Junyan He$^{1}$\corref{mycorrespondingauthor}}
\cortext[mycorrespondingauthor]{Corresponding author}
\ead{jimmy.he@ansys.com}
\author[]{Seid Koric$^{2,3}$}
\author[]{Diab Abueidda$^{3}$}
\author[]{Ali Najafi$^{1}$}
\author[]{Iwona Jasiuk$^2$}

\address{$^1$ Ansys Inc., Canonsburg, PA, USA \\
$^2$ Department of Mechanical Science and Engineering, University of Illinois at Urbana-Champaign, Urbana, IL, USA \\
$^3$ National Center for Supercomputing Applications, University of Illinois at Urbana-Champaign, Urbana, IL, USA \\
}

\begin{abstract}
Modern digital engineering design process commonly involves expensive repeated simulations on varying three-dimensional (3D) geometries. The efficient prediction capability of neural networks (NNs) makes them a suitable surrogate to provide design insights. Nevertheless, few available NNs can handle solution prediction on varying 3D shapes. We present a novel deep operator network (DeepONet) variant called Geom-DeepONet, which encodes parameterized 3D geometries and predicts full-field solutions on an arbitrary number of nodes. To the best of the authors' knowledge, this is the first attempt in the literature and is our primary novelty. In addition to expressing shapes using mesh coordinates, the signed distance function for each node is evaluated and used to augment the inputs to the trunk network of the Geom-DeepONet, thereby capturing both explicit and implicit representations of the 3D shapes. The powerful geometric encoding capability of a sinusoidal representation network (SIREN) is also exploited by replacing the classical feedforward neural networks in the trunk with SIREN. Additional data fusion between the branch and trunk networks is introduced by an element-wise product. A numerical benchmark was conducted to compare Geom-DeepONet to PointNet and vanilla DeepONet, where results show that our architecture trains fast with a small memory footprint and yields the most accurate results among the three with less than 2 MPa stress error. Results show a much lower generalization error of our architecture on unseen dissimilar designs than vanilla DeepONet. Once trained, the model can predict vector solutions, and speed can be over $10^5$ times faster than implicit finite element simulations for large meshes. The ability of the proposed model to perform efficient and accurate field predictions on variable 3D geometries, especially those discretized by different nodes and elements, makes it a valuable tool for preliminary performance evaluation and design optimizations and is the most significant contribution of the current work.
\end{abstract}


\end{frontmatter}

\section{Introduction}
\label{sec:intro}
Modern science and engineering depend heavily on physics-based computational models, and this discipline is expanding quickly. Numerical simulations are widely utilized as a reality model to offer insights for comprehending and forecasting the behavior of various physical phenomena and engineering systems. In addition to loads, boundary and initial conditions, material properties, and other inputs in many problems in engineering and science, the domain geometry is also changing. Due to their complex, multi-physics, multi-scale nature, three-dimensionality, time dependency, and required fidelity, these finite element (FE) simulation models can be computationally expensive, even on the newest high-performance computing platforms. Consequently, relying solely on classical high-fidelity simulation models for tasks like computer-aided design, material discovery, and digital twins might become impractical and often impossible when there is a requirement to explore a vast number of potential design scenarios and/or geometries. On the other hand, a surrogate neural network (NN) model for changing three-dimensional (3D) domain geometry based on mesh data is a promising machine learning method that has the potential to almost instantly infer the solution of a physical problem involving variable 3D domain shapes without the need for expensive traditional numerical simulations consisting of repeated meshing and solver phases. Applying deep learning methods to learn the complexities of 3D variable geometries, represented by its computational mesh, is an active area of research with significant potential in various application domains like real-time simulations for predictions and controls as well as designs, topology and shape optimizations, sensitivity analysis, and uncertainty quantification that require extensive forward evaluations with changing 3D geometries, loads, material properties and other input parameters.  

The work by \cite{sun2020surrogate} attempted to train a physics-informed, fully connected neural network for parameterized geometries by adding the parametric variable as input to the network for surrogate modeling of incompressible steady flows. However, the work was limited to simple 2D fluid flow geometry with minimal variation of geometry parameters. Even though the convolutional neural networks (CNNs) excel in image voxel shape recognition, the computational meshes originating from numerical methods discretizing 3D domain geometries are often unstructured and have varying vertex counts and connectivity, unlike 2D image grids or 3D voxels, thus making traditional CNNs mostly inapplicable even for 2D meshes with irregular geometry. Several works exploited coordinate transformation from irregular geometry to regular image-like computational domains applicable to CNN models \citep{bao2022physics, sun2020surrogate}. Still, they could not handle complex geometries with many edges or holes often encountered in real-world applications.

Another direction is to treat the 3D mesh data directly as a point cloud, a set of 3D points representing geometric data. The pioneering work in this area is PointNet \citep{qi2017pointnet}, which uses multilayer perceptron neural networks with shared weights and a globally acting “symmetrical” pooling function to construct the lower dimensional representation from a set of points. Later, more complex network architectures were devised and inspired by the PointNet architecture, such as PointNet++ \citep{qi2017pointnet++}, which extended PointNet to extract features from local neighborhoods better, and GAPINN \citep{oldenburg2022geometry} with Variational-Auto-Encoder, which was used to reduce the dimensions of the irregular geometries to a latent representation, which is subsequently input into the physics-informed neural network. These works were also mostly confined to simple geometries due to their inability to explicitly capture edge or pairwise relationships between points.

An alternative class of artificial neural networks often used to model parametric and general variability of domain geometries based on mesh data are graphical neural networks (GNN) \citep{gilmer2017neural}. GNNs operate on graph-structured data and can treat mesh nodes as the graph's vertices, but unlike point clouds, they exploit mesh connectivity (elements), too, as the edges of the graph. In GNN, communication is established between neighboring nodes in the so-called message-passing and aggregation routines, and this allows GNN to handle data from different graphs, promising to generalize data over geometries never seen by the network. \cite{wong2022graph} and \cite{jin2023leveraging} used an encode-process-decode GNN architecture to perform the functional performance of geometrical designs in solid mechanics and fluid applications. \cite{franco2023deep} used another encoder-decoder version of GNN to handle parametric 2D geometric variability. They compared it with vanilla feedforward neural network implementation and showed its advantages. \cite{he2023use} observed similar advantages in the context of physics-informed neural networks. However, to a certain extent, all these GNN-based networks lacked generalizability due to their inherited local nature of learning, particularly in 3D, which was somewhat alleviated with more profound and more advanced GNN architectures. This, in turn, brings a severe computational burden for training for many nodes in the GNN. \cite{gladstone2024mesh} recently addressed these issues with edge-augmented GNN and special treatment of variable domains by a novel coordinate transformation that enables rotation and translation invariance. While that work yielded promising results with 2D solid mechanical domains under elastic and hyper-elastic constitutive laws, it did not address variable 3D geometries or significant material nonlinearities.  

Outside the computational mechanics community, NN architectures for encoding and representing varying 3D (surface) meshes are also heated research topics in the computer graphics/vision community. These networks seek to create a lightweight and implicit encoding of 3D surface meshes. One of the key concepts in this area is the signed distance function (SDF). This mathematical function represents the distance between a specific point and the boundary of a geometric shape. It is extensively utilized in computer graphics, computer vision, and robotics to depict three-dimensional shapes. The SDF assigns a numerical value, which can be positive or negative, to every point in space. The sign of the value denotes whether the point is located inside or outside the shape, while the size of the value represents the distance to the nearest surface. Continuous SDF approximated by deep neural networks has been shown to be an efficient way to represent different 3D shapes in the work of \cite{park2019deepsdf}. Another recent notable contribution to representing 3D shapes is the sinusoidal representation networks (SIREN) \citep{sitzmann2020implicit}, which is a periodic activation function that leverages the sine function to encode complex, highly variable inputs and their derivatives. It has shown to be superior to commonly used activation functions such as Tanh and ReLU. SIREN is part of the broader field of implicit neural representations (INR) \citep{dupont2021coin,sitzmann2020implicit,benbarka2022seeing}, which are a class of NNs that can continuously represent signals (such as images, videos, or 3D shapes).

Typical surrogate NN-based models predict solutions in specific locations, or if they manage to predict the specific field solution with a given set of parameters, a slight change of input parameters, such as loads, boundary conditions, or domain geometry, requires computationally costly retraining on transfer learning. To lower generalization error on unseen cases, the concept of operator learning \citep{kovachki2023neural} was proposed, which aims to approximate the governing operator via the NN. A prominent operator learning architecture is the Fourier neural operator (FNO) first proposed by \cite{li2020fourier}, which encodes input functions via multiple Fourier layers. Each Fourier layer takes the Fast Fourier Transform of its input and filters out high-frequency modes. FNO and its improved versions have been successfully applied to various differential equations \citep{li2020fourier} and elastic-plastic deformation problems in the mechanics field \citep{li2022fourier}. More recently, \cite{lu2021learning} devised a Deep Operator Neural Network architecture called DeepONet for effective operator learning. Due to its dual network nature (trunk and branch network), DeepONet can map an infinite space of parameters to the entire solution field on the computational grid. A trained DeepONet does not need retraining or transfer learning when a set of new parameters is presented, and an entire solution field can be inferred several orders of magnitude faster than classical numerical methods. Since its inception, DeepONets have been used to predict complete solution fields in materially nonlinear solid mechanics \citep{koric2023deep,lu2023deep}, fracture \citep{goswami2022physics}, aerodynamics \citep{zhao2023learning}, acoustics \citep{xu2023training}, heat transfer \citep{sahin2024deep,koric2023data}, and seismology \citep{haghighat2024deeponet}. While the original DeepONet consisted of forward fully connected networks in the branch and the trunk, a few new formulations emerged recently, such as sequential DeepONet \citep{he2024sequential} devised to encode time-dependent inputs as well as the DeepONet using a ResUNet in its trunk \citep{he2023deep} to solve problems with complex and highly disparate 2D input geometries (defined on a regular pixel grid) under parametric loads and elastoplastic material behavior and, to this date, that was the only DeepONet paper with variable 2D geometry. To the best of the authors' knowledge, no DeepONets have been operating in 3D real-world computational domains. Moreover, almost all existing DeepONets receive and encode constant coordinate information during training in their trunk network.  

Therefore, a knowledge gap exists in the literature for an NN architecture capable of handling varying geometries in 3D. The excellent generalizability of the DeepONet renders itself a suitable candidate architecture for this task. In this work, we develop and demonstrate a novel architecture called Geom-DeepONet, which combines the operator learning architecture of a DeepONet with the power of SDF and SIREN to perform field predictions on parameterized geometries. By parameterized geometries, we mean 3D geometries that can be fully defined by a series of geometric parameters (e.g., length, width, radius, etc.), typical for many real-world engineering designs. To the best of our knowledge, this marks the first time that a DeepONet architecture has been proposed and applied to field predictions on variable 3D shapes, and it is the most novel contribution of the current work.  

This paper is organized as follows: \sref{sec:methods} provides an overview of the neural network models and the two familes of designs used to test performance. \sref{sec:results} presents and discusses the performance of the NN models with two numerical experiments. \sref{sec:conc} summarizes the outcomes and limitations and highlights future works.

\section{Methods}
\label{sec:methods}
\subsection{PointNet and vanilla DeepONet for parameterized geometries}
\label{bench_mdls}
\begin{figure}[h!] 
    \centering         \includegraphics[width=0.75\textwidth]{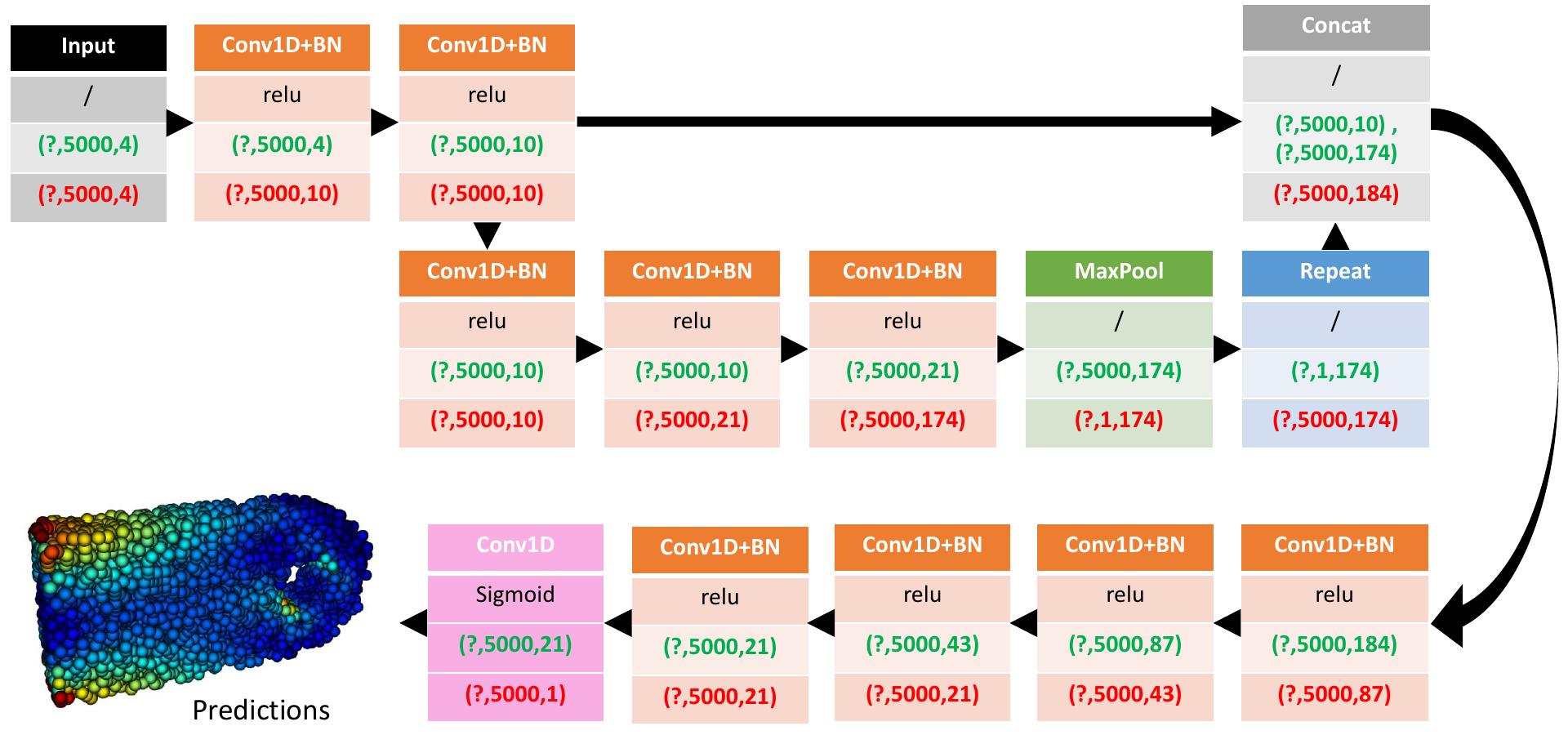}
    \caption{Schematic of the PointNet used for comparison with other networks. Input and output tensor shapes are shown in green and red, respectively. $?$ and $BN$ denote the flexible batch size dimension and batch normalization.}
    \label{ptnet_schematic}
\end{figure}

The PointNet architecture by \cite{qi2017pointnet} had been modified in the work of \cite{kashefi2021point} to predict the flow field on variable and irregular geometries. It leverages a point-cloud description of the underlying geometry and can be easily extended to perform predictions in 3D parameterized geometries. Therefore, implementing the modified PointNet in \cite{point-cfd} was used in this work as a benchmark model.  Five thousand points were used to define the input point cloud, and a model scaling factor of 0.17 was used to generate a model with 25616 trainable parameters, similar to all other networks used in the benchmark. For each point, four input features are provided; they are the X, Y, and Z coordinates of the point and the applied load magnitude (constant for all points in the same case). For implementation details on the PointNet architecture, the readers are referred to the work of \cite{kashefi2021point}. A schematic of the PointNet model used in the benchmark is shown in \fref{ptnet_schematic}.

\begin{figure}[h!] 
    \centering         \includegraphics[width=0.7\textwidth]{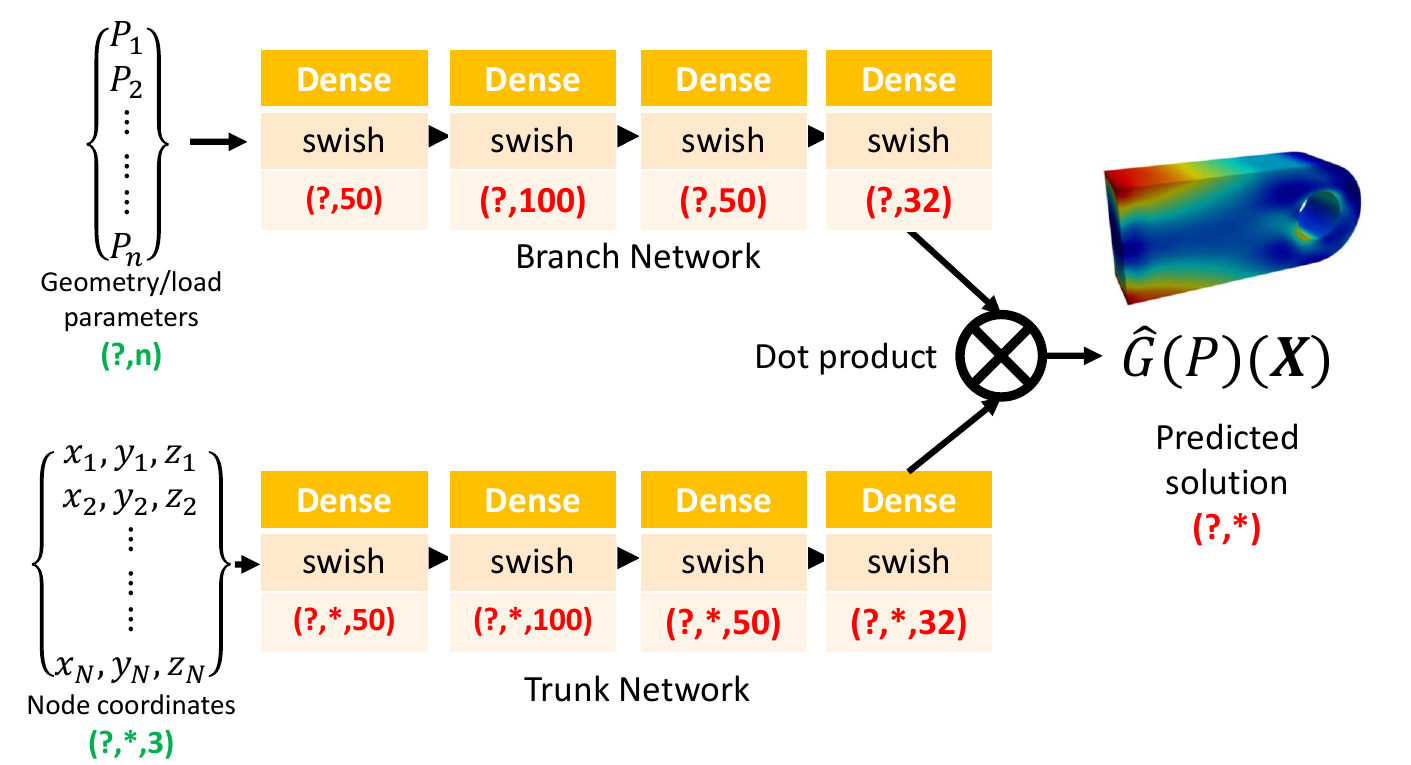}
    \caption{Schematic of the vanilla DeepONet used for comparison with other networks. Input and output shapes are shown in green and red, respectively. $?$ and $*$ denote the flexible batch size and number of nodes in the data. $n$ and $N$ are the number of input parameters and the number of nodes included in the training.}
    \label{don_schematic}
\end{figure}

The original DeepONet architecture \citep{lu2021learning} is limited to a single geometry described by a fixed number of nodal coordinates fed to its trunk network. Since DeepONets has shown great success in various engineering applications and the proposed network is a variant of the DeepONet architecture, it is also included as a performance benchmark model. Since its branch and trunk networks consist of simple dense layers, introducing another flexible dimension in the trunk network input to account for a variable number of nodes during prediction is relatively straightforward; see the schematic shown in \fref{don_schematic}. The encoded data from the branch and trunk networks are combined to form the DeepONet prediction via a dot product along the hidden dimension as follows:
\begin{equation}
    \Hat{G}(P)(\bm{X})_{?i} = \sum_{h=1}^{32} B_{?h} T_{?ih},
\end{equation}
where $B$ and $T$ denote the encoded data from the branch and trunk networks. The hidden dimension $h$ is set to 32 in this work. However, the complete data from different geometries (discretized by different numbers of nodes and elements) inherently have different shapes and cannot be batched together in training\footnote{Tensors of different shapes cannot be concatenated together, and zero-padding to obtain a consistent shape is wasteful in memory.}, a technique commonly used in modern-day machine learning to improve efficiency. To allow batched training, all the nodal coordinates and the corresponding output fields are randomly resampled (with repeats if the resample size is greater than the total mesh node count) to a fixed quantity $N$ \emph{only} during training, while the flexible dimension $*$ in \fref{don_schematic} allows for predictions on arbitrary node counts during prediction time\footnote{To accomplish this, predictions are not batched and are instead computed one after another at the expense of longer prediction times.}. Similar resampling of the field data was done in some previous works \citep{koric2023data,koric2023deep}, but sub-sampling was only a means to reduce the input dataset size to reduce memory consumption and the geometries studied in those works were constant. However, in the current work, it is used specifically to tackle the challenge of a variable number of nodes in different input meshes. The concept of resampling the data for efficient training on different meshes is essential in this work and is used in the vanilla DeepONet and our proposed DeepONet to be introduced in \sref{our_mdl}. It will be investigated further in \sref{resample}. The vanilla DeepONet model used here for the benchmark has 24014 trainable parameters.

\subsection{Geom-DeepONet based on SDF and SIREN}
\label{our_mdl}
\begin{figure}[h!] 
    \centering
         \includegraphics[width=0.9\textwidth]{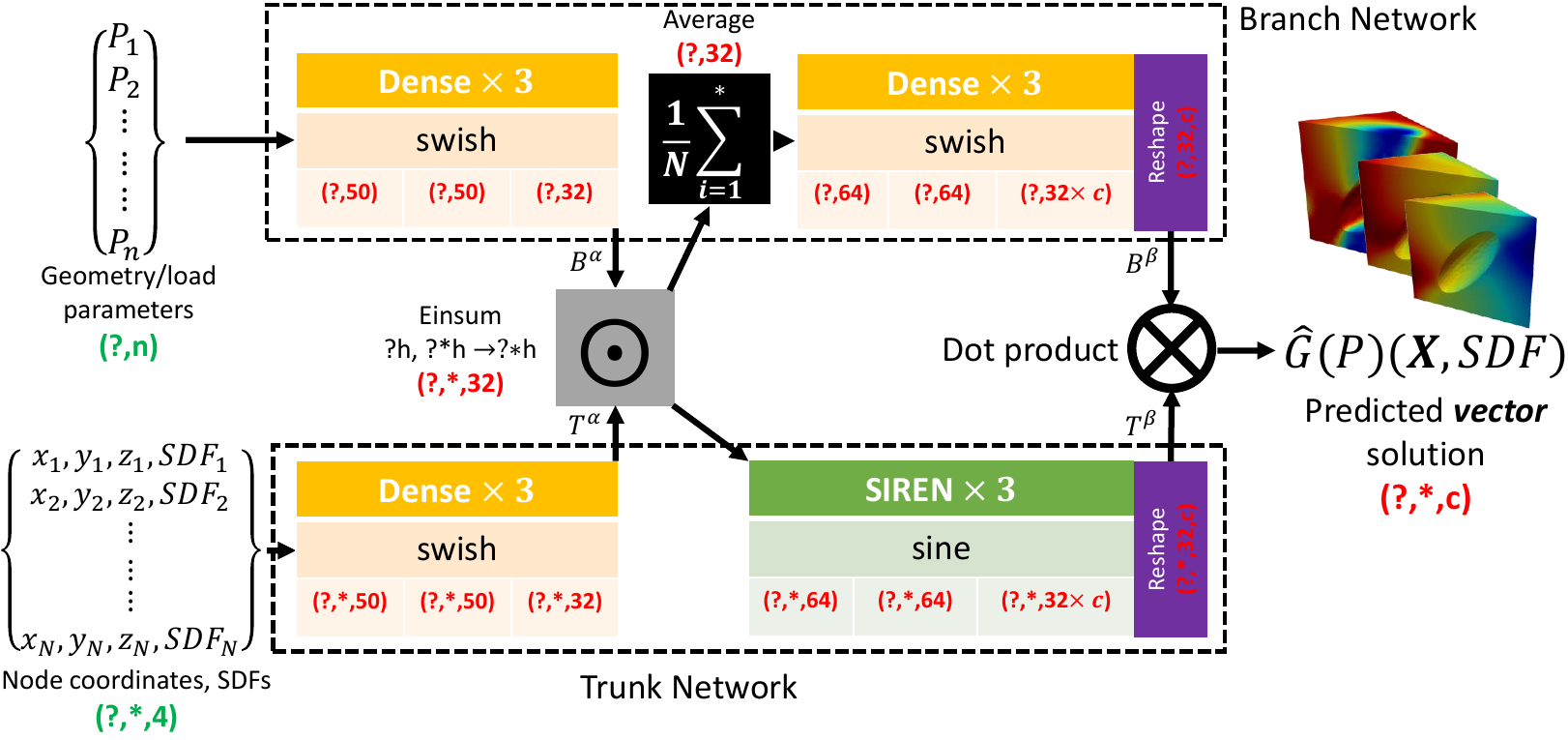}
    \caption{Schematic of the proposed Geom-DeepONet. Input and output shapes are shown in green and red. $?$ and $*$ denote the flexible batch size and number of nodes in the data. $n$, $N$, and $c$ are the number of input parameters and nodes included in training and output vector components, respectively.}
    \label{gdon_schematic}
\end{figure}

The PointNet in \cite{kashefi2021point} and the vanilla DeepONet models have respective limitations when dealing with variable input geometries. First, the PointNet architecture in \cite{kashefi2021point} does not allow varying the number of points (i.e., 5000) in the input point cloud once the model is trained, posing significant limitations for different geometries typically described by different numbers of nodes and elements. Although the vanilla DeepONet architecture can handle an arbitrary number of nodes in prediction with minor modifications, it suffers from the shortcoming that it only leverages coordinate information in the trunk network. While the X, Y, and Z coordinates are sufficient to describe a constant input geometry, it is insufficient when the geometry is changing since a point $(x^*,y^*,z^*)$ that is within the geometry $i$ might not be part of another geometry. As surveyed in \sref{sec:intro}, similar problems of encoding and rendering variable geometries have been studied by the computer vision/graphics community. SDF and the SIREN architecture are two critical tools that stand out, both of which found success in the implicit neural representation of shapes. Inspired by these findings, we proposed to augment the inputs to the trunk network with the SDF of nodes with respect to the external surfaces of the input geometry, such that all the nodes on the external geometric surface will have an SDF of 0. In contrast, all interior nodes have a negative SDF whose magnitude equals its closest distance to the external geometric surfaces. When a point $(x^*,y^*,z^*)$ is provided to the NN, the additional input $SDF^*$ will help to differentiate the positional relationship of this point with respect to different geometries, thus providing additional geometry awareness to the model. We also replace selected dense layers in the trunk network with SIREN layers to leverage the powerful encoding capability of the sinusoidal activation functions. Finally, \cite{wang2022improved} pointed out that the DeepONet architecture may lead to insufficient information fusion since there is no data fusion between the branch and trunk networks until the dot product. Previous works \citep{wang2022improved,HE2023116277} have shown improved performance with intermediate data fusions in the form of an element-wise product. In this work, we leveraged a similar concept to compute an intermediate data fusion $\bm{F}$ via Einstein summation\footnote{Einsum is used instead of a direct element-wise summation to account for the different shapes of the input tensors, see \fref{gdon_schematic}}:
\begin{equation}
    F_{?ih} = B^\alpha_{?h} T^\alpha_{?ih},
\end{equation}
where $B^\alpha$ and $T^\alpha$ denote the intermediate encoded data from the branch and trunk networks. The final encoded data from the two networks are further combined to produce the model output:
\begin{equation}
    \Hat{G}(P)(\bm{X}, SDF)_{?ic} = \sum_{h=1}^{32} B^\beta_{?hc} T^\beta_{?ihc},
\end{equation}
where $B^\beta$ and $T^\beta$ denote the final encoded data from the respective networks, and $c$ denote the number of vector components in a vector solution field. Similar to the vanilla DeepONet, the input nodes in the meshes are resampled to a constant length before they are fed into the trunk network of the proposed model so that batched training is possible. The proposed architecture, termed Geom-DeepONet for its specific capability of performing field predictions on variable parameterized geometries, is shown in \fref{gdon_schematic}. SIREN blocks were only applied to the second portion of the trunk network, a setting that the authors found empirically to give the best performance. We designed the Geom-DeepONet specifically to be able to predict all $c$ components of a vector solution. The total numbers of trainable parameters of the Geom-DeepONet are 25568 (used in benchmark with other models) and 38298 when $c=1$ and $c=4$, respectively. 

The vanilla DeepONet and the proposed Geom-DeepONet were implemented in the DeepXDE framework \citep{lu2021deepxde}. The SIREN implementation from \cite{siren_github} was used. All three NN implementations use a TensorFlow backend \citep{tensorflow2015-whitepaper}. All models were trained with a batch size of 16 and a learning rate of $2\times 10^{-3}$. An inverse time decay was used to adjust the learning rate during training with a coefficient of $2\times 10^{-4}$. The Adam optimizer \citep{kingma2014adam} was used, and the scaled mean squared error (MSE) was used as the loss function.

\subsection{Training data generation}
\label{sec:data_gen}
Two different families of parameterized geometries, (1) beam with circular hole and (2) cuboid with randomly oriented ellipsoidal void, are considered in this work. The first family of geometries is used to benchmark the performance of the three different NN models introduced above. The second family of geometries, bearing much more significant geometry variations, is used to demonstrate the capability of the Geom-DeepONet to handle significant geometry change and predict vector solutions. The following sections introduce the settings of both examples.

\subsubsection{Parameterized beam with a circular hole}
\label{model1}
\begin{figure}[h!] 
    \centering
     \subfloat[Beam example 1]{
         \includegraphics[trim={2cm 2cm 2cm 3.5cm},clip,width=0.19\textwidth]{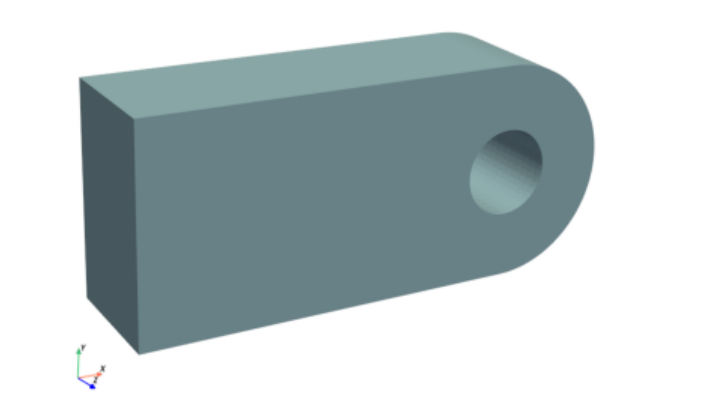}
         \label{d0}
     }
     \subfloat[Beam example 2]{
         \includegraphics[trim={2cm 2cm 2cm 3.5cm},clip,width=0.19\textwidth]{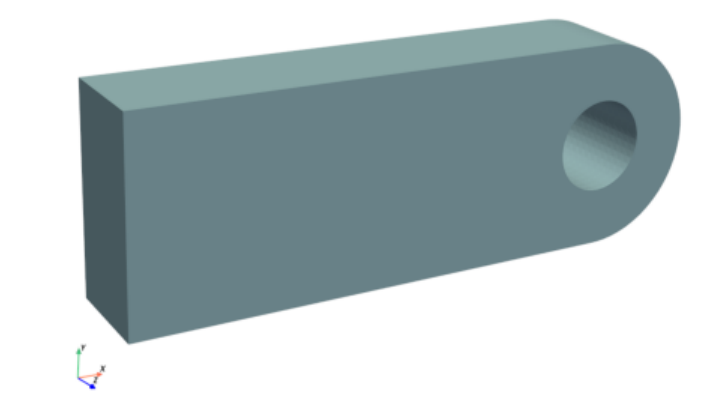}
         \label{d1}
     }
     \subfloat[Beam example 3]{
         \includegraphics[trim={2cm 2cm 2cm 3.5cm},clip,width=0.19\textwidth]{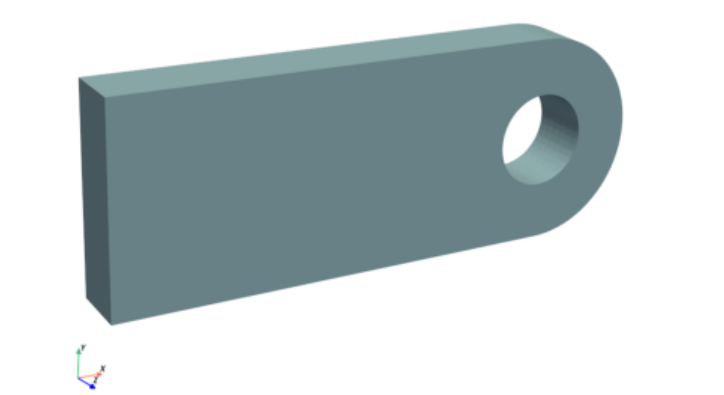}
         \label{d2}
     }
     \subfloat[Beam example 4]{
         \includegraphics[trim={2cm 2cm 2cm 3.5cm},clip,width=0.19\textwidth]{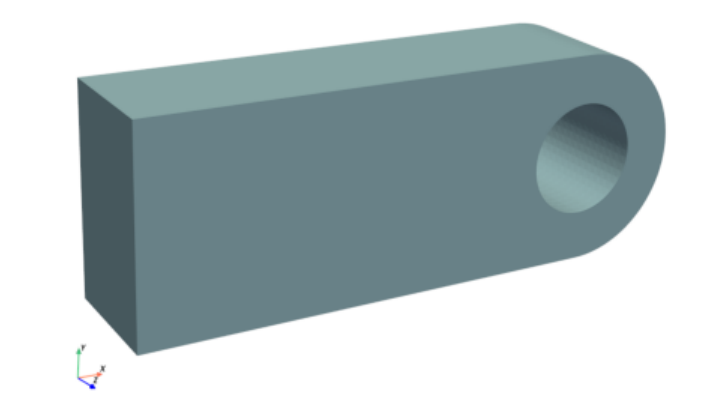}
         \label{d3}
     }
     \subfloat[Beam example 5]{
         \includegraphics[trim={2cm 2cm 2cm 3.5cm},clip,width=0.19\textwidth]{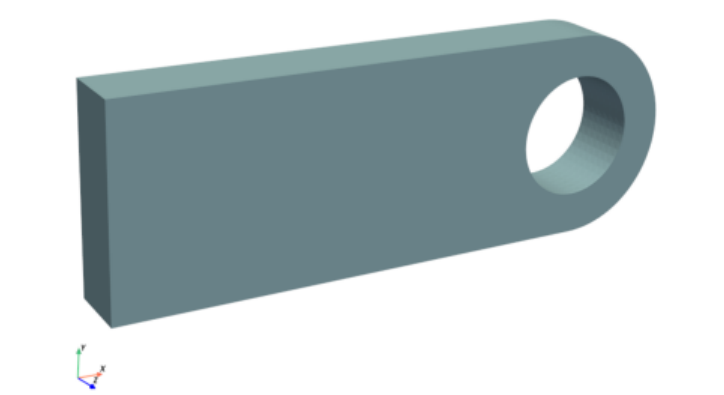}
         \label{d4}
     }
    \caption{Randomly selected beam designs.}
    \label{beam_design}
\end{figure}

\begin{figure}[h!] 
    \centering
     \subfloat[]{
         \includegraphics[trim={0cm 0cm 0cm 1cm},clip,width=0.4\textwidth]{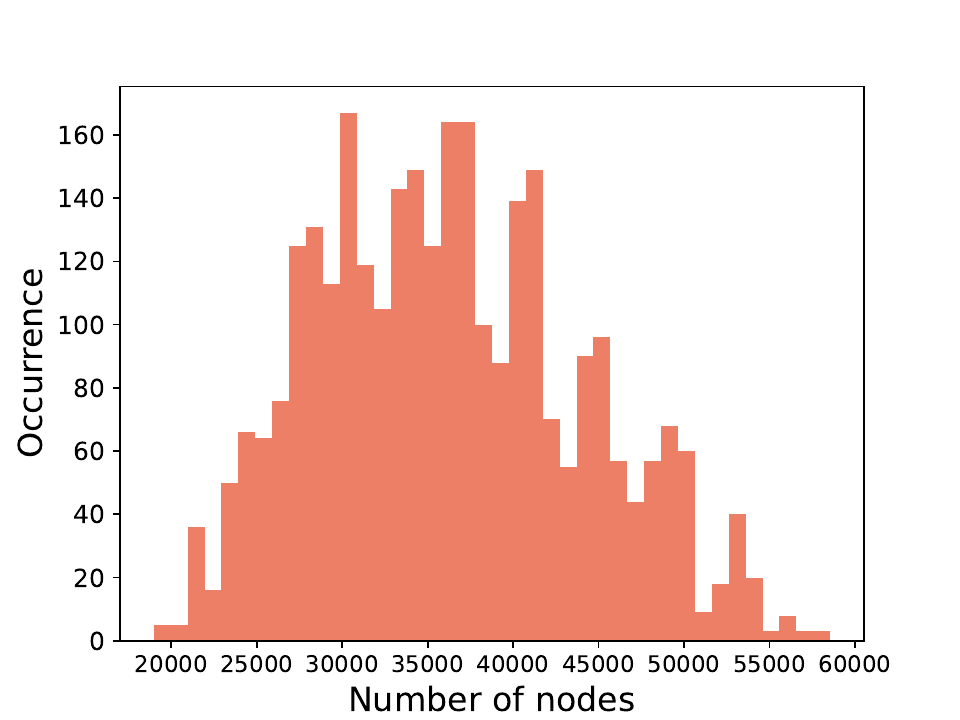}
         \label{beam_nodes}
     }
     \subfloat[]{
         \includegraphics[trim={0cm 0cm 0cm 1cm},clip,width=0.4\textwidth]{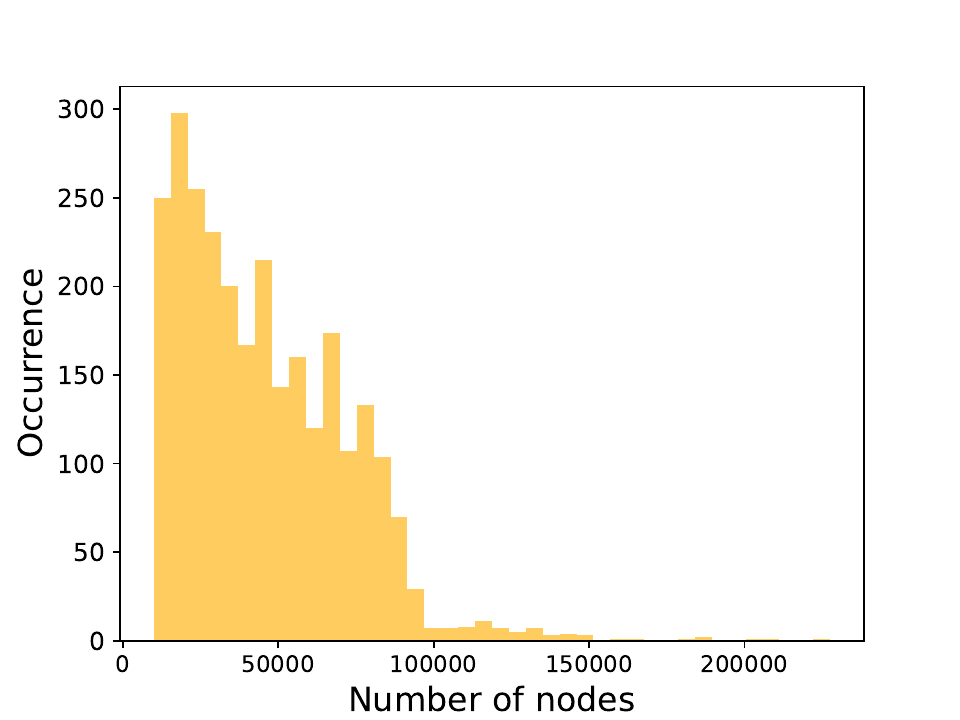}
         \label{cube_nodes}
     }
    \caption{Histograms of number of mesh nodes in different geometries for \psubref{beam_nodes} beams and \psubref{cube_nodes} cuboids with random void. }
    \label{node_hist}
\end{figure}

The beam geometry is parameterized by three geometric parameters, namely the length, the beam's thickness, and the circular hole's radius. The left end of the beam is held fixed. A pressure load is applied to the bottom half of the circular hole on the right end, inducing bending in the beam. The magnitude of the applied pressure is also considered a variable (albeit not geometric). The parameters and their respective ranges are listed in \tref{para1}. An elastic-plastic material with linear isotropic hardening is considered, with the following material properties: Young's modulus of 200 GPa, Poisson's ratio of 0.3, yield stress of 380 MPa, and a hardening modulus of 571.4 MPa. A total of 3000 unique beam designs were generated by randomly sampling from the input parameter space. Quadratic hexahedral elements were used to mesh the geometries. Selected beam designs are shown in \fref{beam_design} to showcase geometry variability. As the geometries vary, so does the corresponding mesh discretization. A histogram is shown in \fref{beam_nodes} to depict the distribution of the number of mesh nodes in all 3000 geometries. All FE simulations were quasi-static and performed under small deformation assumption. For this example, the nodal von Mises equivalent stresses $\sigma_{vM}$ were stored and used as the ground-truth labels in NN training, which is defined as:

\begin{equation}
    \sigma_{vM} = \sqrt{ \frac{3}{2} \bm{S} : \bm{S} },
\end{equation}
where $\bm{S} = \bm{\sigma} - \frac{1}{3}tr(\bm{\sigma})\bm{I}$ is the deviatoric part of the stress tensor.
\begin{table}[h]
\caption{Geometric and load parameters used in the beam example\tablefootnote{Unit in mm unless otherwise noted} }
\centering
\begin{tabular}{ccccccccccc}
\hline
Parameter  &  \vline & Length & Thickness & Radius & Pressure magnitude  \\
\hline
Min. value  &  \vline & 80 & 15 & 10 & 50 MPa \\
Max. value  &  \vline & 120 & 30 & 15 & 100 MPa \\
\hline
\end{tabular}
\label{para1}
\end{table}

\subsubsection{Cuboid with a randomly oriented ellipsoidal void}
\label{model2}
\begin{figure}[h!] 
    \centering
     \subfloat[Cuboid example 1]{
         \includegraphics[trim={2cm 1cm 2cm 2cm},clip,width=0.19\textwidth]{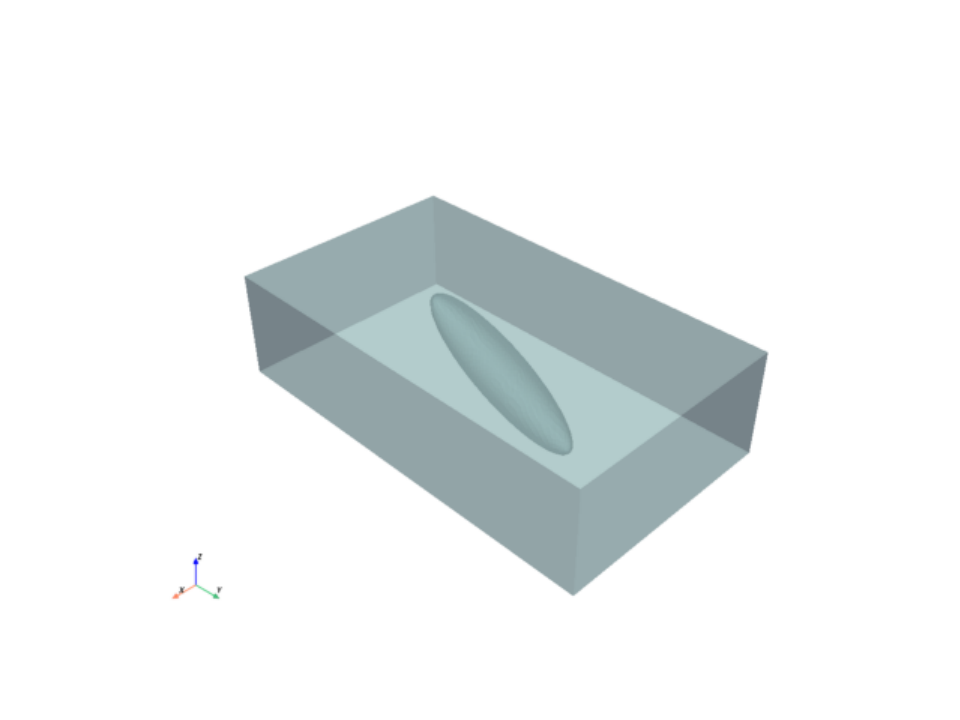}
         \label{cd0}
     }
     \subfloat[Cuboid example 2]{
         \includegraphics[trim={2cm 1cm 2cm 2cm},clip,width=0.19\textwidth]{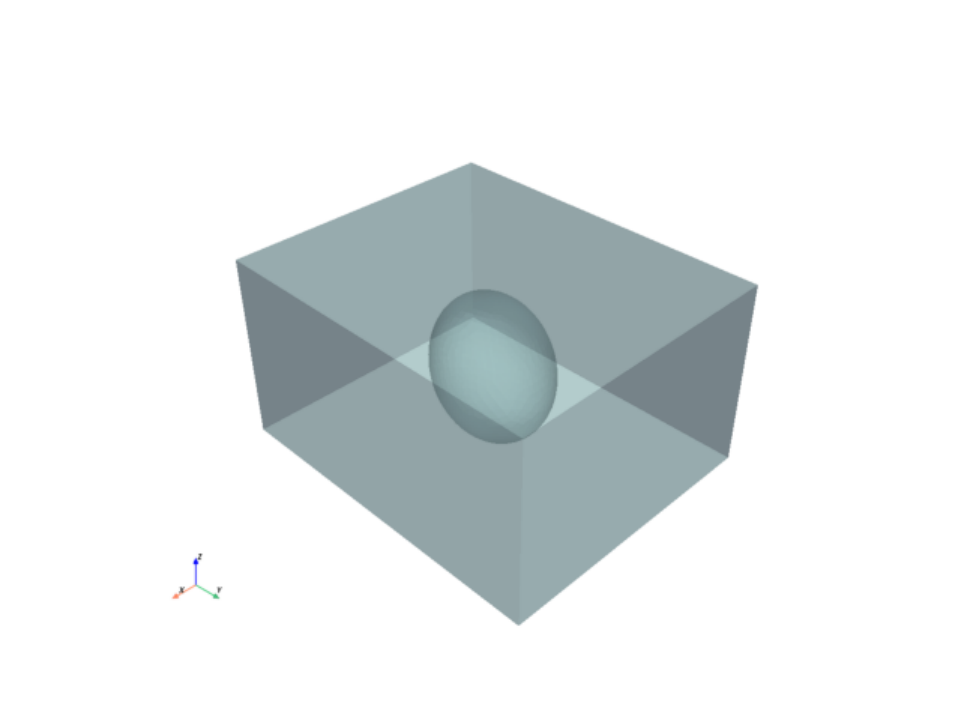}
         \label{cd1}
     }
     \subfloat[Cuboid example 3]{
         \includegraphics[trim={2cm 1cm 2cm 2cm},clip,width=0.19\textwidth]{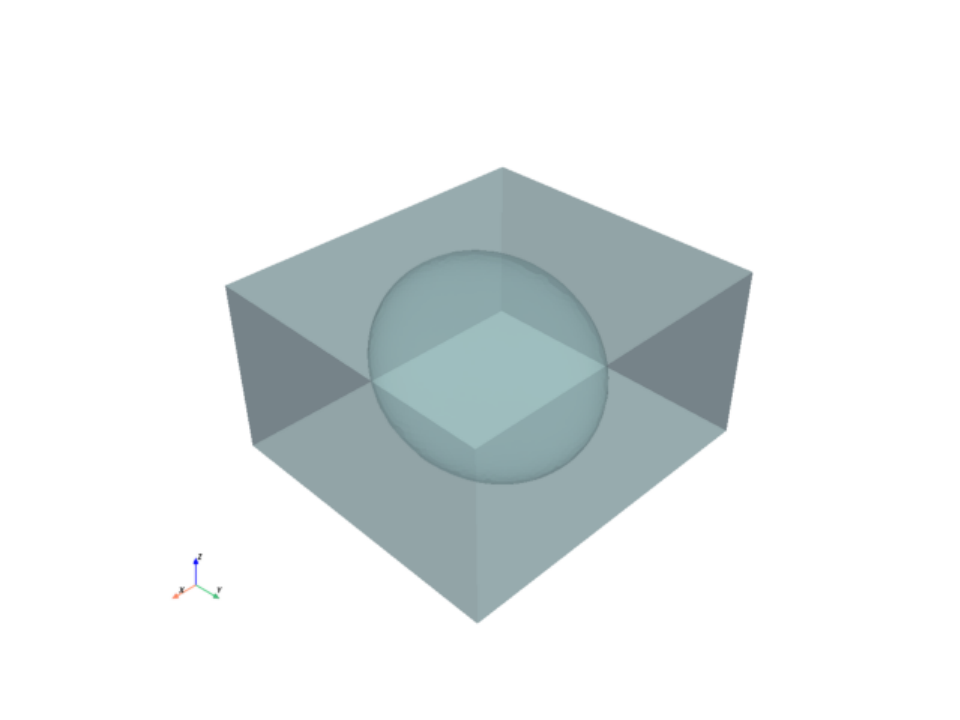}
         \label{cd2}
     }
     \subfloat[Cuboid example 4]{
         \includegraphics[trim={2cm 1cm 2cm 2cm},clip,width=0.19\textwidth]{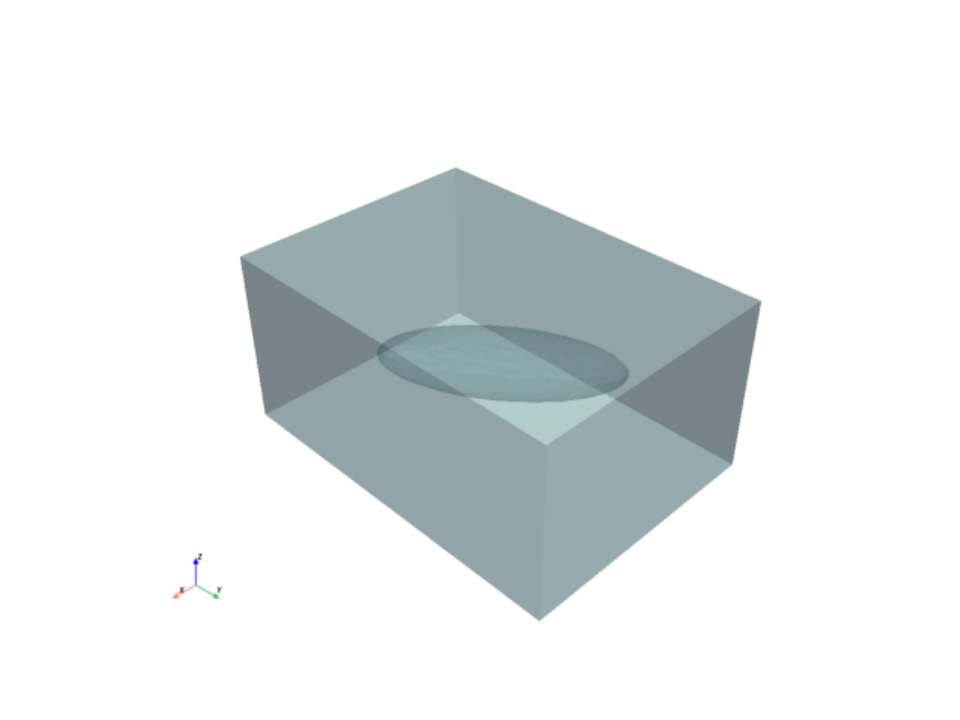}
         \label{cd3}
     }
     \subfloat[Cuboid example 5]{
         \includegraphics[trim={2cm 1cm 2cm 2cm},clip,width=0.19\textwidth]{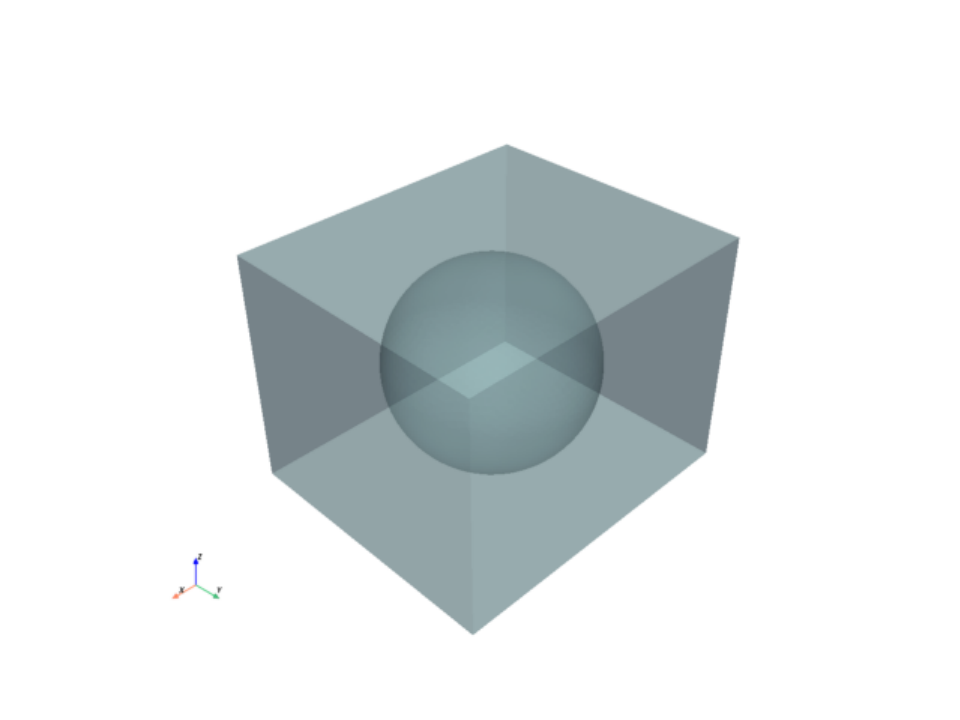}
         \label{cd4}
     }
    \caption{Randomly selected cuboid designs. The bodies are set to opaque to visualize the internal void.}
    \label{cube_design}
\end{figure}

The cuboid geometry is parameterized by eight geometric parameters. At the center of each cuboid, there is a randomly oriented ellipsoidal void, which is characterized by its in-plane major and minor axes (denoted $r_{major}$ and $r_{minor}$, the void is generated by revolving the 2D cross-section) as well as three extrinsic Euler angles specifying the orientation of the void (denoted $\theta_X$, $\theta_Y$ and $\theta_Z$, in X-Y-Z rotation order). The size of the cuboid is determined by the size of the bounding box of the rotated ellipsoidal void as well as three random offset values in the X, Y, and Z directions (denoted $d_X$, $d_Y$ and $d_Z$), respectively. The parameters' ranges are listed in \tref{para2}, where 2500 unique designs were generated. Quadratic tetrahedral elements were used to mesh the geometries. Selected cuboid designs are shown in \fref{cube_design}, and the histogram of all mesh node counts in different meshes is shown in \fref{cube_nodes}. The cuboids are simply supported at the $Y=0$ face to prevent movement in the Y direction only. The amount of applied tensile strain on the $Y=Y_{max}$ surface ($\epsilon_Y$) is also a parameter for increased solution diversity, varying from 0.1\% to 0.15\%. A similar elastic-plastic material is considered, with the following material properties: Young's modulus of 209 GPa, Poisson's ratio of 0.3, yield stress of 235 MPa, and a hardening modulus of 800 MPa. Due to the presence of the internal void and the stress concentration around it, a small amount of strain, such as those in the range of 0.1\% to 0.15\%, is sufficient to induce large-scale yielding in the cuboid. For this example, the von Mises stress and the displacement vector are stored as the four solution fields to be predicted by the NN.

\begin{table}[h]
\caption{Geometric and load parameters used in the cuboid with void example\tablefootnote{Unit in mm unless otherwise noted} }
\centering
\begin{tabular}{ccccccccccc}
\hline
Parameter   &  \vline & $r_{major}$ & $r_{minor}$ & $\theta_X$ & $\theta_Y$ & $\theta_Z$ & $d_X$ & $d_Y$ & $d_Z$ & $\epsilon_Y$  \\
\hline
Min. value  &  \vline & 0.5 & 0.5 & 0$^\circ$ & 0$^\circ$ & 0$^\circ$ & 1 & 1 & 1 & 0.1\% \\
Max. value  &  \vline & 5 & 5 & 90$^\circ$ & 90$^\circ$ & 90$^\circ$ & 5 & 5 & 5 & 0.15\% \\
\hline
\end{tabular}
\label{para2}
\end{table}

\section{Results and discussion}
\label{sec:results}
All simulations were conducted with six high-end AMD EPYC 7763 Milan CPU cores. All NN training and inference were conducted using a single Nvidia A100 GPU card on Delta, an HPC cluster hosted at the National Center for Supercomputing Applications (NCSA). Two quantitative metrics were used to evaluate the model performance in the test set. They are the relative error and mean absolute error:
\begin{equation}
\begin{aligned}
    {\rm{Relative \; error}} = \left| \frac{ f_{FE} - f_{Pred} }{ f_{FE} } \right | \times 100\%,\\
     {\rm{MAE}} = \frac{1}{N_T} \sum_{i=1}^{N_T} \left |  f_{FE} - f_{Pred} \right |,
\end{aligned}
\end{equation}
where $f_{FE}$, $f_{Pred}$, and $N_T$ denote the finite element (FE) simulated field values, NN-predicted field values, and the number of test cases, respectively. For \sref{resample} - \sref{generalization}, the parametric beam dataset was used, while for \sref{model2_result}, the cuboid with void dataset was used.

\subsection{Effect of number of resample points}
\label{resample}
\begin{figure}[h!]
\newcommand\x{0.25}
    \centering
    \begin{tabular}{ c c c c }
    \begin{minipage}[c]{\x\textwidth}
       \centering 
        \subfloat[$N$=250]{\includegraphics[trim={1cm 0cm 1cm 1cm},clip,width=\textwidth]{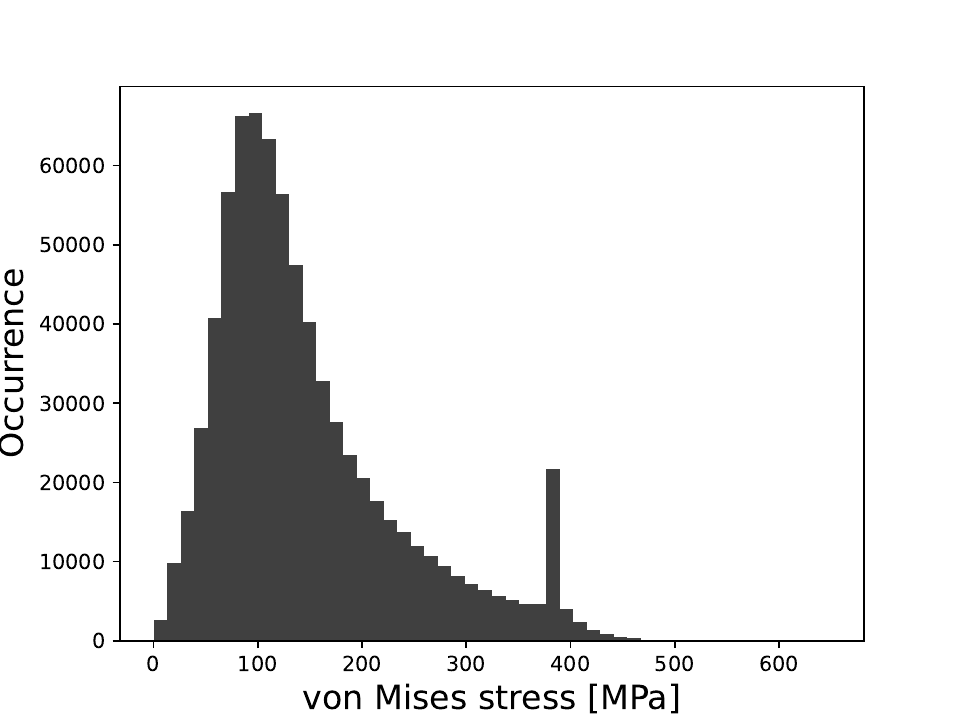}
        \label{p31}}
    \end{minipage} &
    \begin{minipage}[c]{\x\textwidth}
       \centering 
        \subfloat[$N$=1000]{\includegraphics[trim={1cm 0cm 1cm 1cm},clip,width=\textwidth]{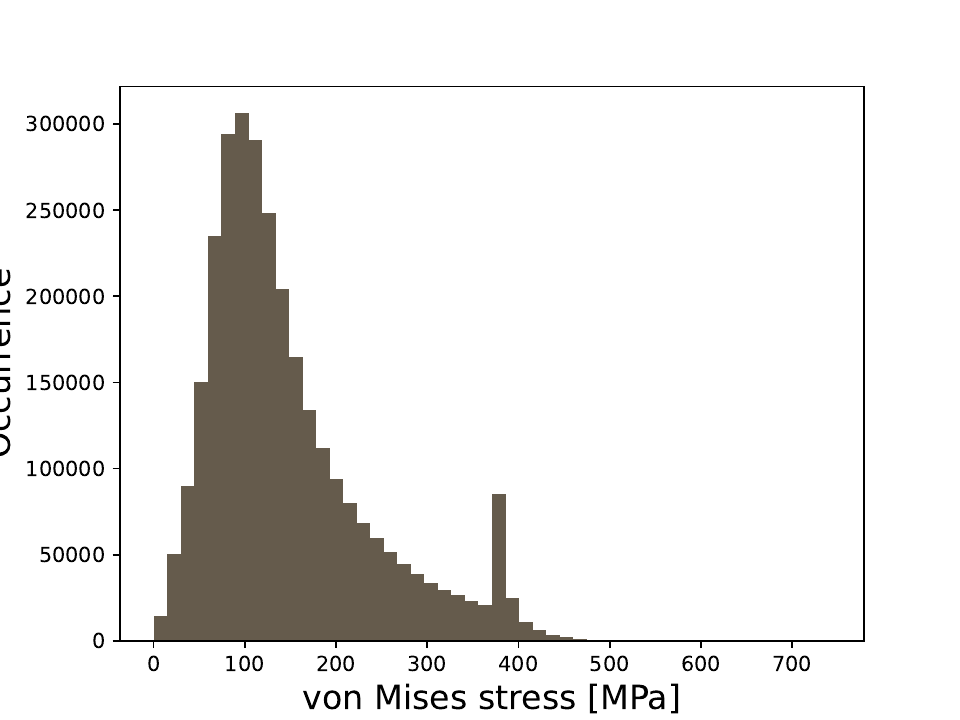}
        \label{p32}}
    \end{minipage} &
    \begin{minipage}[c]{\x\textwidth}
       \centering 
        \subfloat[$N$=2000]{\includegraphics[trim={1cm 0cm 1cm 1cm},clip,width=\textwidth]{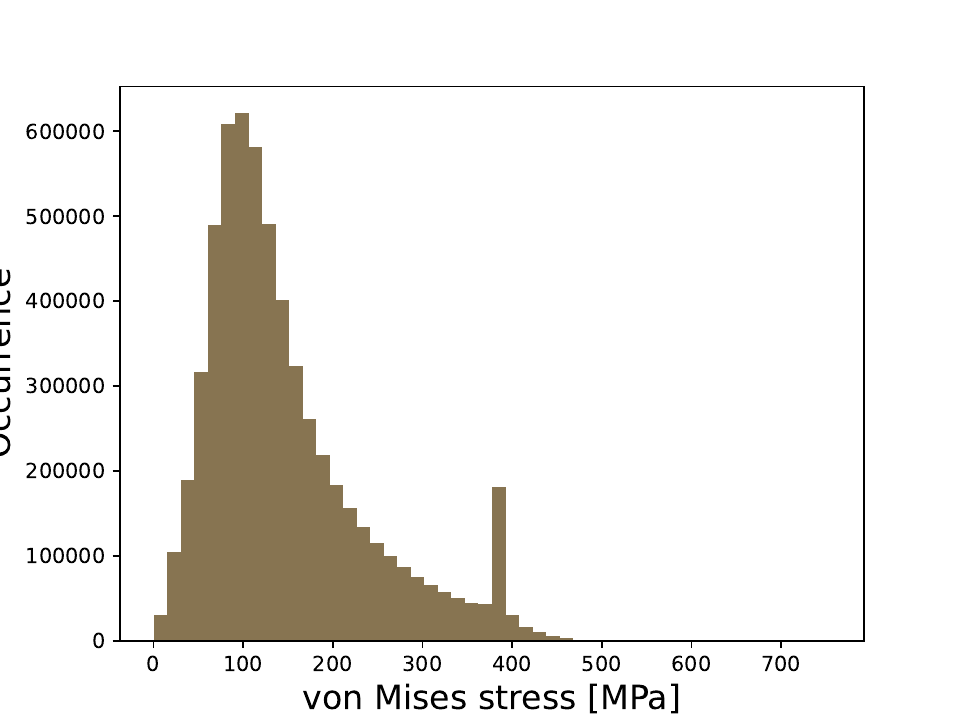}
        \label{p34}}
    \end{minipage} \\

    \begin{minipage}[c]{\x\textwidth}
       \centering 
        \subfloat[$N$=5000]{\includegraphics[trim={1cm 0cm 1cm 1cm},clip,width=\textwidth]{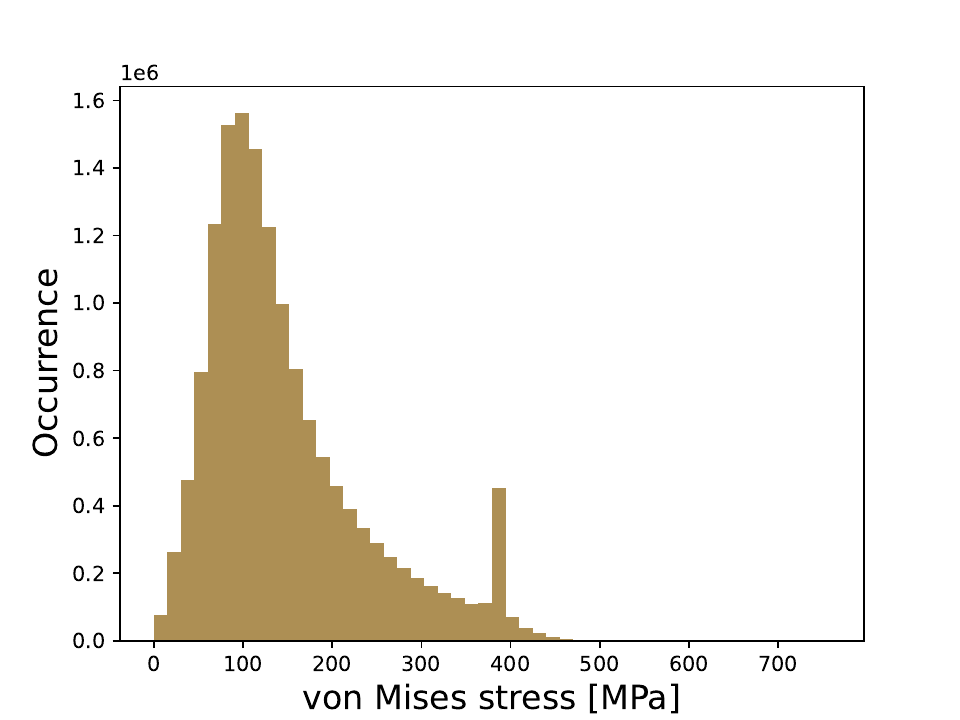}
        \label{p41}}
    \end{minipage} &
    \begin{minipage}[c]{\x\textwidth}
       \centering 
        \subfloat[$N$=10000]{\includegraphics[trim={1cm 0cm 1cm 1cm},clip,width=\textwidth]{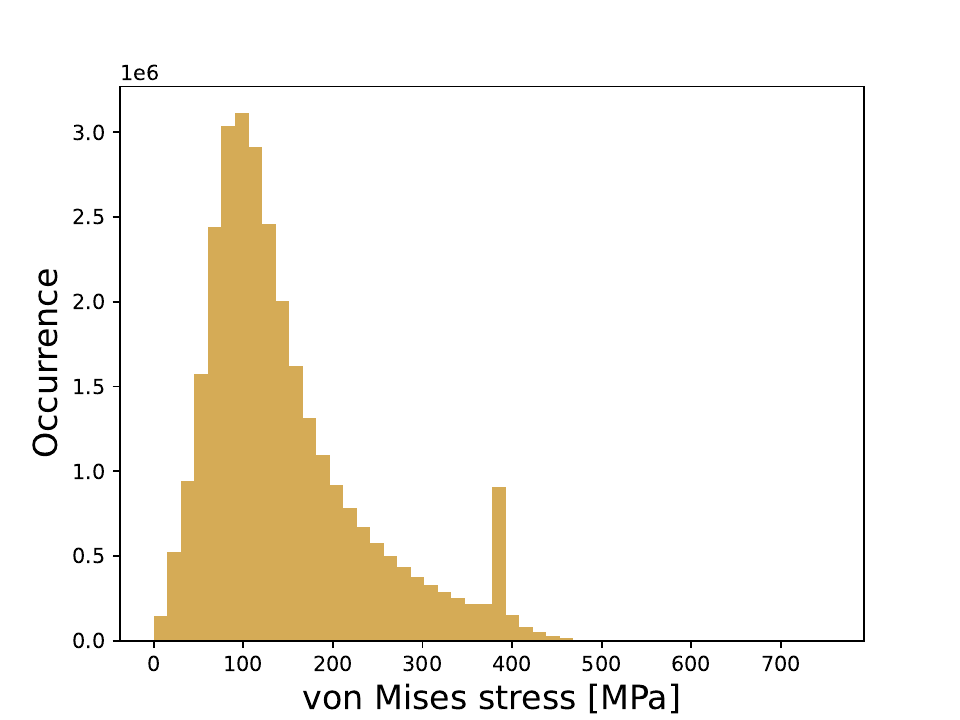}
        \label{p42}}
    \end{minipage} &
    \begin{minipage}[c]{\x\textwidth}
       \centering 
        \subfloat[$N$=25000]{\includegraphics[trim={1cm 0cm 1cm 1cm},clip,width=\textwidth]{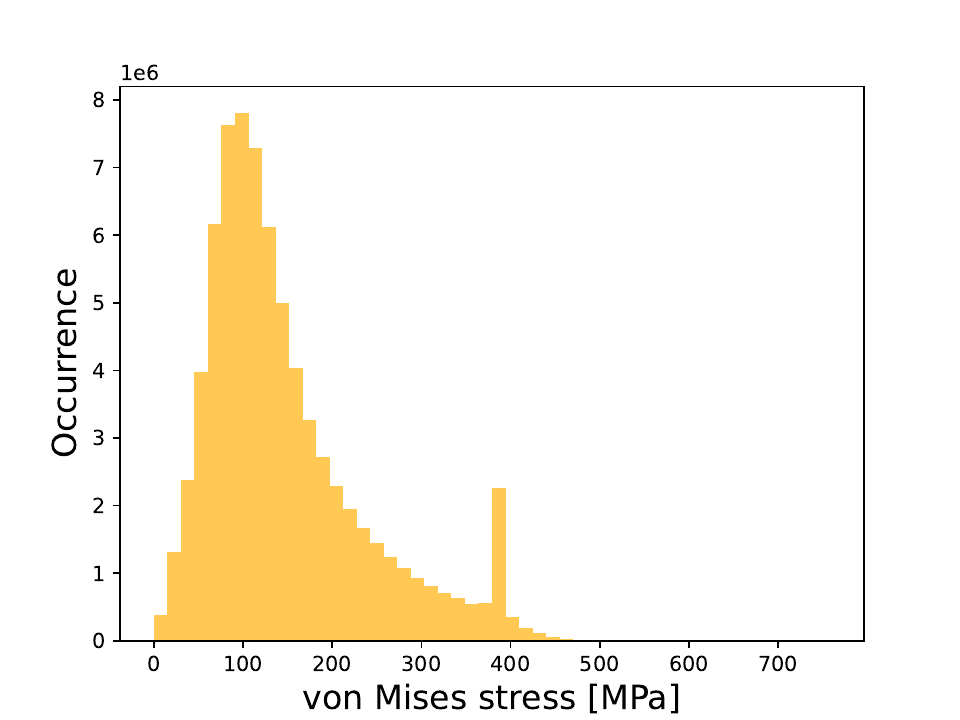}
        \label{p44}}
    \end{minipage} \\
    \end{tabular}
    \caption{Histograms of the output von Mises stress with different numbers of resampled points.}
    \label{data_hist}
\end{figure}

\begin{figure}[h!] 
    \centering
     \subfloat[]{
         \includegraphics[trim={0cm 0cm 0cm 0cm},clip,width=0.4\textwidth]{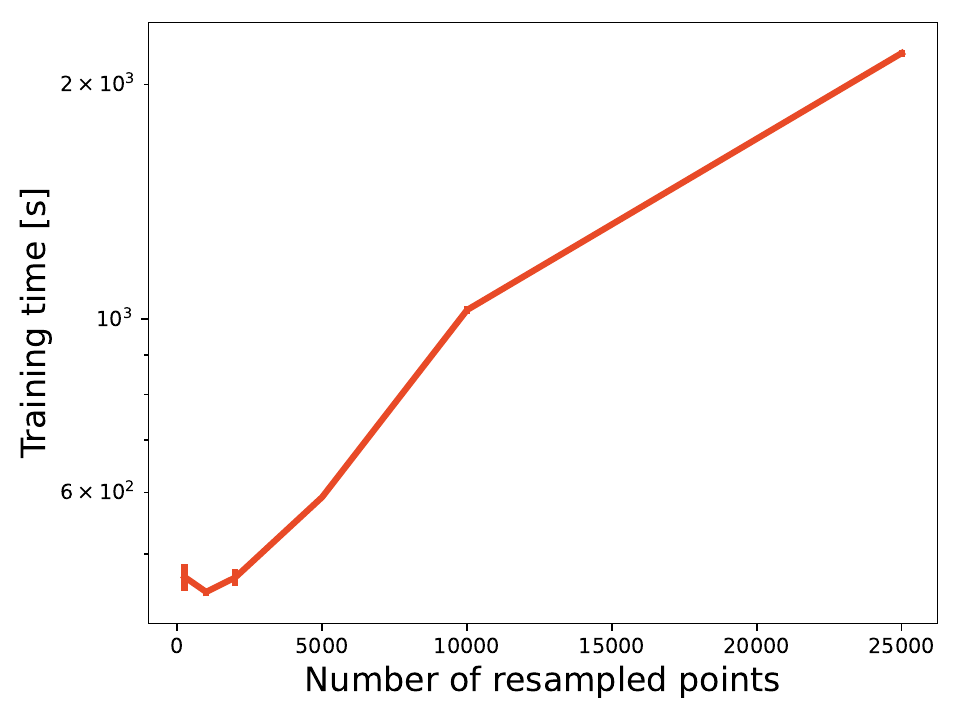}
         \label{ttvn}
     }
     \subfloat[]{
         \includegraphics[trim={0cm 0cm 0cm 0cm},clip,width=0.4\textwidth]{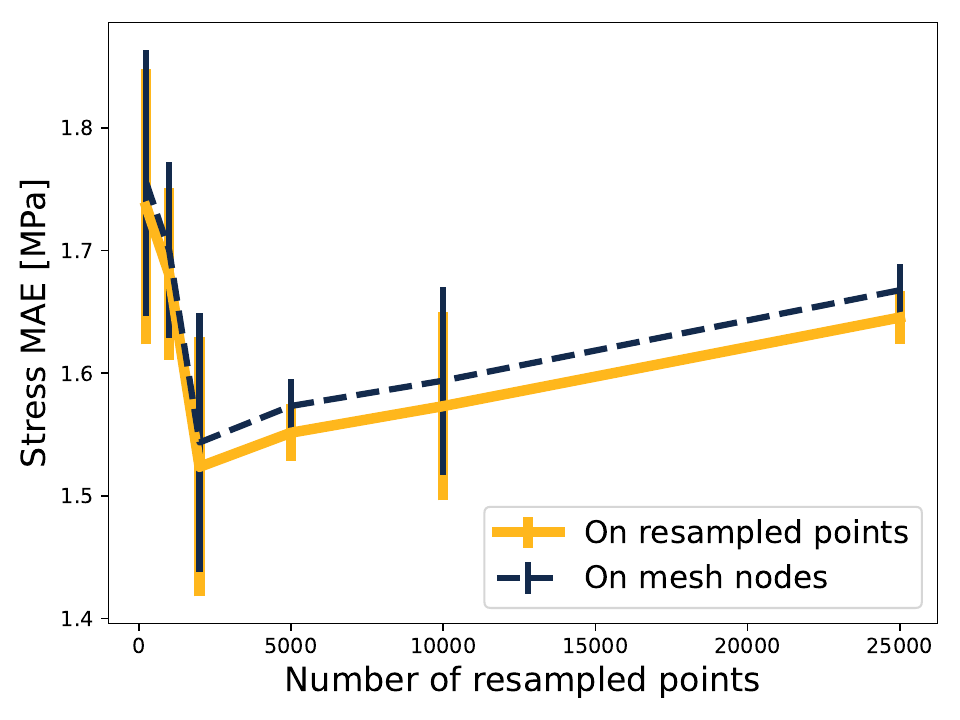}
         \label{maevn}
     }
    \caption{Plots of model training time and prediction error at different $N$ values.}
    \label{resample_err}
\end{figure}

As introduced in \sref{bench_mdls}, one of the critical steps in the vanilla DeepONet and Geom-DeepONet to allow for efficient batched training on variable geometries is to resample the input mesh nodes to a constant number ($N$) of points. Hence, the input nodes and output fields have identical shapes for all geometries. However, it is not immediately obvious what the value of $N$ should be and how that affects the prediction and generalization (from resampled points back to the full mesh) capabilities. To this end, 6 different $N$ values were studied; they are 250, 1000, 2000, 5000, 10000, and 25000. For each $N$ value, the random data resampling and Geom-DeepONet training were repeated 3 times to collect statistics. The Geom-DeepONet models were trained for 150000 iterations. Data from 2400 geometries are used in training, while 600 are reserved for training, leading to an 80/20 data split. First, it is of interest to compare the evolution of the output data distribution as $N$ increases. Histograms of the resampled data at different $N$ values are shown in \fref{data_hist}. To evaluate model performance, MAE in von Mises stress was evaluated on the resampled points in the test set and all mesh nodes of the corresponding geometry. The model training time and stress MAE are depicted in \fref{resample_err}.

From \fref{data_hist}, it is obvious that distributions of the output stress do not vary significantly with the number of resampled points, yielding similar distributions once more than 1000 points were randomly picked from all mesh nodes of each geometry. However, \fref{maevn} clearly shows that although the data distributions are similar, the $N$ value can significantly influence the model performance. A general observation is that the prediction errors on the full mesh nodes are larger than those on the resampled set. This is reasonable as the resampling does not cover all the possible locations where a high stress value may occur. However, we noticed that the difference in prediction errors between the resampled points and the mesh nodes is small, indicating that a model trained \emph{only} on a subset of nodes can indeed generalize predictions to all nodes in the full mesh, although the node counts are distinct for each case, a powerful finding of the current study. More importantly, the results indicate that an optimal number of $N$ exists and is around 2000 points for the beam problem. However, the variability of the model performance is relatively high at $N$=2000, while $N$=5000 seems to balance model prediction accuracy, performance variability, and training time. However, it is worth mentioning that the optimal value of $N$ is likely problem-dependent and should be fine-tuned for each specific problem for best performance. To demonstrate baseline performance without any problem-specific hyper-parameter optimization, $N$=5000 is used for all subsequent model training presented in this work. We also highlight that \emph{all} meshes in the beam geometries have more than 5000 nodes, indicating that the Geom-DeepONet is extremely data efficient in terms of capturing the variable geometries with a sample point cloud.

\subsection{Comparison with PointNet and vanilla DeepONet}
\label{benchmark}
\begin{figure}[h!] 
    \centering
     \subfloat[PointNet]{
         \includegraphics[trim={0cm 0cm 0cm 0cm},clip,width=0.31\textwidth]{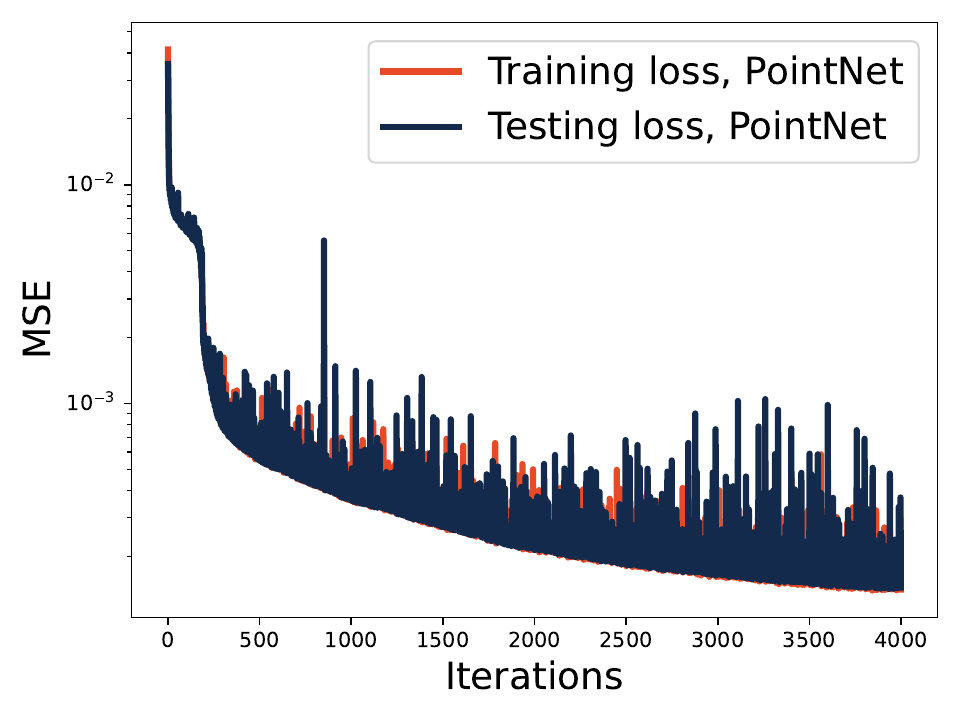}
         \label{pn_tr}
     }
     \subfloat[Vanilla DeepONet]{
         \includegraphics[trim={0cm 0cm 0cm 0cm},clip,width=0.31\textwidth]{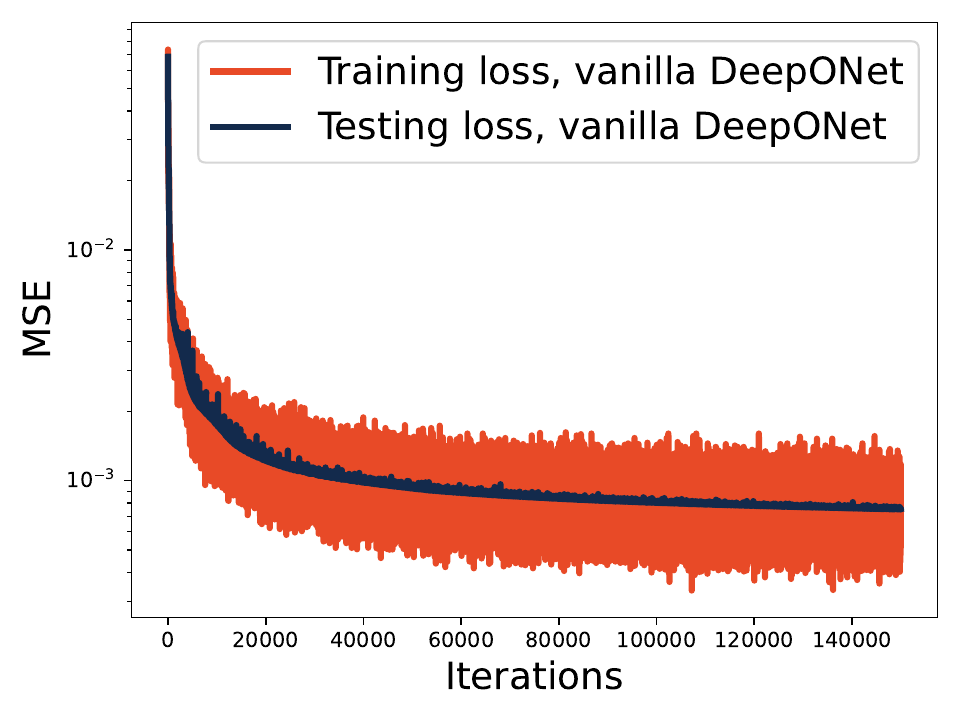}
         \label{vd_tr}
     }
     \subfloat[Geom-DeepONet]{
         \includegraphics[trim={0cm 0cm 0cm 0cm},clip,width=0.31\textwidth]{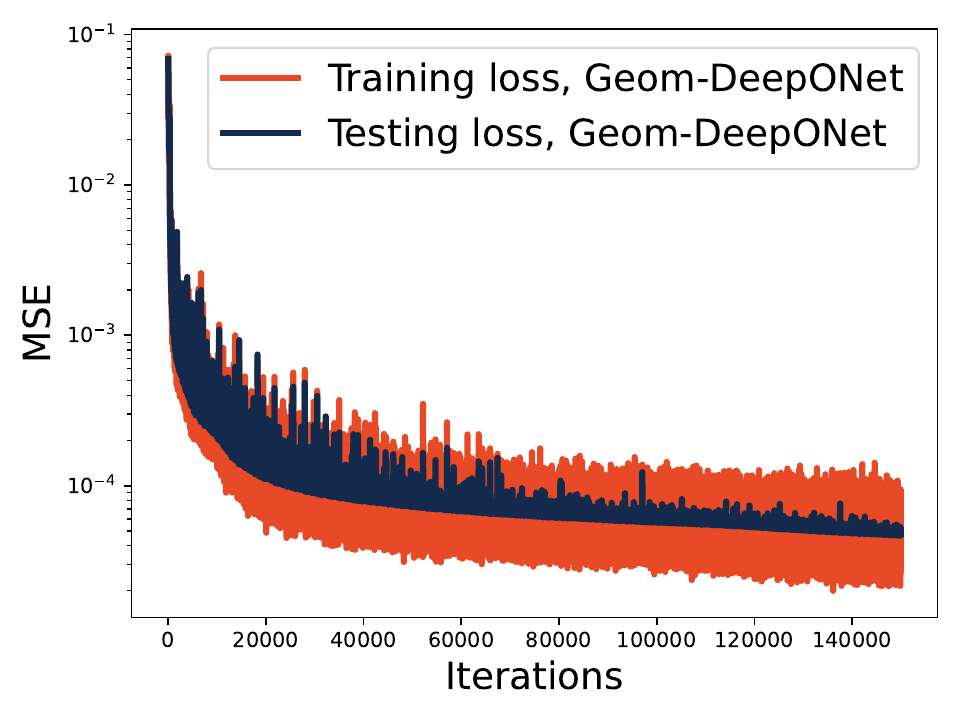}
         \label{gd_tr}
     }
    \caption{Training and testing loss history for different models.}
    \label{train_history}
\end{figure}

In this example, we compare the performance of PointNet, vanilla DeepONet, and the proposed Geom-DeepONet using the parametric beam dataset. Each model was trained 3 times with randomly generated 80/20 data split and random data resampling to obtain statistics. The PointNet was trained for 4000 iterations with a batch size of 32 (i.e., following the original implementation as in \cite{point-cfd}). The two DeepONet models were trained for 150000 iterations with a batch size of 16. The training histories for the three models are depicted in \fref{train_history}.

\begin{figure}[h!]
\newcommand\x{0.2}
    \centering
    \begin{tabular}{ c c c c }
    \begin{minipage}[c]{\x\textwidth}
       \centering 
        \subfloat[FE, best]{\includegraphics[trim={1.5cm 0cm 1.8cm 3cm},clip,width=\textwidth]{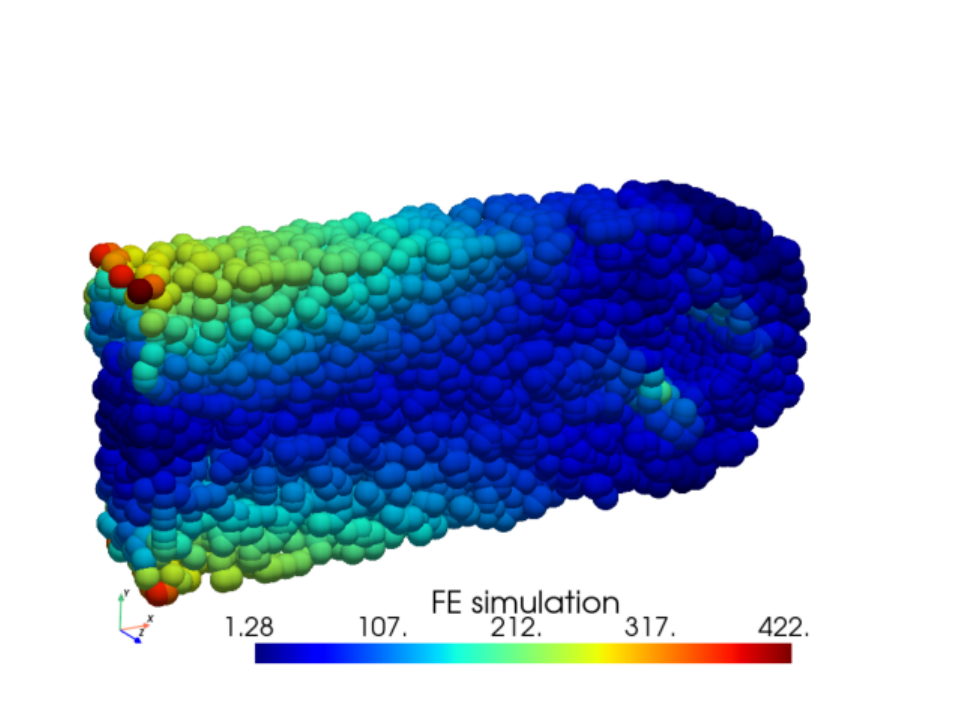}
        }
    \end{minipage} &
    \begin{minipage}[c]{\x\textwidth}
       \centering 
        \subfloat[FE, 50$^{th}$ pct.]{\includegraphics[trim={1.5cm 0cm 1.8cm 3cm},clip,width=\textwidth]{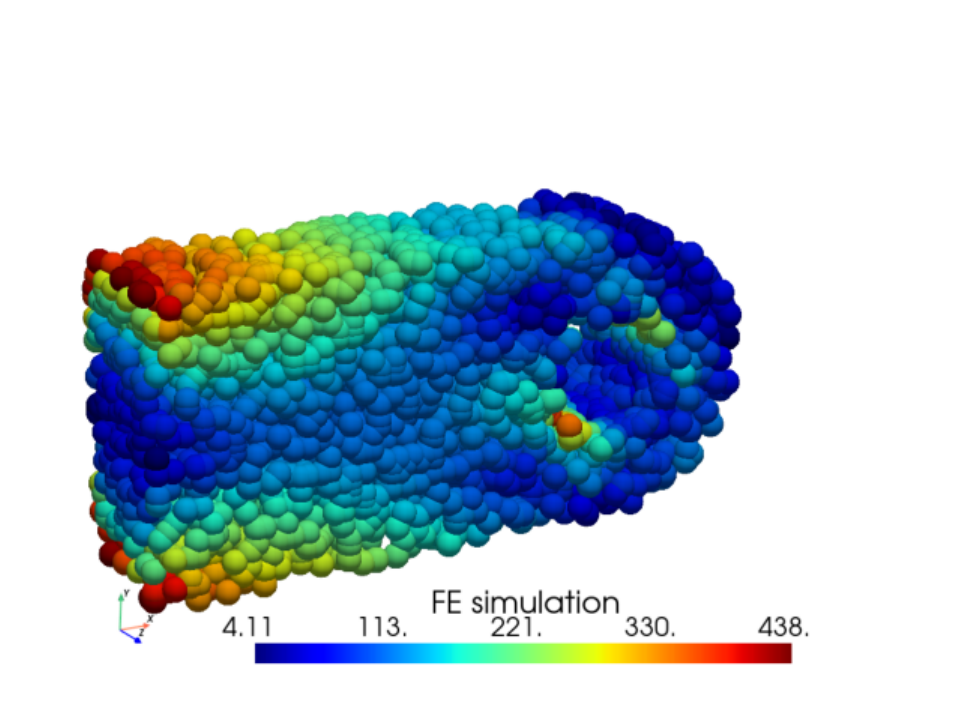}
        }
    \end{minipage} &
    \begin{minipage}[c]{\x\textwidth}
       \centering 
        \subfloat[FE, 75$^{th}$ pct.]{\includegraphics[trim={1.5cm 0cm 1.8cm 3cm},clip,width=\textwidth]{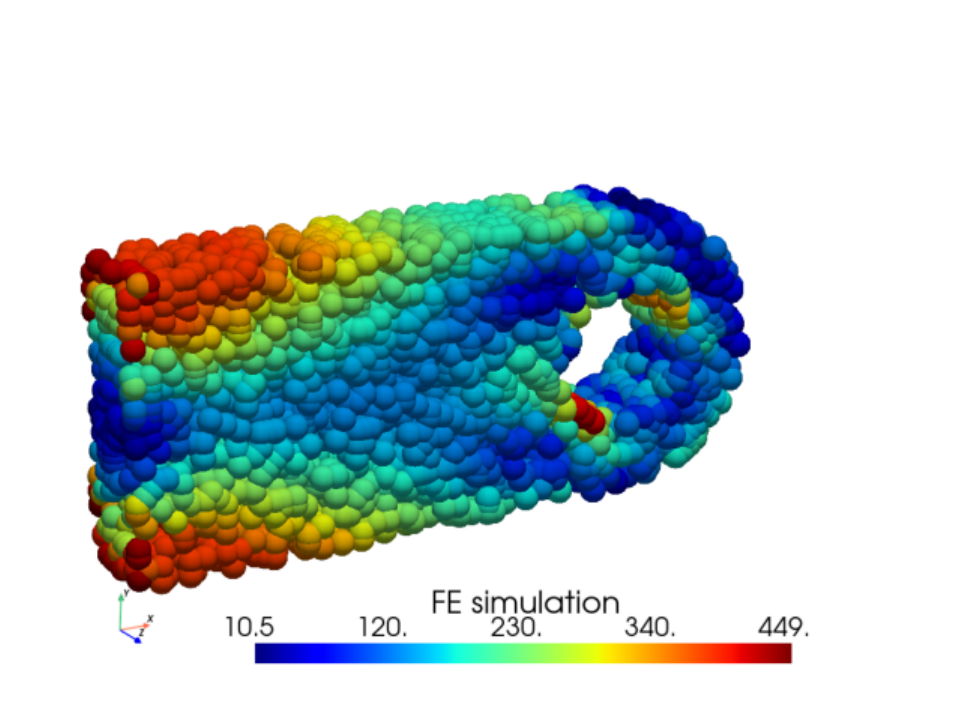}
        }
    \end{minipage} &
    \begin{minipage}[c]{\x\textwidth}
       \centering 
        \subfloat[FE, worst]{\includegraphics[trim={1.5cm 0cm 1.8cm 3cm},clip,width=\textwidth]{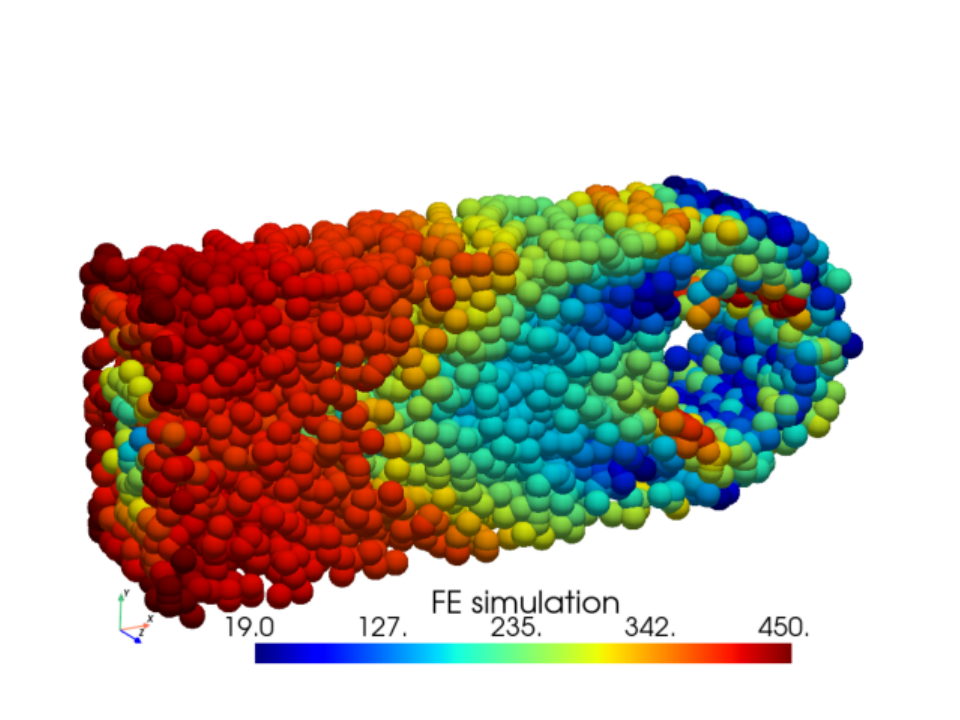}
        }
    \end{minipage} \\

    \begin{minipage}[c]{\x\textwidth}
       \centering 
        \subfloat[Pred., best]{\includegraphics[trim={1.5cm 0cm 1.8cm 3cm},clip,width=\textwidth]{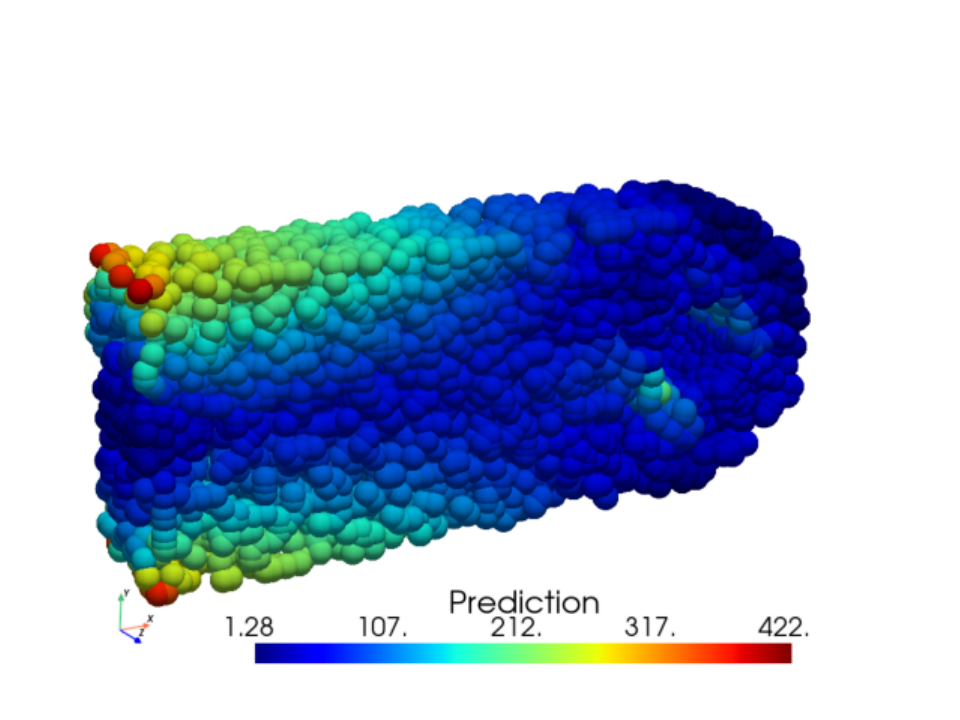}
        }
    \end{minipage} &
    \begin{minipage}[c]{\x\textwidth}
       \centering 
        \subfloat[Pred., 50$^{th}$ pct.]{\includegraphics[trim={1.5cm 0cm 1.8cm 3cm},clip,width=\textwidth]{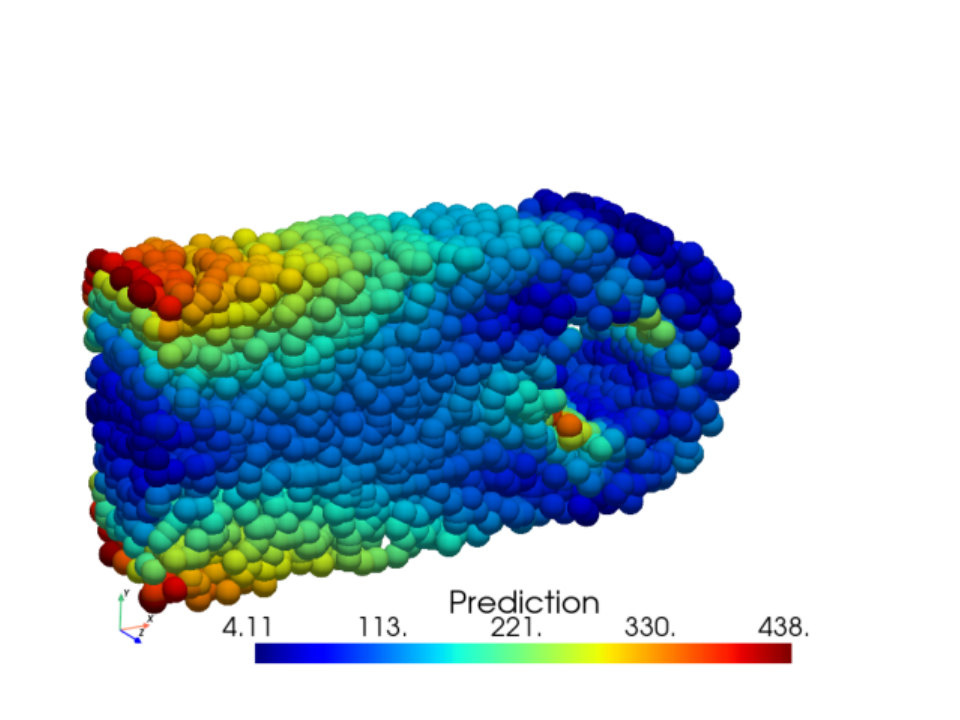}
        }
    \end{minipage} &
    \begin{minipage}[c]{\x\textwidth}
       \centering 
        \subfloat[Pred., 75$^{th}$ pct.]{\includegraphics[trim={1.5cm 0cm 1.8cm 3cm},clip,width=\textwidth]{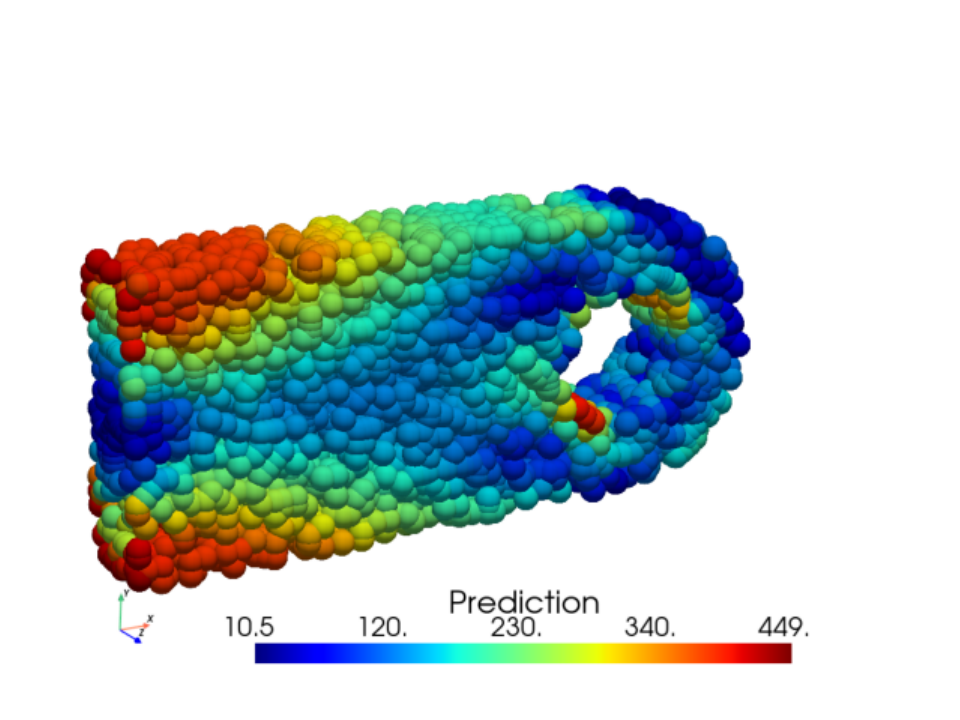}
        }
    \end{minipage} &
    \begin{minipage}[c]{\x\textwidth}
       \centering 
        \subfloat[Pred., worst]{\includegraphics[trim={1.5cm 0cm 1.8cm 3cm},clip,width=\textwidth]{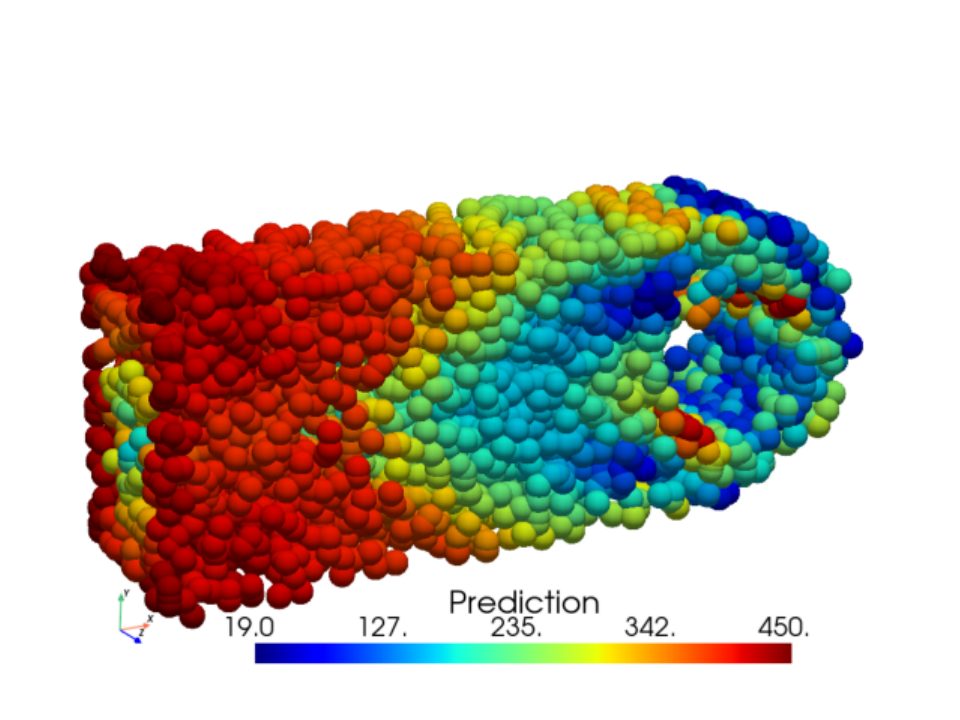}
        }
    \end{minipage} \\

    \begin{minipage}[c]{\x\textwidth}
       \centering 
        \subfloat[MAE, best]{\includegraphics[trim={1.5cm 0cm 1.8cm 2cm},clip,width=\textwidth]{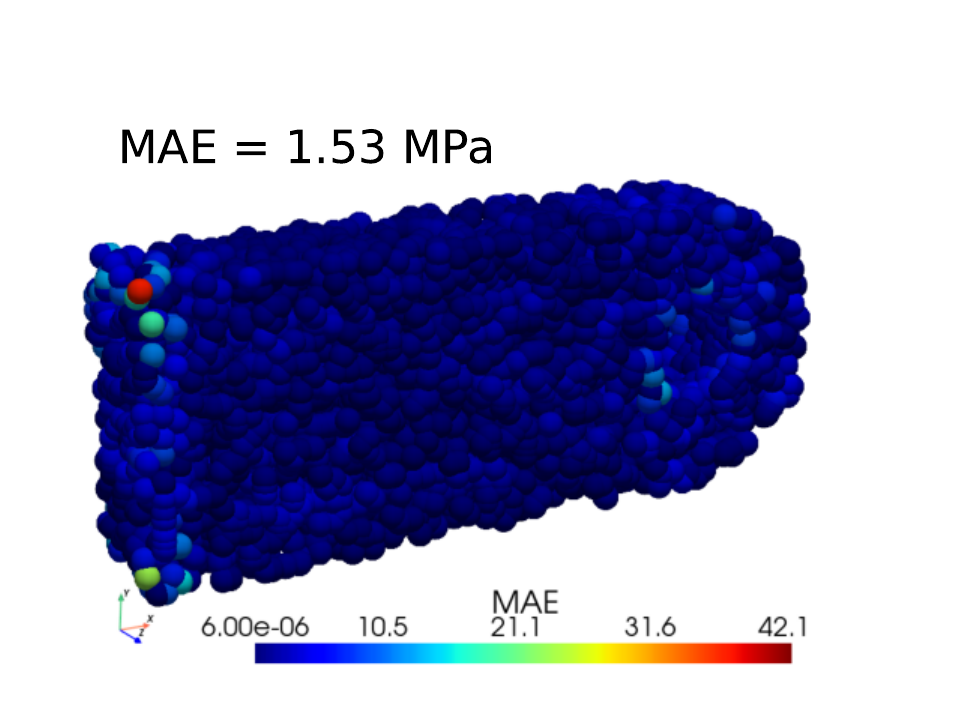}
        }
    \end{minipage} &
    \begin{minipage}[c]{\x\textwidth}
       \centering 
        \subfloat[MAE, 50$^{th}$ pct.]{\includegraphics[trim={1.5cm 0cm 1.8cm 2cm},clip,width=\textwidth]{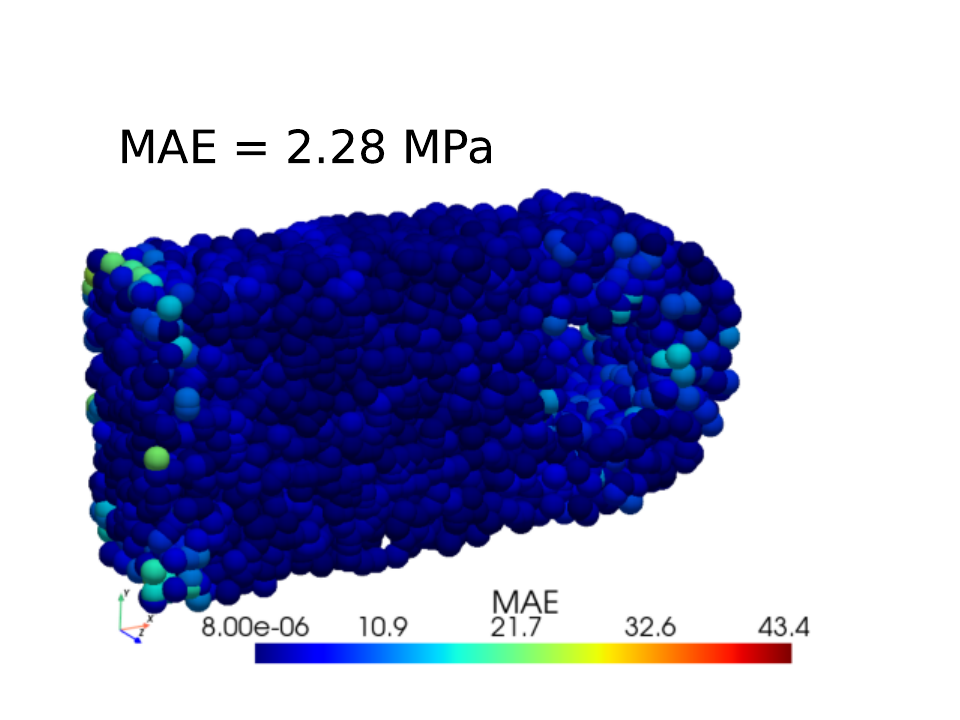}
        }
    \end{minipage} &
    \begin{minipage}[c]{\x\textwidth}
       \centering 
        \subfloat[MAE, 75$^{th}$ pct.]{\includegraphics[trim={1.5cm 0cm 1.8cm 2cm},clip,width=\textwidth]{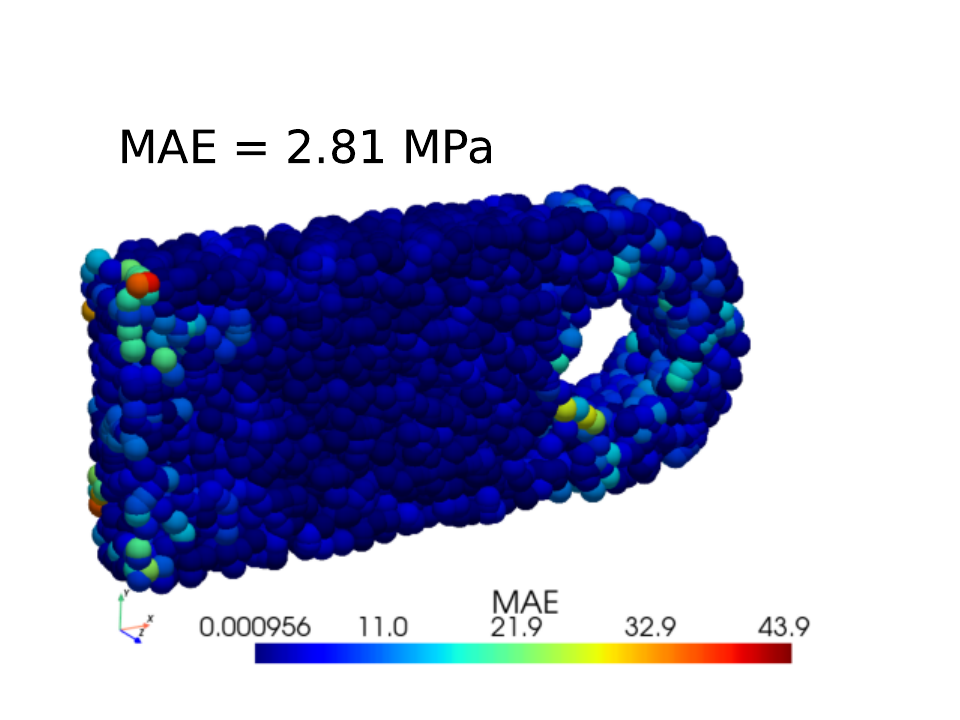}
        }
    \end{minipage} &
    \begin{minipage}[c]{\x\textwidth}
       \centering 
        \subfloat[MAE, worst]{\includegraphics[trim={1.5cm 0cm 1.8cm 2cm},clip,width=\textwidth]{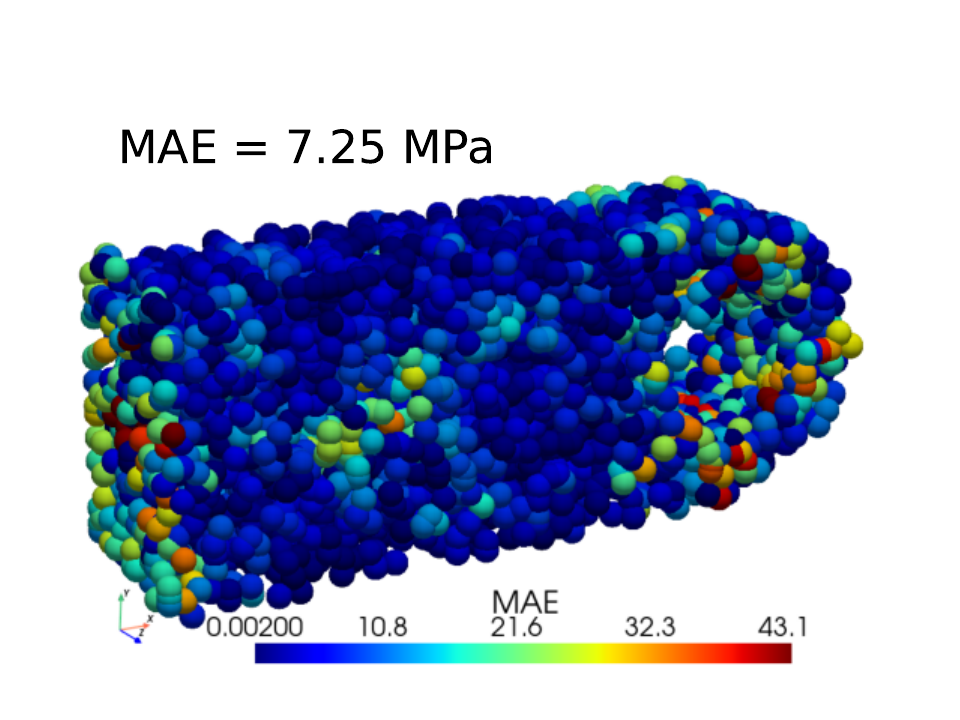}
        }
    \end{minipage} \\

    \end{tabular}
    \caption{Volume predictions by PointNet, ranked by different percentiles in stress MAE. The first and second rows show the FE ground truth and the model predictions, and they share identical color scales. The third row shows the MAE, and the color range is set to $1/10$ of that in the FE ground truth. The 5000 points in each test case are rendered as spheres to create a volume plot.}
    \label{point_net_pred}
\end{figure}

\begin{figure}[h!]
\newcommand\x{0.2}
    \centering
    \begin{tabular}{ c c c c }
    \begin{minipage}[c]{\x\textwidth}
       \centering 
        \subfloat[FE, best]{\includegraphics[trim={1.5cm 0cm 1.3cm 3cm},clip,width=\textwidth]{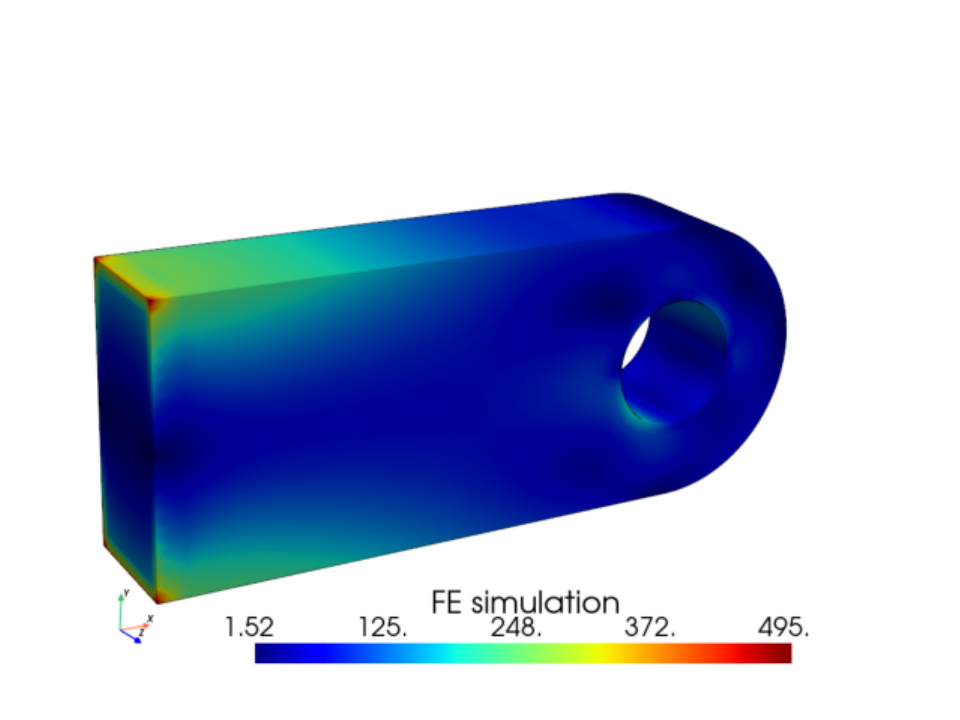}
        }
    \end{minipage} &
    \begin{minipage}[c]{\x\textwidth}
       \centering 
        \subfloat[FE, 50$^{th}$ pct.]{\includegraphics[trim={1.5cm 0cm 1.3cm 3cm},clip,width=\textwidth]{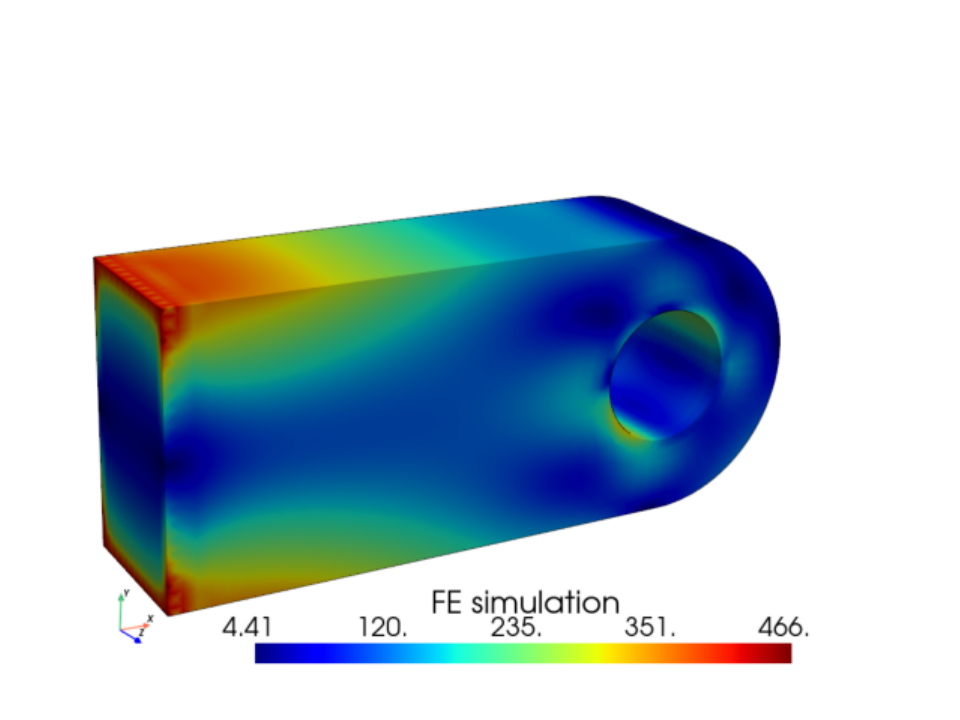}
        }
    \end{minipage} &
    \begin{minipage}[c]{\x\textwidth}
       \centering 
        \subfloat[FE, 75$^{th}$ pct.]{\includegraphics[trim={1.5cm 0cm 1.3cm 3cm},clip,width=\textwidth]{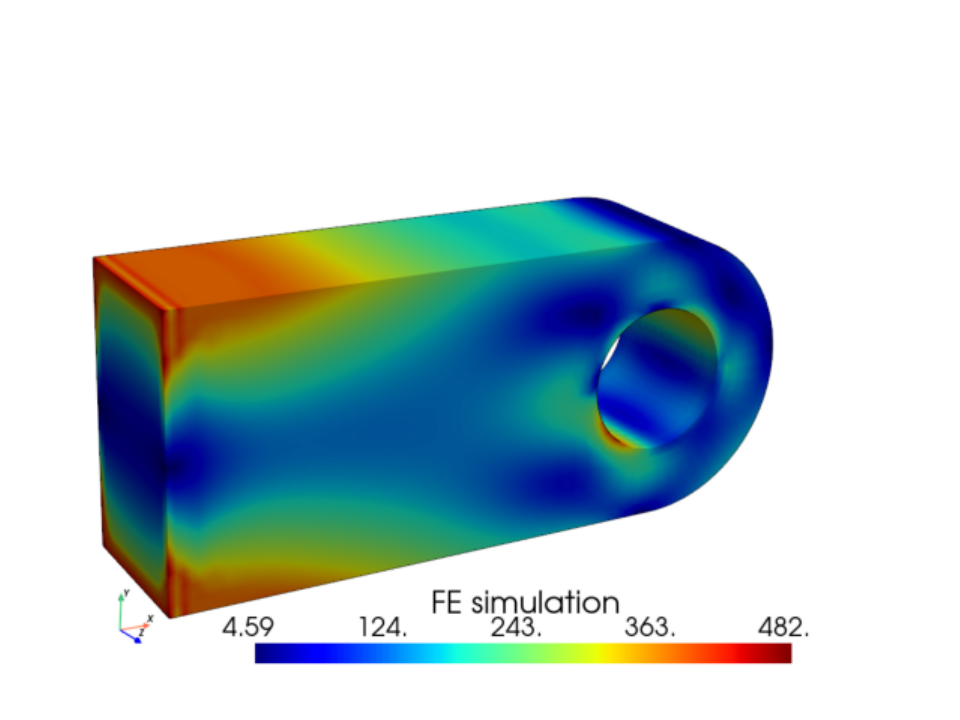}
        }
    \end{minipage} &
    \begin{minipage}[c]{\x\textwidth}
       \centering 
        \subfloat[FE, worst]{\includegraphics[trim={1.5cm 0cm 1.3cm 3cm},clip,width=\textwidth]{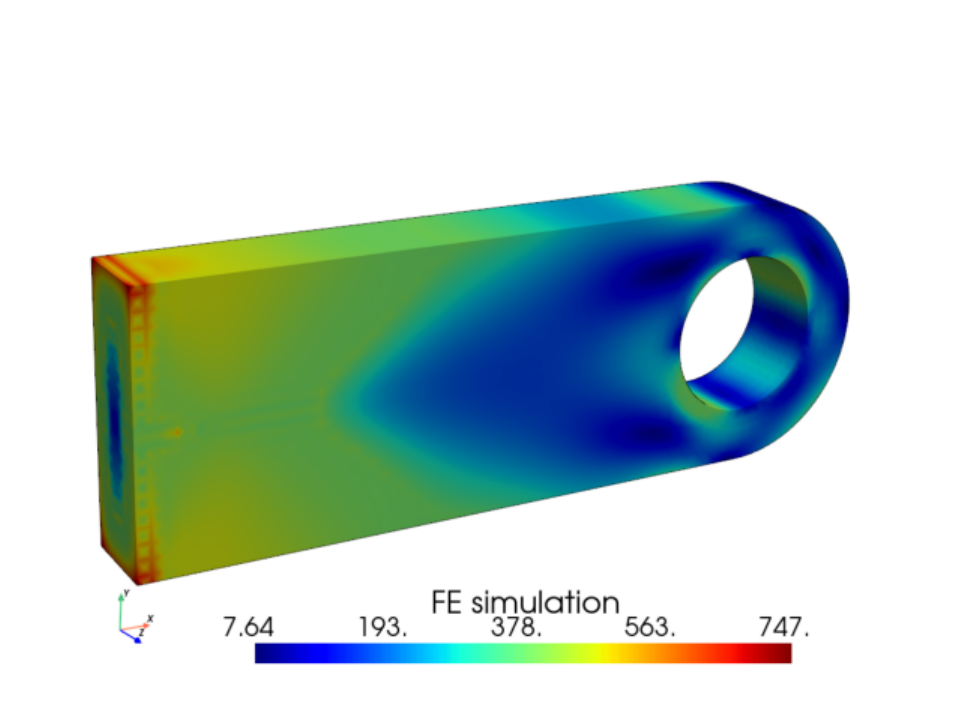}
        }
    \end{minipage} \\

    \begin{minipage}[c]{\x\textwidth}
       \centering 
        \subfloat[Pred., best]{\includegraphics[trim={1.5cm 0cm 1.3cm 3cm},clip,width=\textwidth]{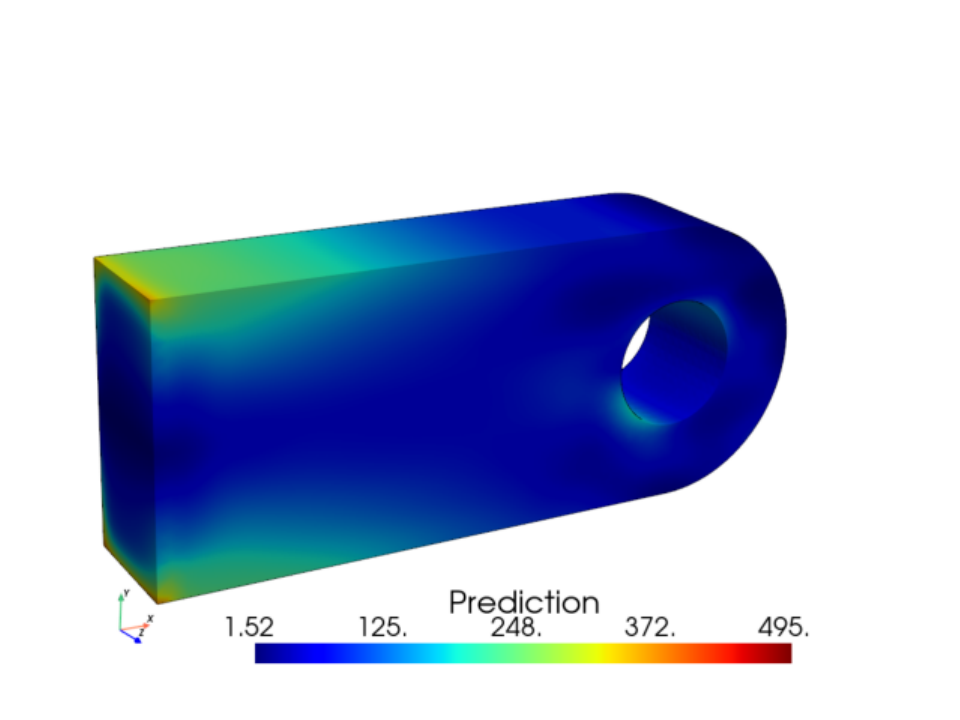}
        }
    \end{minipage} &
    \begin{minipage}[c]{\x\textwidth}
       \centering 
        \subfloat[Pred., 50$^{th}$ pct.]{\includegraphics[trim={1.5cm 0cm 1.3cm 3cm},clip,width=\textwidth]{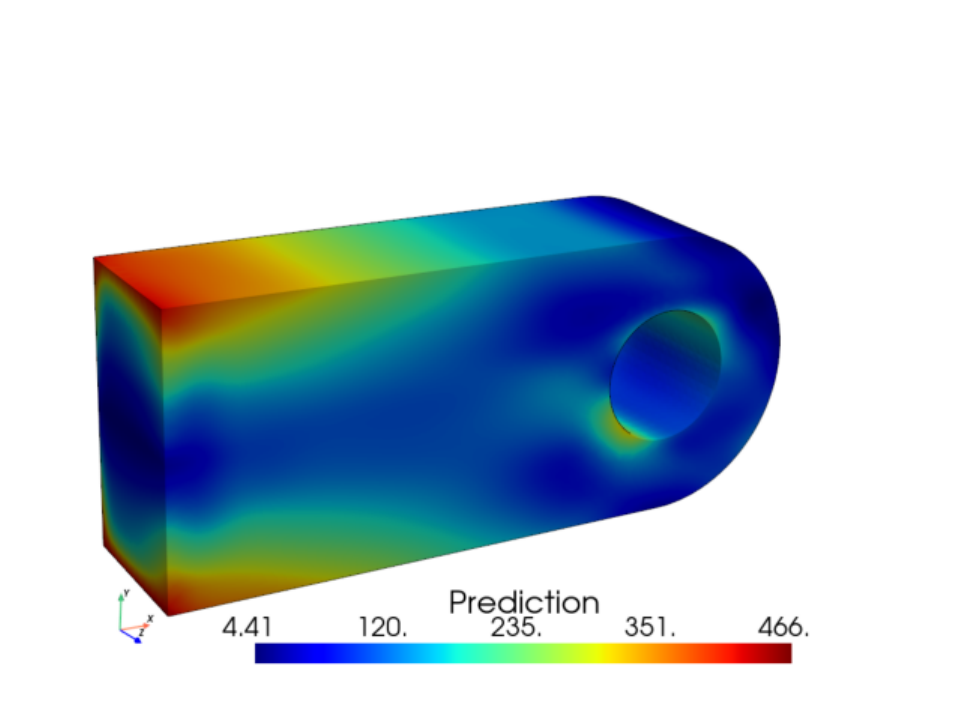}
        }
    \end{minipage} &
    \begin{minipage}[c]{\x\textwidth}
       \centering 
        \subfloat[Pred., 75$^{th}$ pct.]{\includegraphics[trim={1.5cm 0cm 1.3cm 3cm},clip,width=\textwidth]{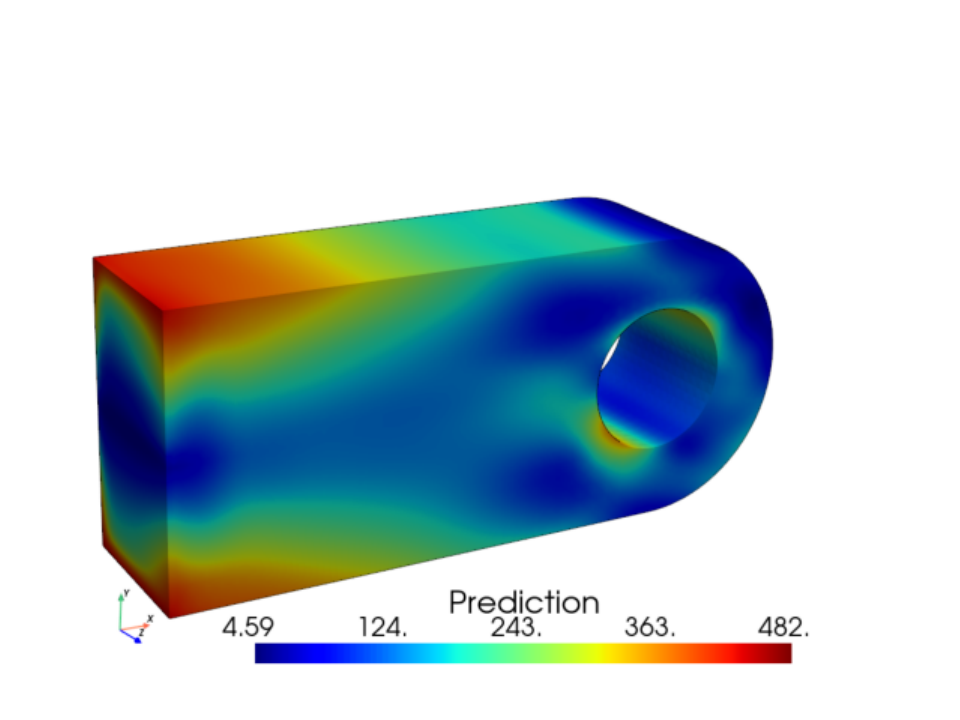}
        }
    \end{minipage} &
    \begin{minipage}[c]{\x\textwidth}
       \centering 
        \subfloat[Pred., worst]{\includegraphics[trim={1.5cm 0cm 1.3cm 3cm},clip,width=\textwidth]{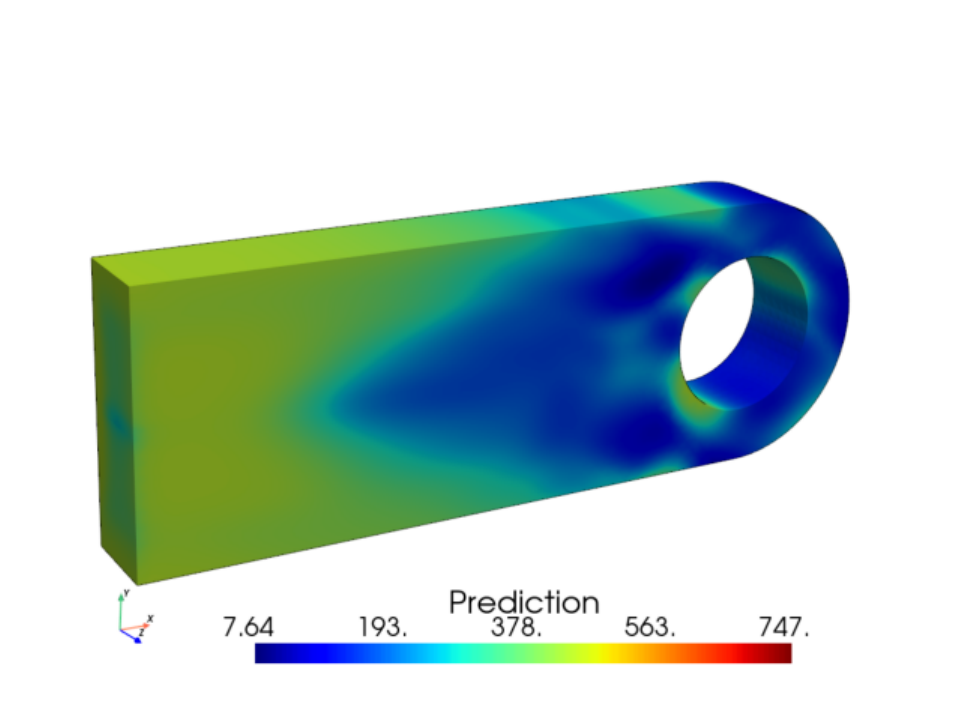}
        }
    \end{minipage} \\

    \begin{minipage}[c]{\x\textwidth}
       \centering 
        \subfloat[MAE, best]{\includegraphics[trim={1.5cm 0cm 1.3cm 2cm},clip,width=\textwidth]{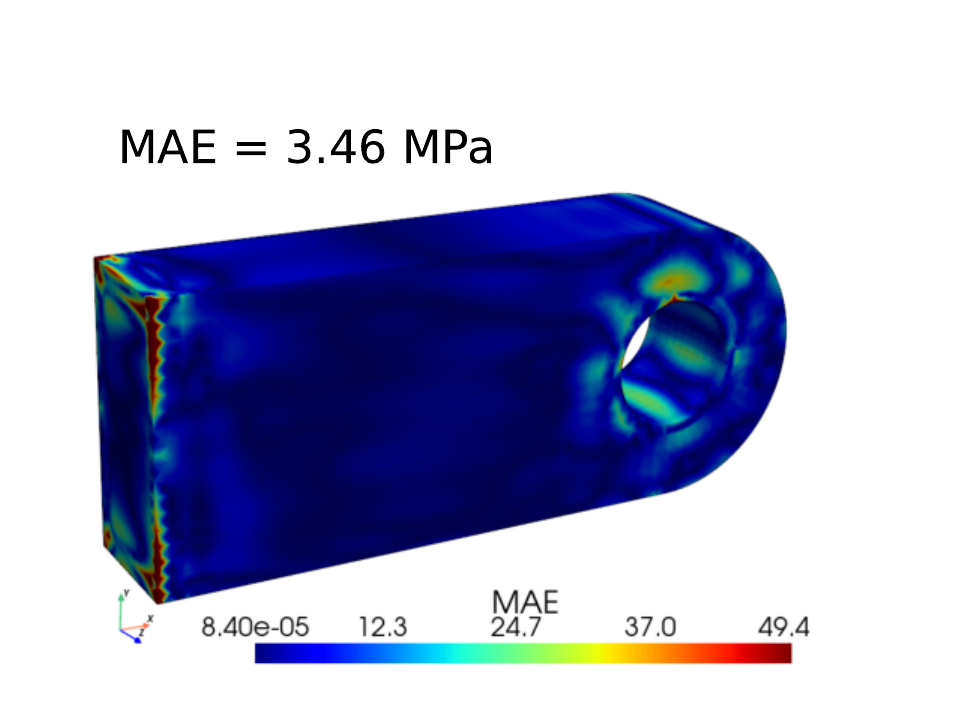}
        }
    \end{minipage} &
    \begin{minipage}[c]{\x\textwidth}
       \centering 
        \subfloat[MAE, 50$^{th}$ pct.]{\includegraphics[trim={1.5cm 0cm 1.3cm 2cm},clip,width=\textwidth]{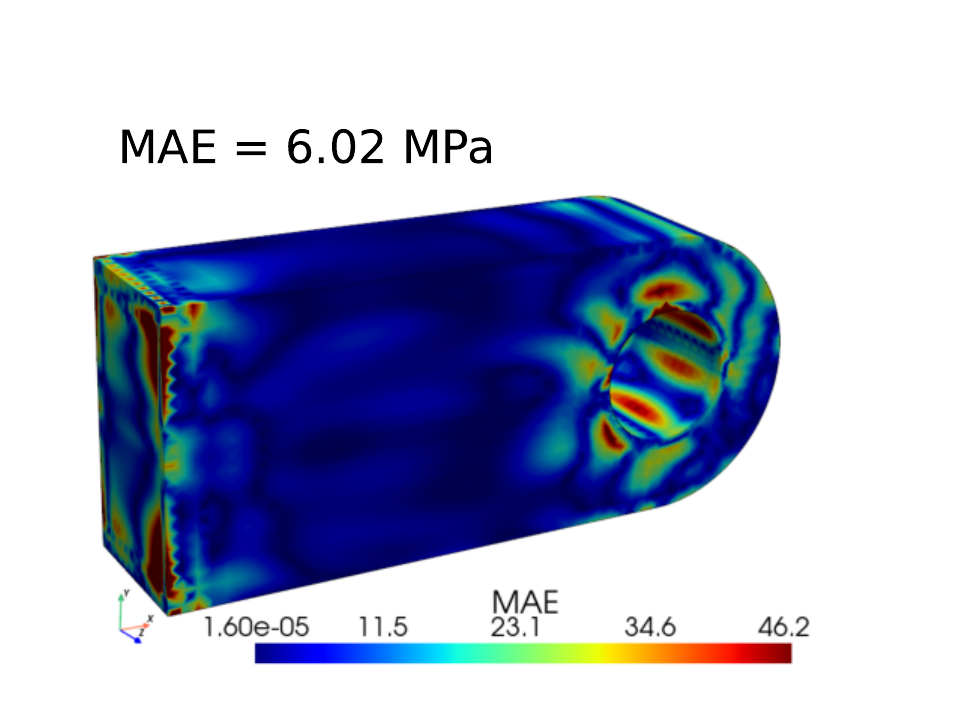}
        }
    \end{minipage} &
    \begin{minipage}[c]{\x\textwidth}
       \centering 
        \subfloat[MAE, 75$^{th}$ pct.]{\includegraphics[trim={1.5cm 0cm 1.3cm 2cm},clip,width=\textwidth]{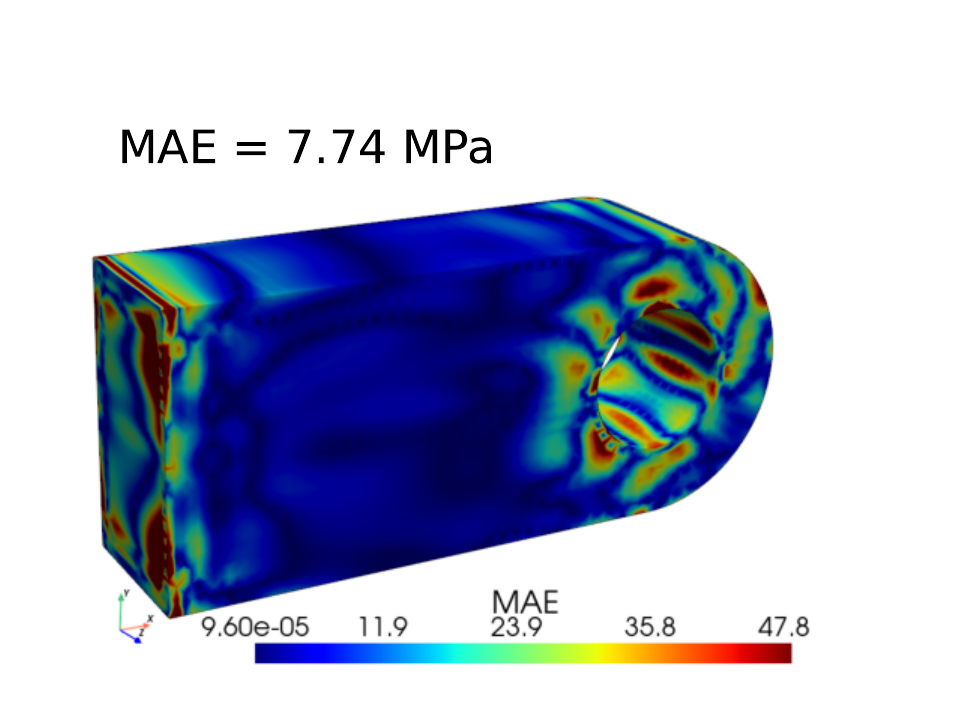}
        }
    \end{minipage} &
    \begin{minipage}[c]{\x\textwidth}
       \centering 
        \subfloat[MAE, worst]{\includegraphics[trim={1.5cm 0cm 1.3cm 2cm},clip,width=\textwidth]{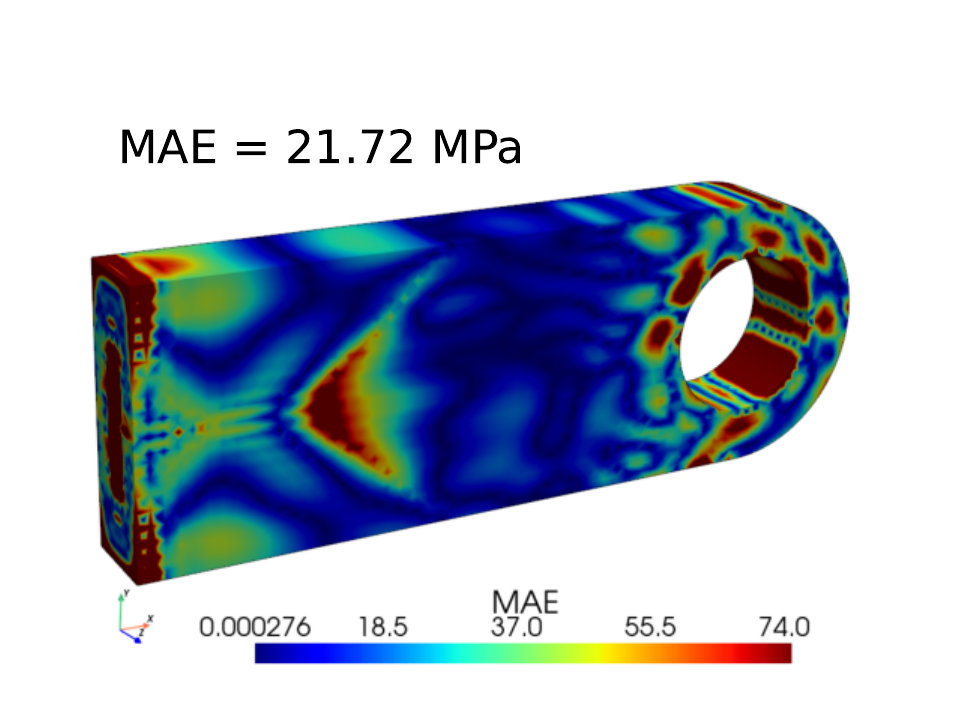}
        }
    \end{minipage} \\

    \end{tabular}
    \caption{Volume predictions by the vanilla DeepONet, ranked by different percentiles in stress MAE. The first and second rows show the FE ground truth and the model predictions, and they share identical color scales. The third row shows the MAE and the color range is set to $1/10$ of that in the FE ground truth to make any error concentration visible.}
    \label{don_pred}
\end{figure}

\begin{figure}[h!]
\newcommand\x{0.2}
    \centering
    \begin{tabular}{ c c c c }
    \begin{minipage}[c]{\x\textwidth}
       \centering 
        \subfloat[FE, best]{\includegraphics[trim={1.5cm 0cm 1.3cm 3cm},clip,width=\textwidth]{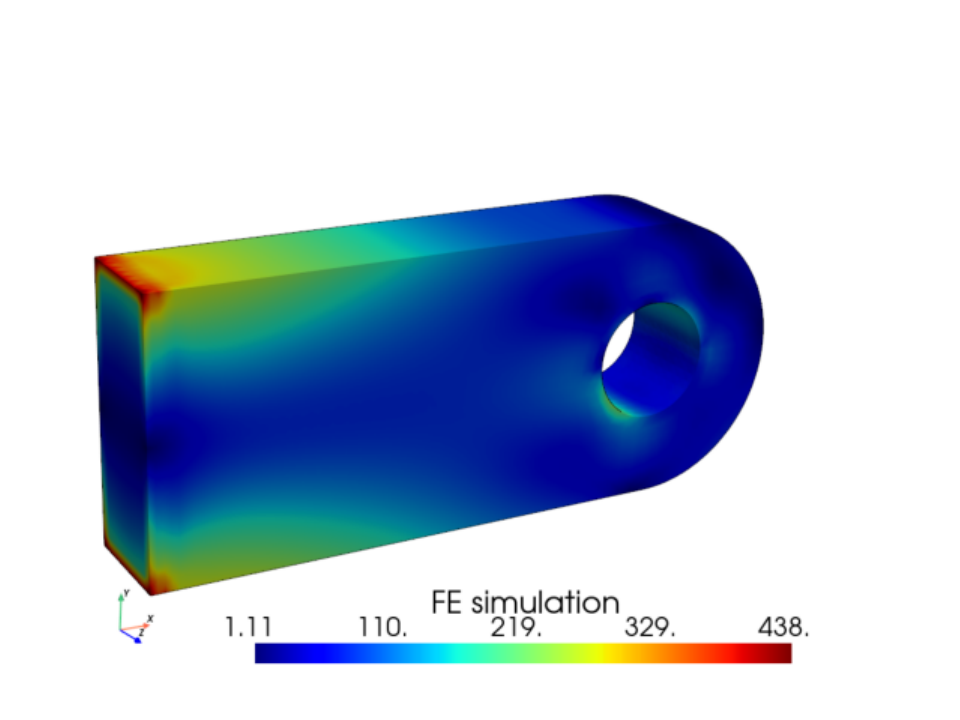}
        }
    \end{minipage} &
    \begin{minipage}[c]{\x\textwidth}
       \centering 
        \subfloat[FE, 50$^{th}$ pct.]{\includegraphics[trim={1.5cm 0cm 1.3cm 3cm},clip,width=\textwidth]{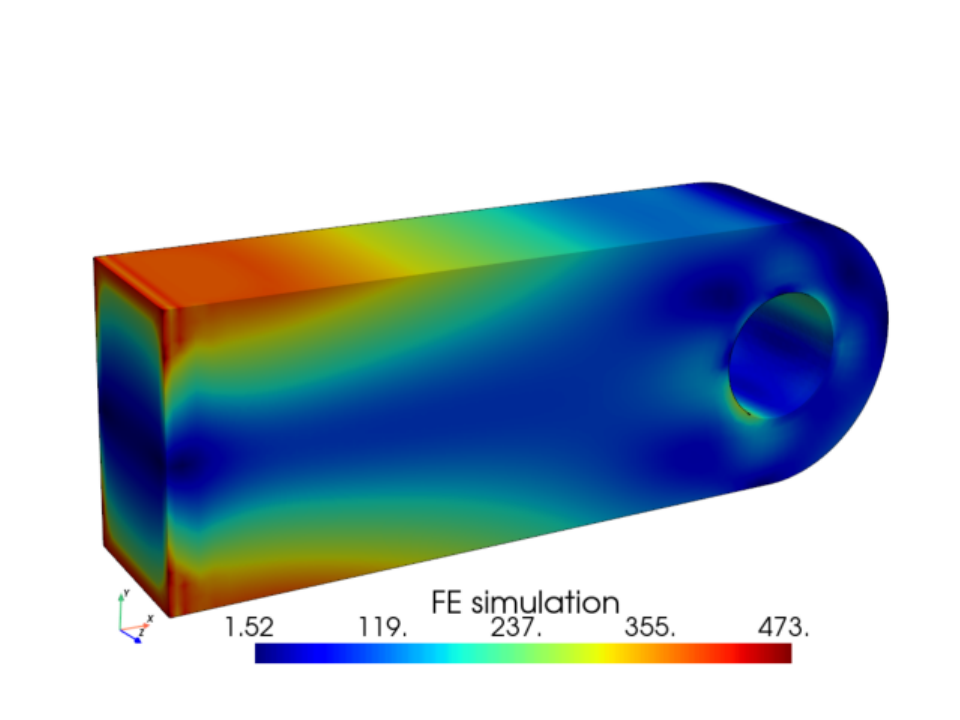}
        }
    \end{minipage} &
    \begin{minipage}[c]{\x\textwidth}
       \centering 
        \subfloat[FE, 75$^{th}$ pct.]{\includegraphics[trim={1.5cm 0cm 1.3cm 3cm},clip,width=\textwidth]{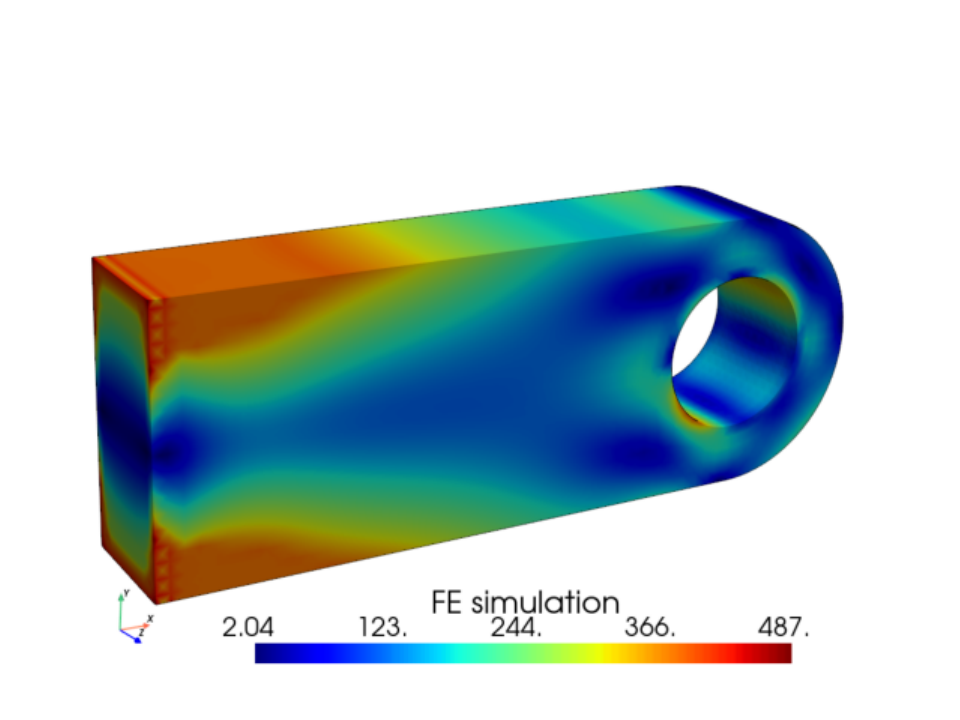}
        }
    \end{minipage} &
    \begin{minipage}[c]{\x\textwidth}
       \centering 
        \subfloat[FE, worst]{\includegraphics[trim={1.5cm 0cm 1.3cm 3cm},clip,width=\textwidth]{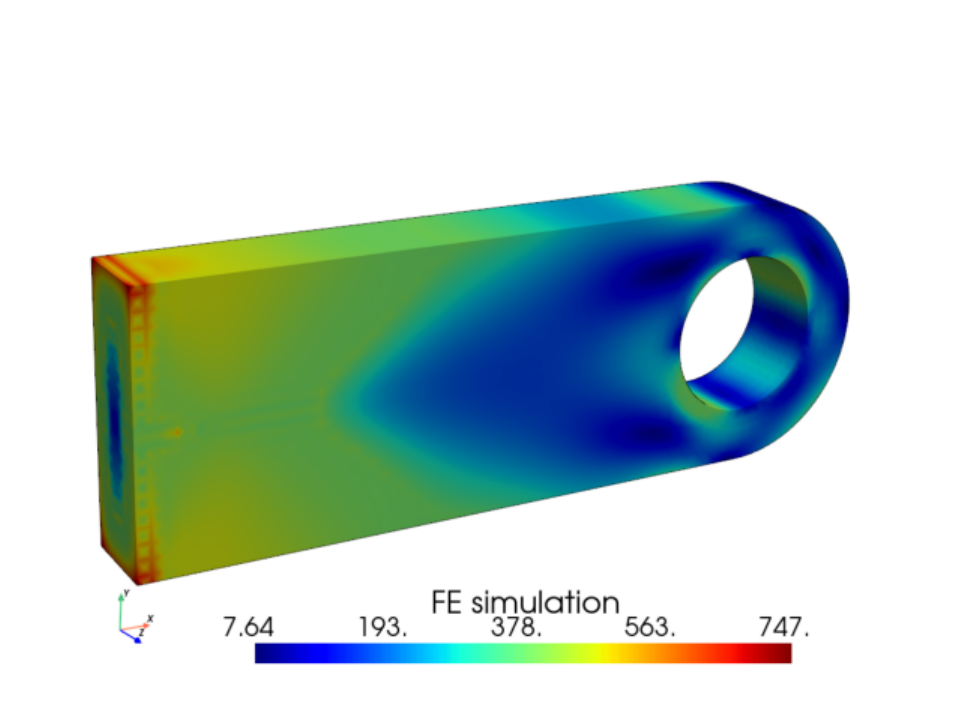}
        }
    \end{minipage} \\

    \begin{minipage}[c]{\x\textwidth}
       \centering 
        \subfloat[Pred., best]{\includegraphics[trim={1.5cm 0cm 1.3cm 3cm},clip,width=\textwidth]{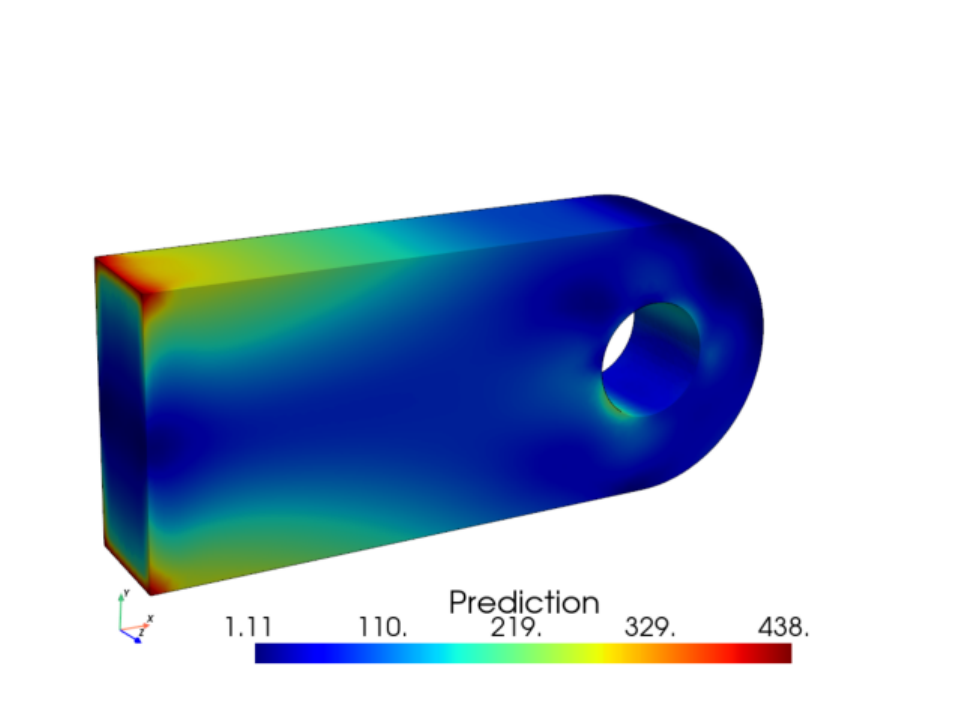}
        }
    \end{minipage} &
    \begin{minipage}[c]{\x\textwidth}
       \centering 
        \subfloat[Pred., 50$^{th}$ pct.]{\includegraphics[trim={1.5cm 0cm 1.3cm 3cm},clip,width=\textwidth]{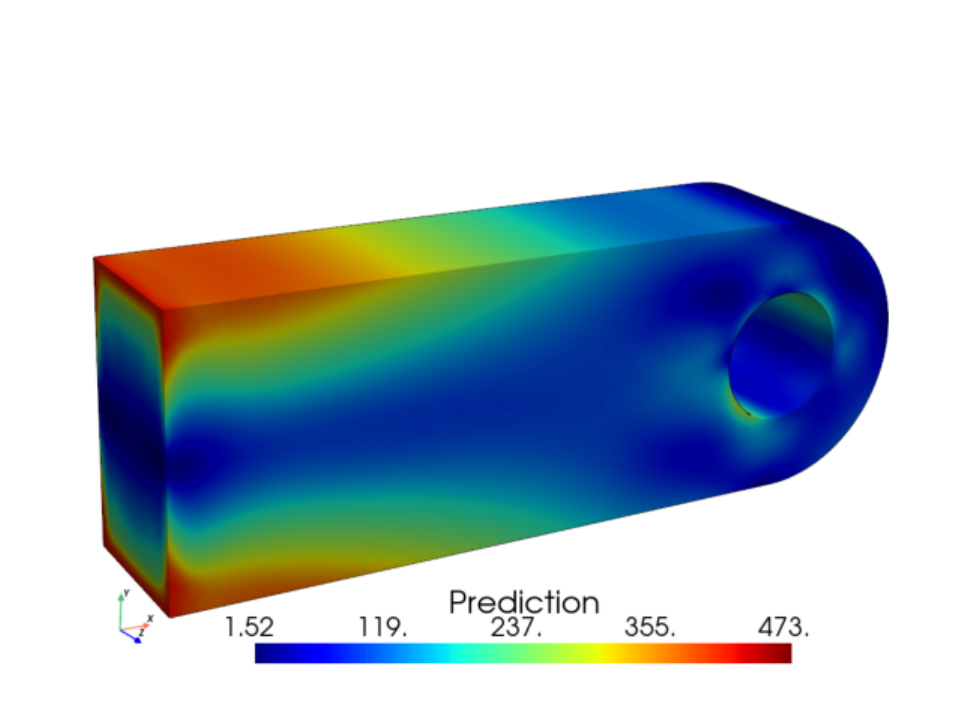}
        }
    \end{minipage} &
    \begin{minipage}[c]{\x\textwidth}
       \centering 
        \subfloat[Pred., 75$^{th}$ pct.]{\includegraphics[trim={1.5cm 0cm 1.3cm 3cm},clip,width=\textwidth]{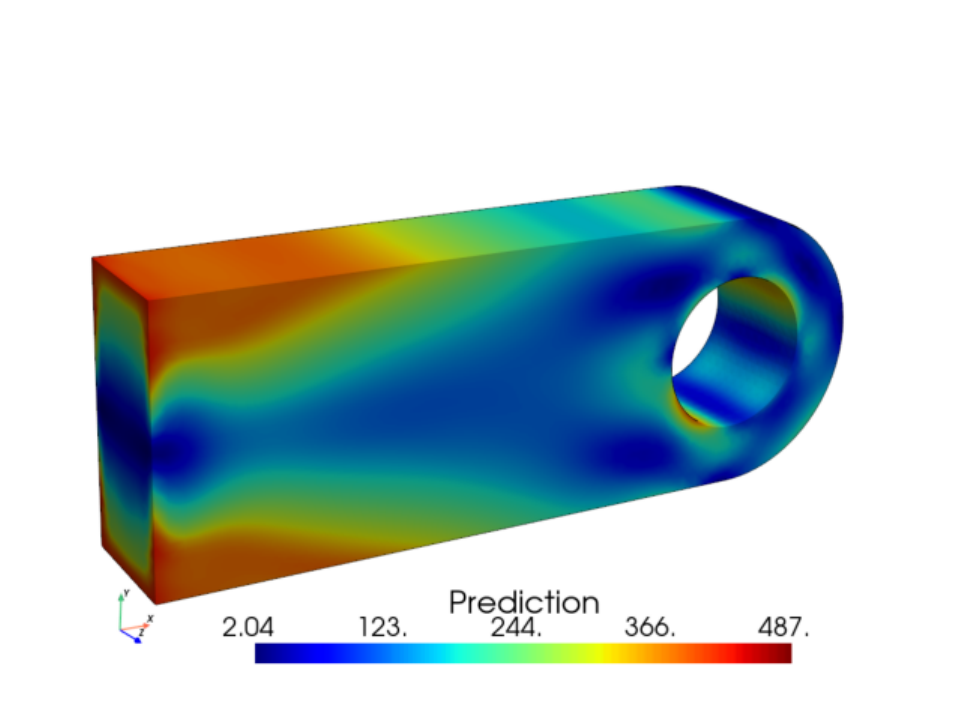}
        }
    \end{minipage} &
    \begin{minipage}[c]{\x\textwidth}
       \centering 
        \subfloat[Pred., worst]{\includegraphics[trim={1.5cm 0cm 1.3cm 3cm},clip,width=\textwidth]{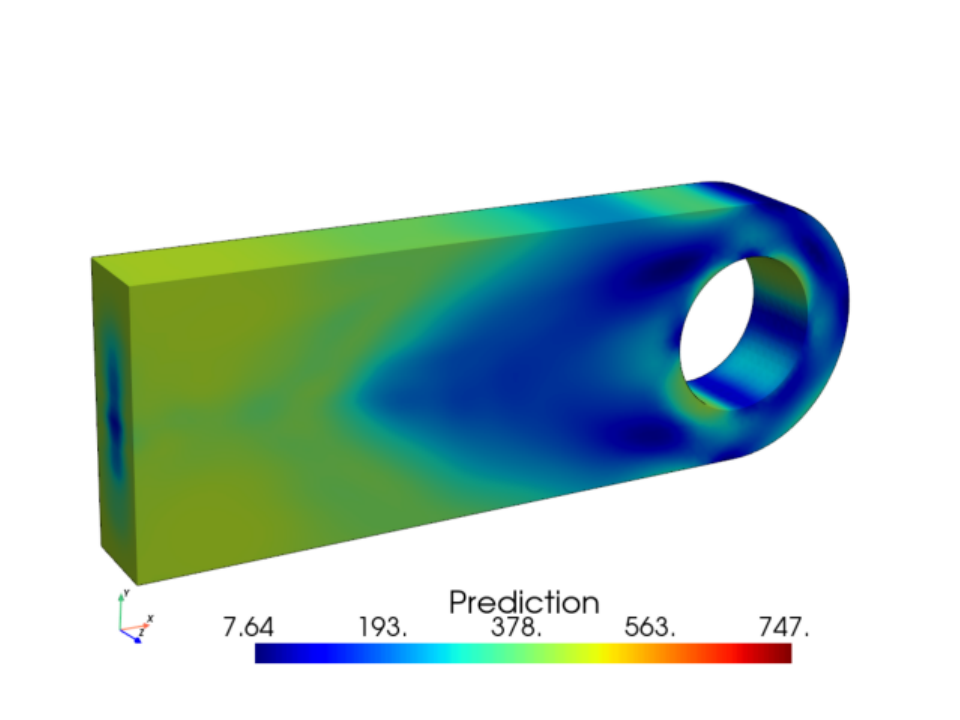}
        }
    \end{minipage} \\

    \begin{minipage}[c]{\x\textwidth}
       \centering 
        \subfloat[MAE, best]{\includegraphics[trim={1.5cm 0cm 1.3cm 2cm},clip,width=\textwidth]{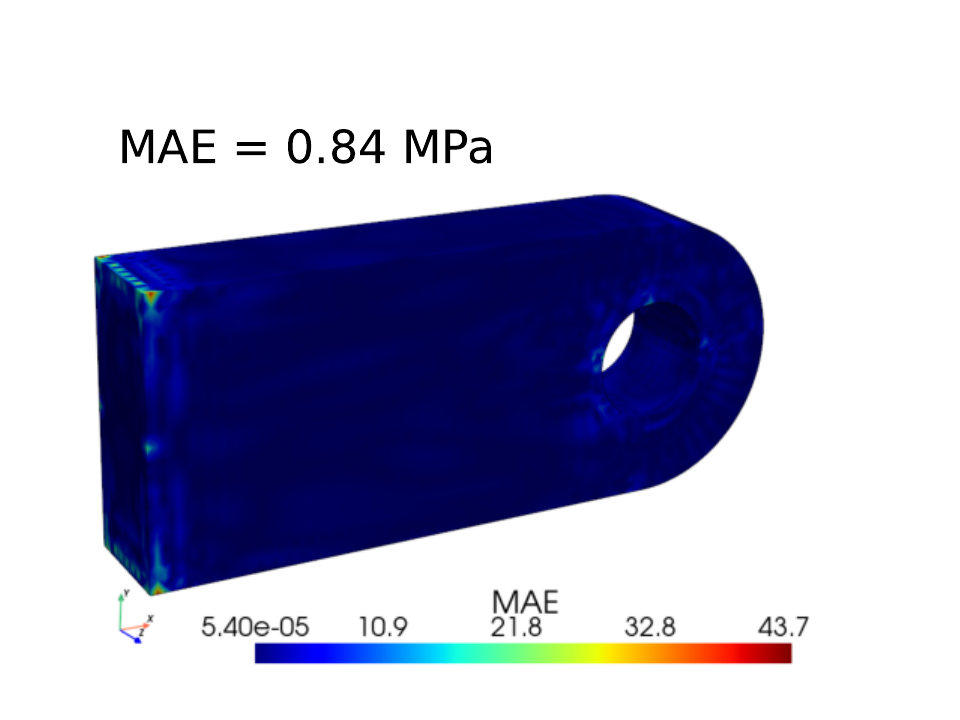}
        }
    \end{minipage} &
    \begin{minipage}[c]{\x\textwidth}
       \centering 
        \subfloat[MAE, 50$^{th}$ pct.]{\includegraphics[trim={1.5cm 0cm 1.3cm 2cm},clip,width=\textwidth]{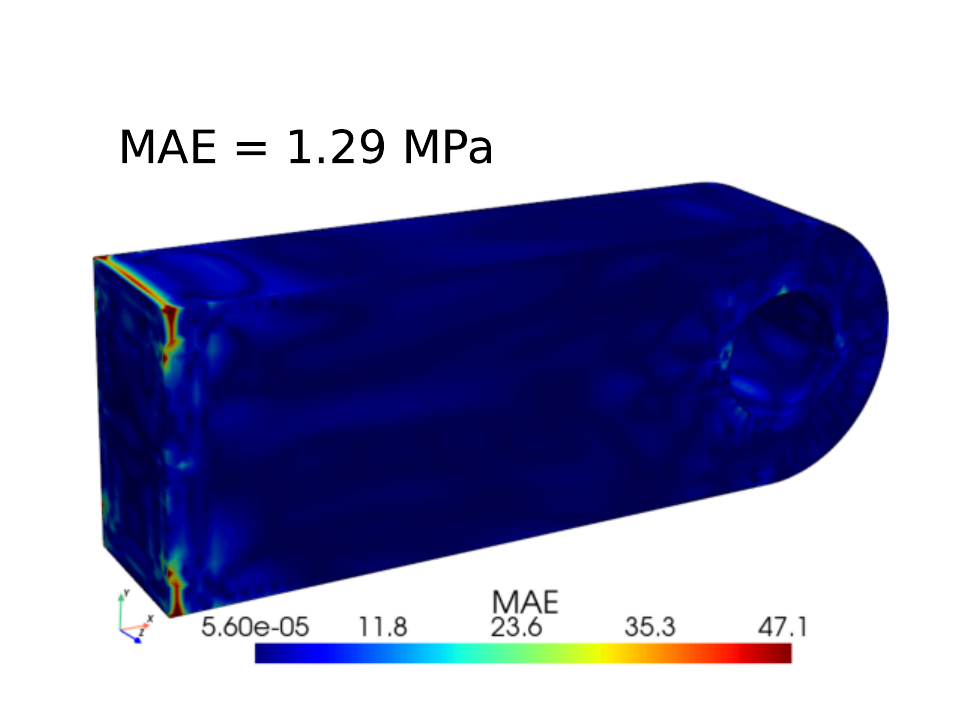}
        }
    \end{minipage} &
    \begin{minipage}[c]{\x\textwidth}
       \centering 
        \subfloat[MAE, 75$^{th}$ pct.]{\includegraphics[trim={1.5cm 0cm 1.3cm 2cm},clip,width=\textwidth]{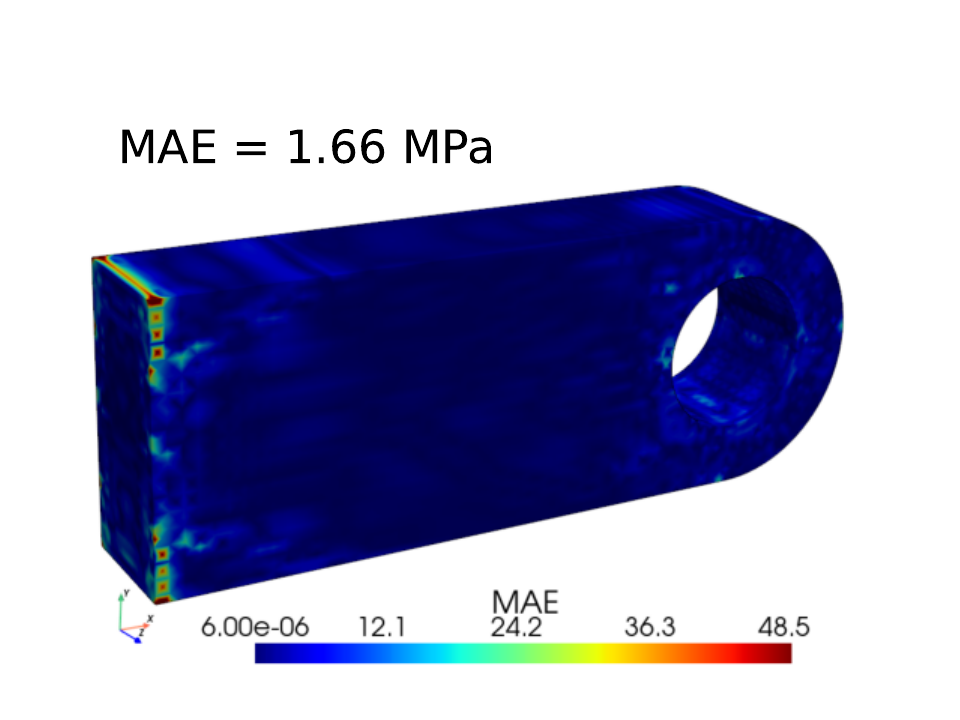}
        }
    \end{minipage} &
    \begin{minipage}[c]{\x\textwidth}
       \centering 
        \subfloat[MAE, worst]{\includegraphics[trim={1.5cm 0cm 1.3cm 2cm},clip,width=\textwidth]{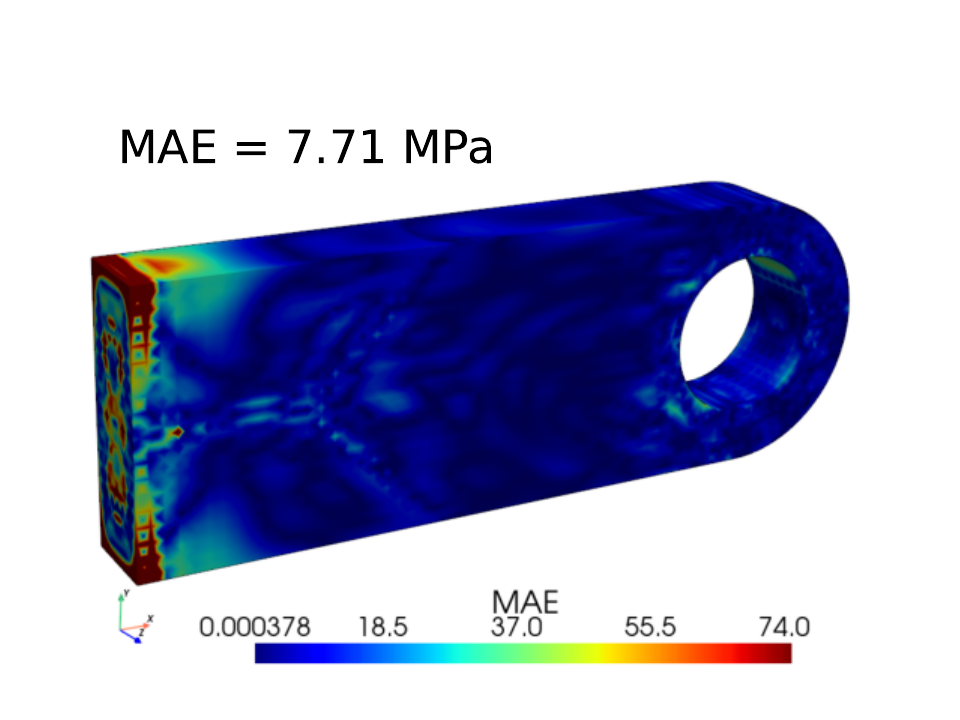}
        }
    \end{minipage} \\

    \end{tabular}
    \caption{Volume predictions by Geom-DeepONet, ranked by different percentiles in stress MAE. The first and second rows show the FE ground truth and the model predictions, and they share identical color scales. The third row shows the MAE and the color range is set to $1/10$ of that in the FE ground truth to make any error concentration visible.}
    \label{gdon_pred}
\end{figure}

The average training time, GPU memory usage during training, average (over all 3 training trials) stress MAE on the resampled points and all mesh nodes for all three models are listed in \tref{benchmark_tbl}. The numbers in the table are color-coded to indicate ranking: green for the best model, yellow for the medium model, and red for the worst model. To showcase the stress field predictions on different geometries, volume rendering plots for PointNet, vanilla DeepONet, and Geom-DeepONet are shown in Figs.\ref{point_net_pred}-\ref{gdon_pred}, respectively. The predictions are ranked by the stress MAE, and the best, median, 75$^{th}$ percentile, and the worst case are shown in the plot. The best-performing instance from the 3 repeated runs of each model was used to generate the plots. Since the PointNet architecture does not allow for varying the input dataset size once the model is trained, the 5000 resampled points are rendered as spheres to emulate a volume plot. For the two DeepONets, volume rendering was done by predicting on all mesh nodes of each geometry and rendered as a volume using the corresponding mesh connectivity.
\begin{table}[h]
\caption{Training time and stress MAE (in MPa) for three different models}
\centering
\begin{tabular}{cccccccccc}
\hline
Model  &  \vline &  Training time [s] & Memory [GB] & MAE,subset & MAE,mesh nodes  \\
\hline
PointNet &  \vline & \textcolor{red}{7450} & \textcolor{red}{37.93}\tablefootnote{Measured using a batch size of 16 to be consistent across all three models. When using a batch size of 32, memory usage was similar.} & \textcolor{yellow}{3.35} & / \\
Vanilla DeepONet &  \vline & \textcolor{green}{ \bf{347} } & \textcolor{green}{ \bf{0.96} } & \textcolor{red}{6.96} & \textcolor{red}{6.98} \\
Geom-DeepONet &  \vline & \textcolor{yellow}{591} & \textcolor{yellow}{1.20} & \textcolor{green}{ \bf{1.55}} & \textcolor{green}{ \bf{1.57}} \\
\hline
\end{tabular}
\label{benchmark_tbl}
\end{table}

From \fref{pn_tr}, it is obvious that the test loss of the PointNet model, although highly oscillatory, remained similar to that in the training set. For the DeepONets shown in \fref{vd_tr} and \fref{gd_tr}, notable oscillation in the training loss is visible even in later stage of the training process, but the test loss is stable at approximately the mean value of the training loss at different iterations. Compared to vanilla DeepONet, the training history of the Geom-DeepONet showed more oscillations, especially towards the early stage of training, but was able to reach a much lower test loss value. In general, no significant over-fitting has occurred for any of the three models. Although it can be argued that the training iterations in PointNet are much less than the two DeepONet models, therefore seemingly giving an advantage to the DeepONet models, we highlight that the training time for PointNet already exceeded 10 times of the DeepONet training times, while the proposed Geom-DeepONet delivered over 50\% lower prediction error. Therefore, although having more iterations might increase the PointNet performance, it is simply more efficient to use the proposed Geom-DeepONet to obtain more accurate predictions in a shorter training time. As shown in \tref{benchmark_tbl}, Geom-DeepONet has the lowest stress MAE on both the resampled subset and on all mesh nodes, having prediction errors less than 2 MPa while requiring an acceptable training time of less than 10 mins for 150000 iterations. In addition, we highlight that the proposed network is also memory efficient, requiring only 1.2GB of memory during training, while the PointNet required almost 38GB. Therefore, we can conclude that the Geom-DeepONet is superior to PointNet and vanilla DeepONet of similar sizes. 

Comparing the volume renderings in Figs.\ref{point_net_pred}-\ref{gdon_pred} provides more direct insights into the model performance. For PointNet, since the number of points in the input point cloud is fixed (to 5000 points) once the model is trained, predictions cannot be made easily on meshes with varying numbers of nodes. Hence, the results had to be rendered only at the 5000 resampled points as point clouds. While the two DeepONet models can handle flexible input dimensions during prediction, enabling them to make predictions on meshes of various node counts and render the results as volume contours, a powerful improvement from the PointNet model. Comparing the predictions of the vanilla and the Geom-DeepONets, it is obvious that Geom-DeepONet predictions are much more accurate, especially near the circular hole of the beam, where most vanilla DeepONet predictions show significant errors. Compared to vanilla DeepONet, the prediction errors of the Geom-DeepONet are concentrated in the fixed end of the beam, where the stress gradient is high (can be seen in the FE stress contours in the worst case, last column) due to the imposed boundary conditions. Therefore, it is evident that employing SDF and SIREN in the model architecture and allowing intermediate data fusion from the branch and trunk networks have significantly enhanced the spatial geometric awareness of the resulting network.

\subsection{Generalization in the parameter space}
\label{generalization}
\begin{figure}[h!] 
    \centering
     \subfloat[]{
         \includegraphics[trim={0cm 0cm 0cm 0cm},clip,width=0.32\textwidth]{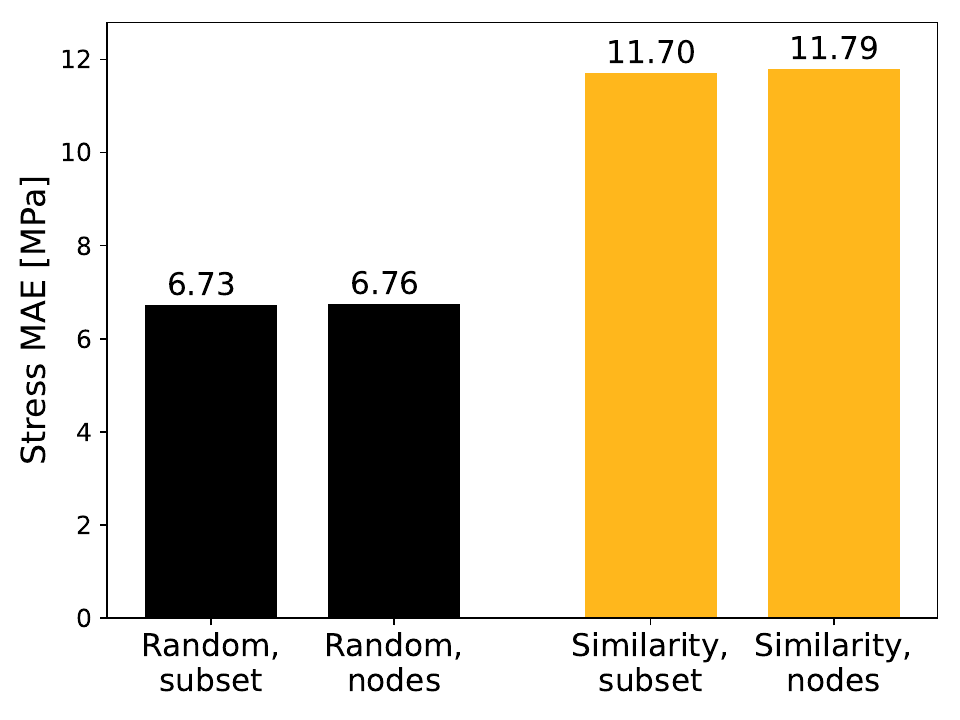}
         \label{bar1}
     }
     \subfloat[]{
         \includegraphics[trim={0cm 0cm 0cm 0cm},clip,width=0.32\textwidth]{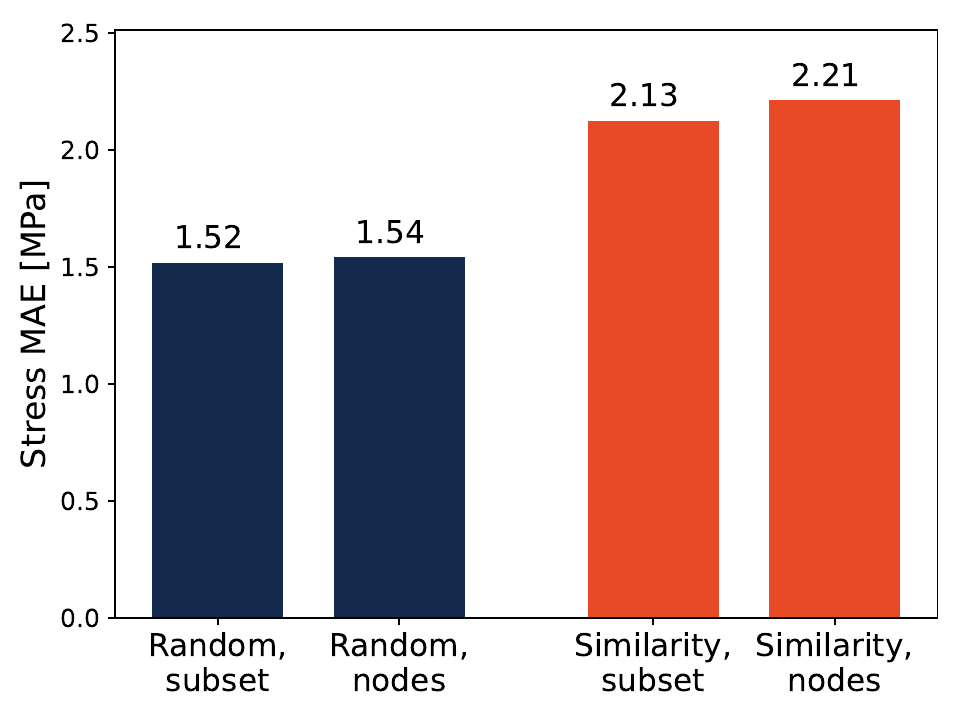}
         \label{bar2}
     }
     \subfloat[]{
         \includegraphics[trim={0cm 0cm 0cm 0cm},clip,width=0.32\textwidth]{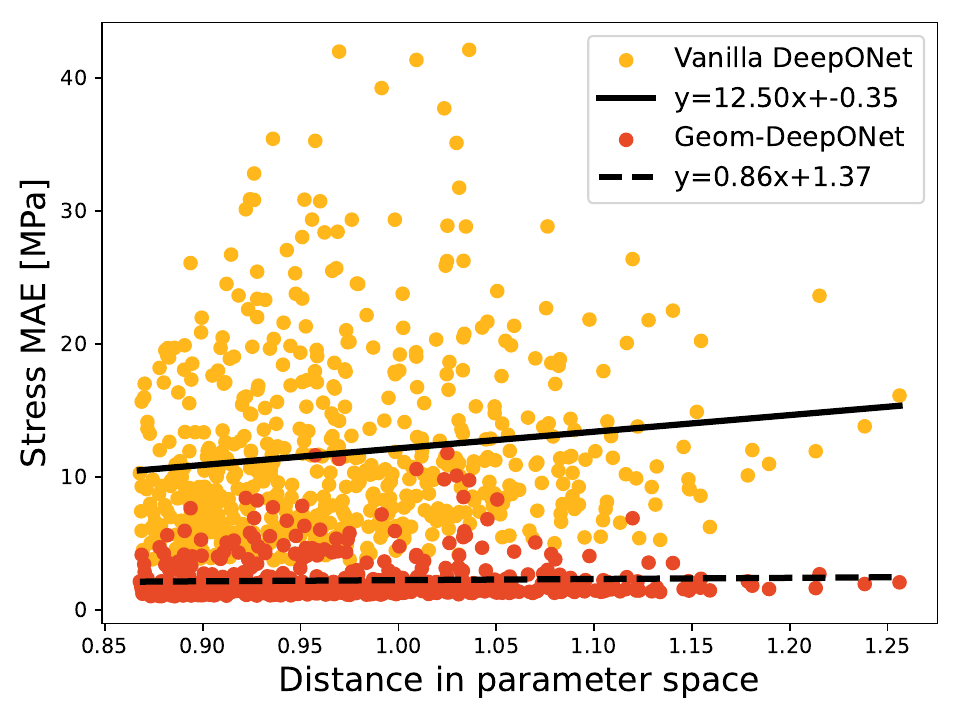}
         \label{dist_v_err}
     }
    \caption{Comparison of generalization error in parameter space: \psubref{bar1} Vanilla DeepONet, random split compared to similarity split. \psubref{bar2} Geom-DeepONet, random split compared to similarity split. \psubref{dist_v_err} Scatter plot showing the correlation of prediction error with design distance. }
    \label{bar_charts}
\end{figure}

In \sref{resample} and \sref{benchmark}, all models were trained with a randomly generated 80/20 data split. In the context of capturing geometry changes, it is of interest to investigate the model performance when the geometries in the test set are dissimilar to those in the training set and thus test the model's generalizability to dissimilar shapes. This work focuses only on parameterized geometries, which naturally provide a means to measure design similarity in the parameter space. We define a normalized similarity between two designs $i$ and $j$ using a simple $L_2$ norm:
\begin{equation}
    S_{ij} = \sqrt{ \sum_{k=1}^n ( \Hat{P^i_k} - \Hat{P^j_k} )^2 },
\end{equation}
where $S_{ij}$ is the similarity, and $\Hat{P^i_k}$ denotes the $k^{th}$ normalized (to the range 0-1) geometric parameter of geometry $i$. Only geometric parameters are considered, and load parameters are excluded from the similarity calculation. Without loss of generality, we can take the first geometry as the reference geometry, compute similarity with respect to this reference, and rank all other geometries based on similarity. A special data split can be formed by considering 80\% of the most similar geometries in the training set while the 20\% of the most dissimilar geometries in the test set. The vanilla DeepONet and the Geom-DeepONet were trained using this similarity-based data splitting. The bar charts comparing similarity-based and random data splitting are shown in \fref{bar1} and \fref{bar2}. Scatter plots showing the correlation between design similarity and prediction errors (on mesh nodes) are shown in \fref{dist_v_err}.

The bar charts in \fref{bar_charts} clearly show the differences in performance between the vanilla DeepONet and Geom-DeepONet. When changing from a random data split to a similarity data split, the test error for vanilla DeepONet increased significantly to around 12 MPa, while that for Geom-DeepONet remained less than 2.5 MPa. Similarly, when inspecting the scatter plot of stress prediction error versus distance in the design space, we see a clear positive correlation for the vanilla DeepONet, evident from the large positive slope of the linear regression fit (solid black trend line). In contrast, the correlation for Geom-DeepONet, although still positive, is much weaker (much smaller slope of the dashed black trend line), indicating the prediction error does not grow significantly as the design becomes more dissimilar with the reference. These findings again highlight the superior spatial geometric awareness of the proposed Geom-DeepONet model compared to vanilla DeepONet based only on nodal coordinates.

\subsection{Extension to vector predictions: Cuboid with ellipsoidal void}
\label{model2_result}
In the beam dataset, we have only trained the model to predict the von Mises equivalent stress. In this section, we trained a Geom-DeepONet to predict both the von Mises stress and the X, Y, and Z components of the displacement vector using the cuboid with random void dataset.

\subsubsection{Effect of Geom-DeepONet model size}
\label{model_size}
\begin{figure}[h!] 
    \centering
     \subfloat[]{
         \includegraphics[trim={0cm 0cm 0cm 0cm},clip,width=0.4\textwidth]{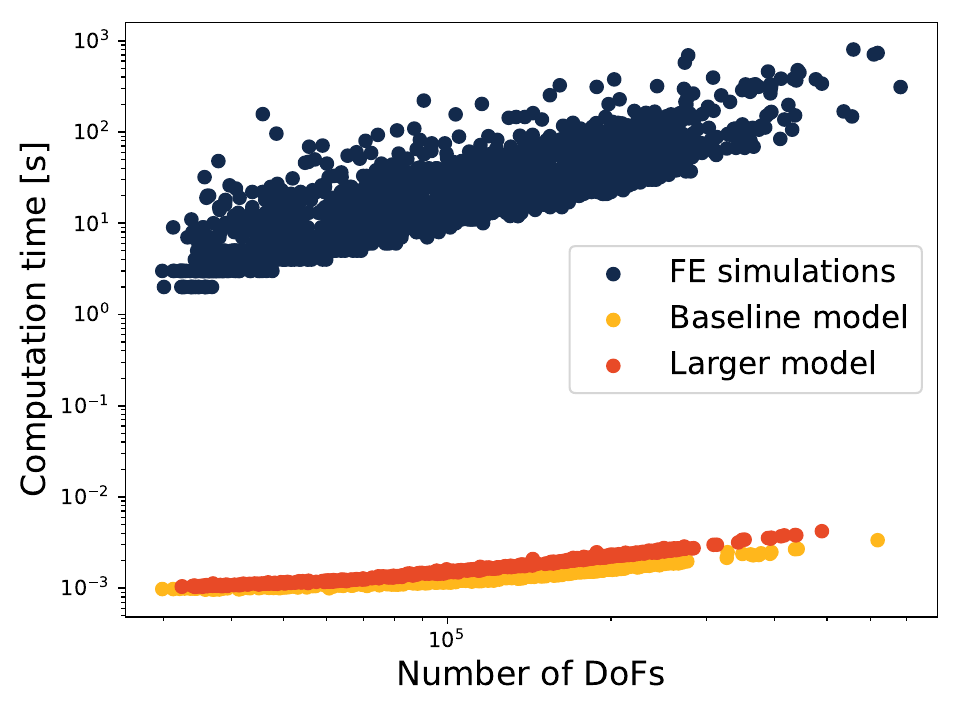}
         \label{pred_time}
     }
     \subfloat[]{
         \includegraphics[trim={0cm 0cm 0cm 0cm},clip,width=0.4\textwidth]{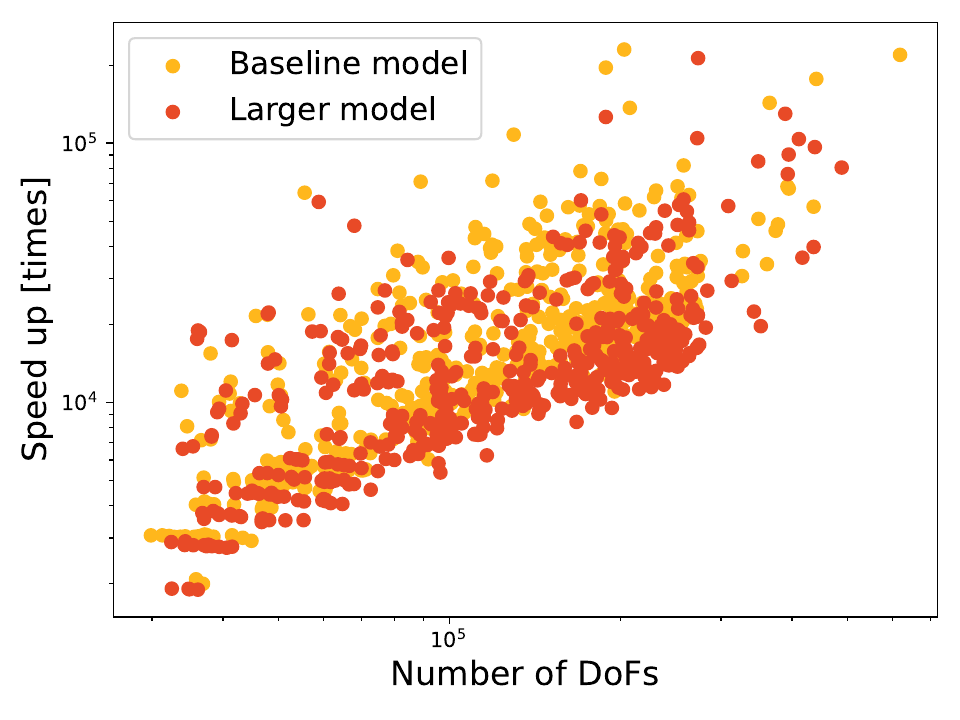}
         \label{pred_speedup}
     }
    \caption{Comparison of finite element simulation and neural network prediction times and the amount of speedup.}
    \label{speedup}
\end{figure}

\begin{figure}[h!]
\newcommand\x{0.17}
    \centering
    \begin{tabular}{ c c c c c }
    \begin{minipage}[c]{\x\textwidth}
       \centering 
        \subfloat[FE, best]{\includegraphics[trim={2.6cm 0cm 2.5cm 2.2cm},clip,width=\textwidth]{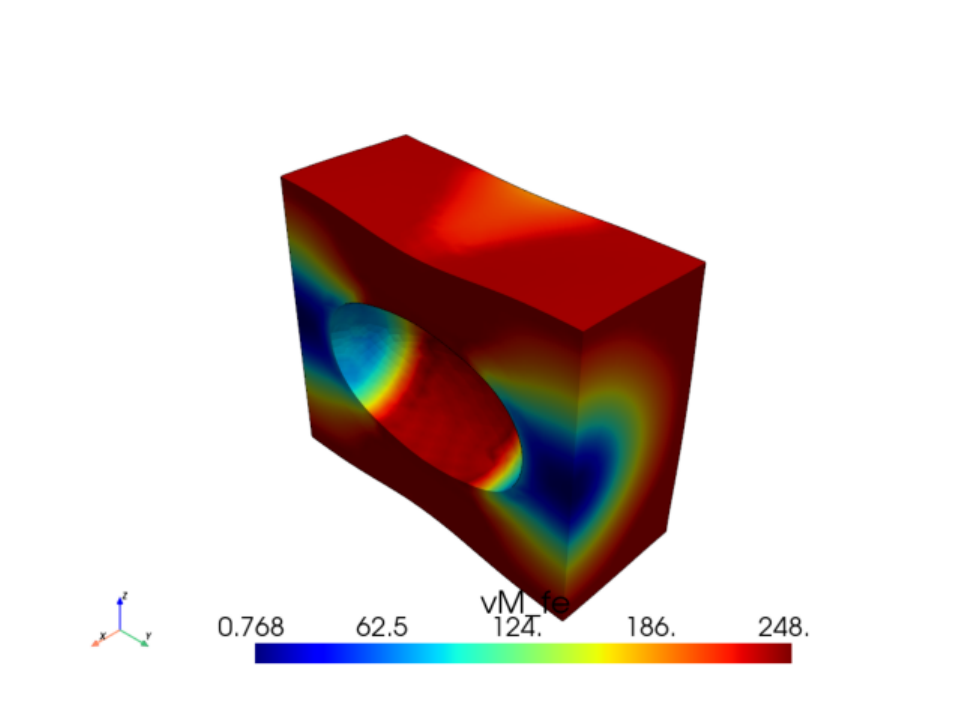}
        }
    \end{minipage} &
    \begin{minipage}[c]{\x\textwidth}
       \centering 
        \subfloat[FE, 70$^{th}$ pct.]{\includegraphics[trim={2.6cm 0cm 2.5cm 2.2cm},clip,width=\textwidth]{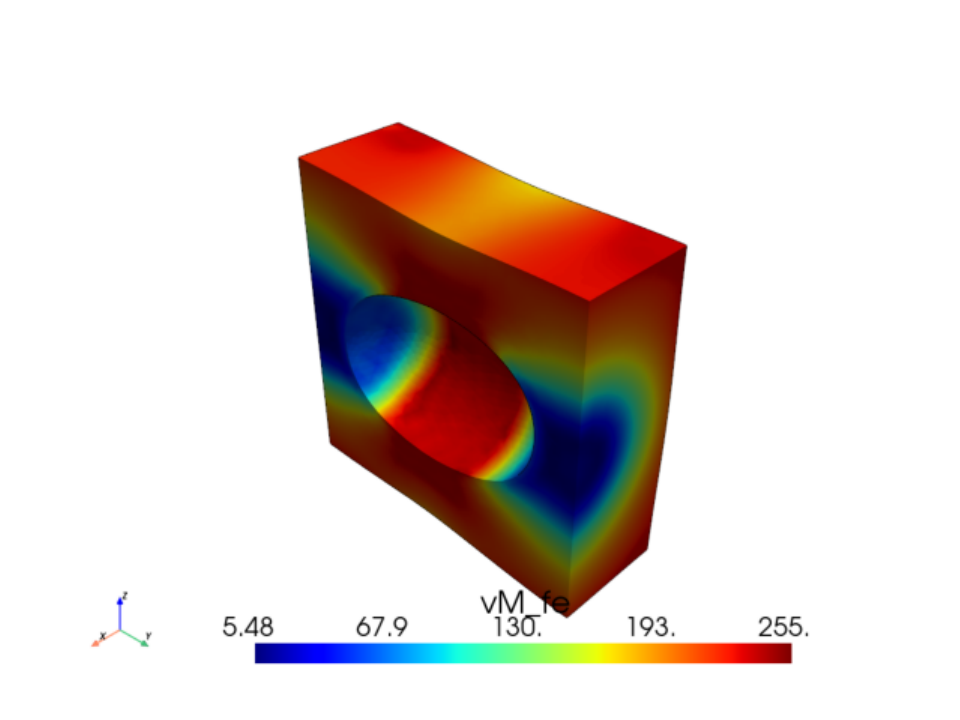}
        }
    \end{minipage} &
    \begin{minipage}[c]{\x\textwidth}
       \centering 
        \subfloat[FE, 80$^{th}$ pct.]{\includegraphics[trim={2.6cm 0cm 2.5cm 2.2cm},clip,width=\textwidth]{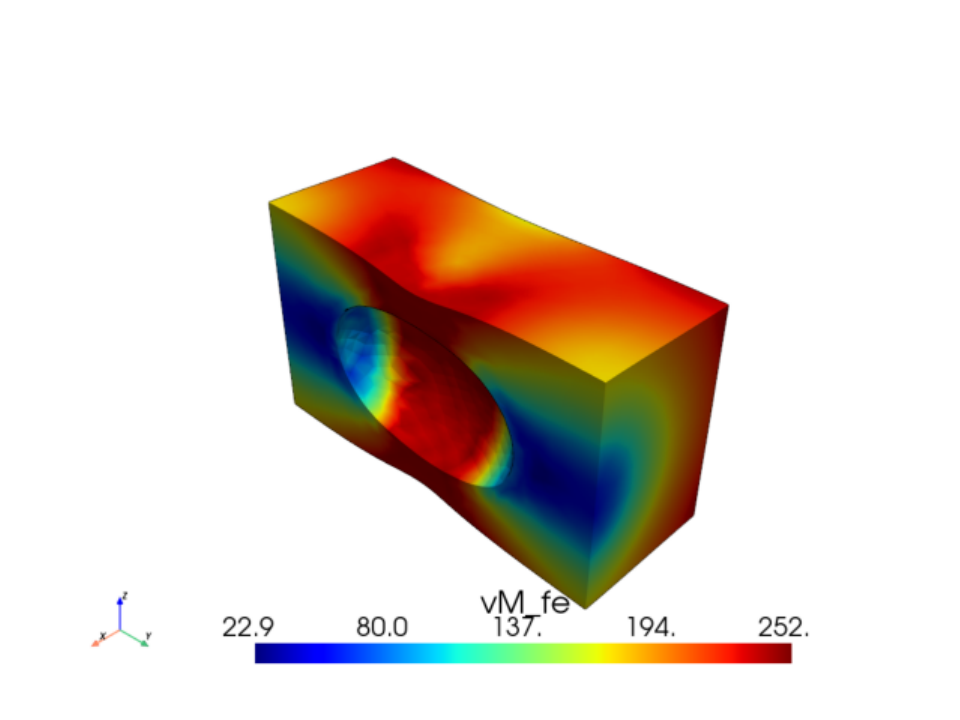}
        }
    \end{minipage} &
    \begin{minipage}[c]{\x\textwidth}
       \centering 
        \subfloat[FE, 90$^{th}$ pct.]{\includegraphics[trim={2.6cm 0cm 2.5cm 2.2cm},clip,width=\textwidth]{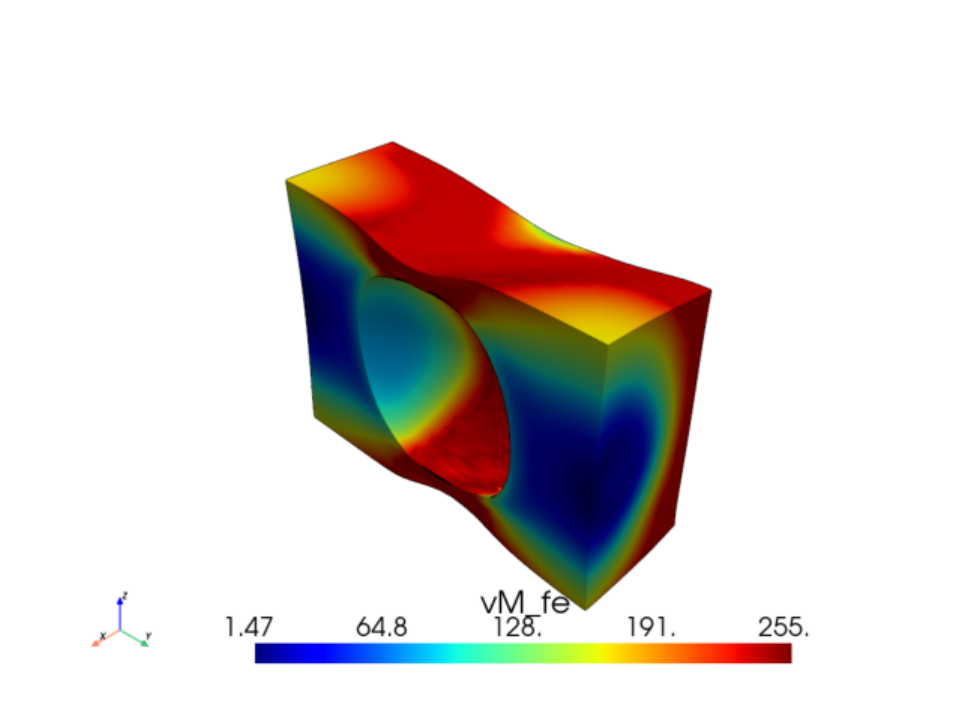}
        }
    \end{minipage} &
    \begin{minipage}[c]{\x\textwidth}
       \centering 
        \subfloat[FE, worst]{\includegraphics[trim={2.6cm 0cm 2.5cm 2.2cm},clip,width=\textwidth]{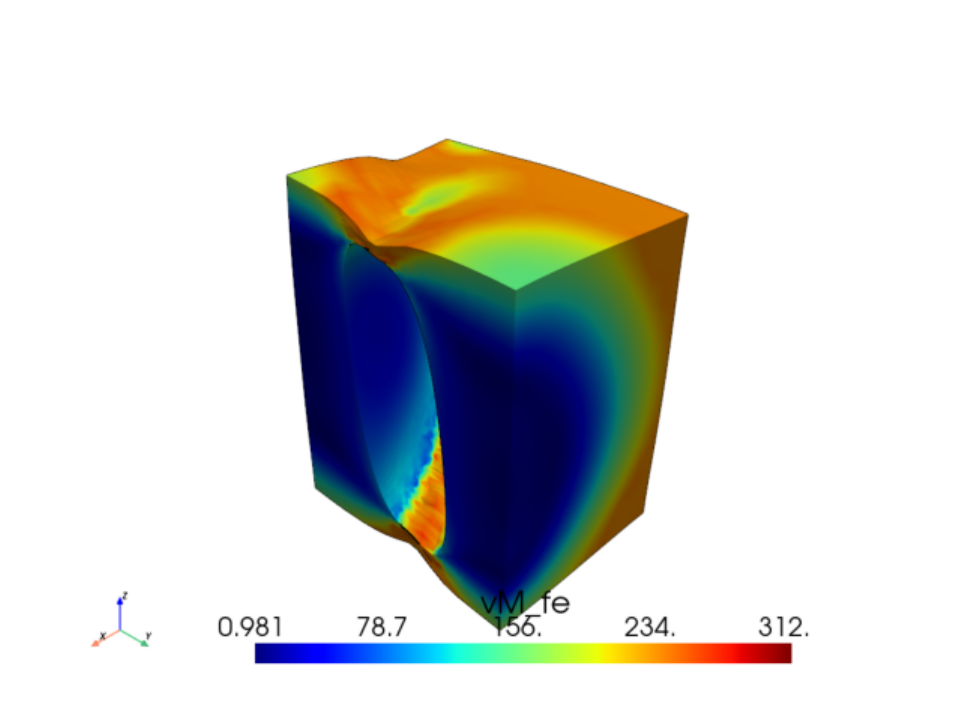}
        }
    \end{minipage} \\

    \begin{minipage}[c]{\x\textwidth}
       \centering 
        \subfloat[Pred., best]{\includegraphics[trim={2.6cm 0cm 2.5cm 2.2cm},clip,width=\textwidth]{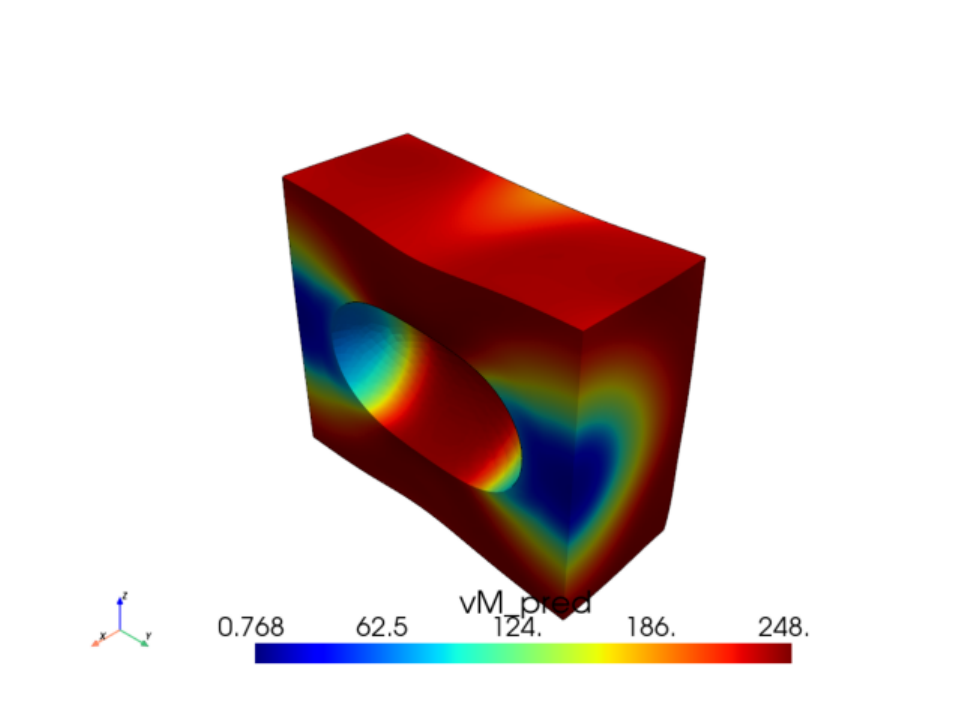}
        }
    \end{minipage} &
    \begin{minipage}[c]{\x\textwidth}
       \centering 
        \subfloat[Pred., 70$^{th}$ pct.]{\includegraphics[trim={2.6cm 0cm 2.5cm 2.2cm},clip,width=\textwidth]{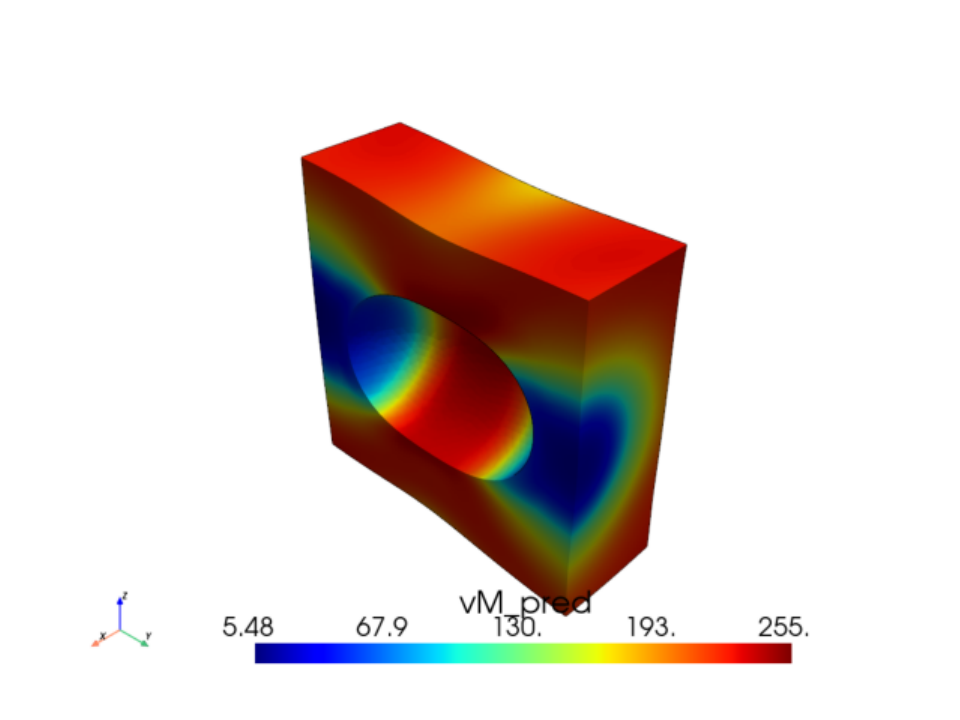}
        }
    \end{minipage} &
    \begin{minipage}[c]{\x\textwidth}
       \centering 
        \subfloat[Pred., 80$^{th}$ pct.]{\includegraphics[trim={2.6cm 0cm 2.5cm 2.2cm},clip,width=\textwidth]{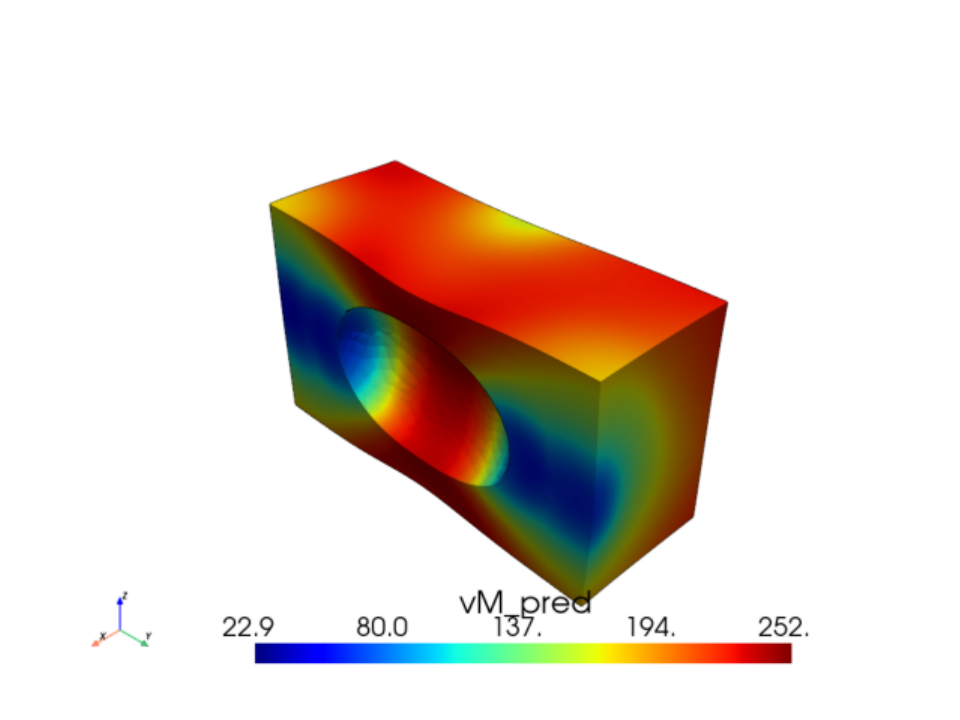}
        }
    \end{minipage} &
    \begin{minipage}[c]{\x\textwidth}
       \centering 
        \subfloat[Pred., 90$^{th}$ pct.]{\includegraphics[trim={2.6cm 0cm 2.5cm 2.2cm},clip,width=\textwidth]{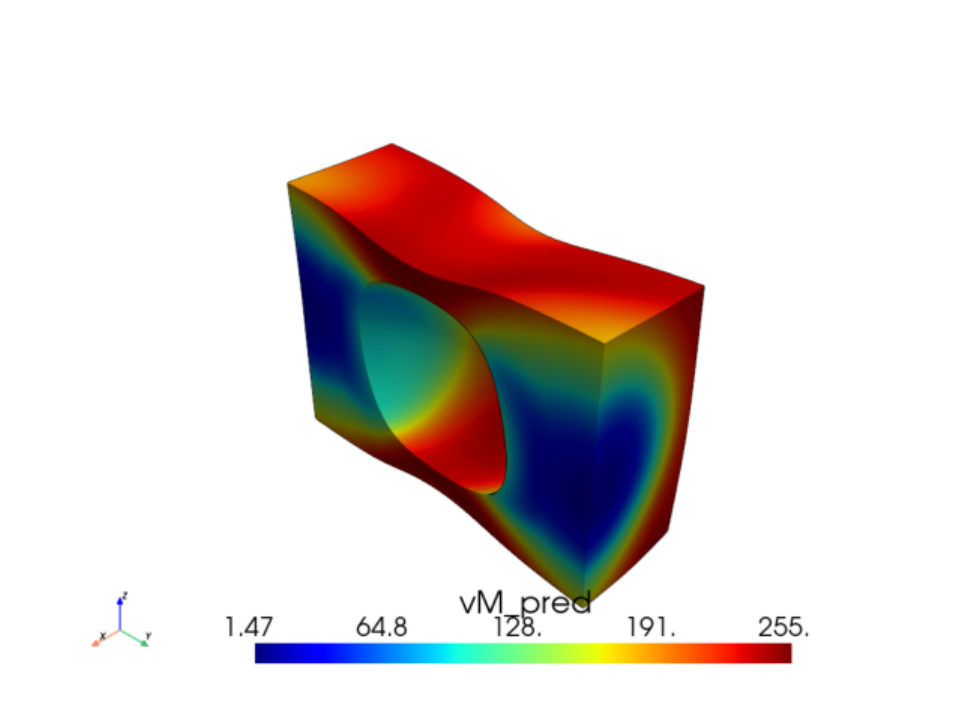}
        }
    \end{minipage} &
    \begin{minipage}[c]{\x\textwidth}
       \centering 
        \subfloat[Pred., worst]{\includegraphics[trim={2.6cm 0cm 2.5cm 2.2cm},clip,width=\textwidth]{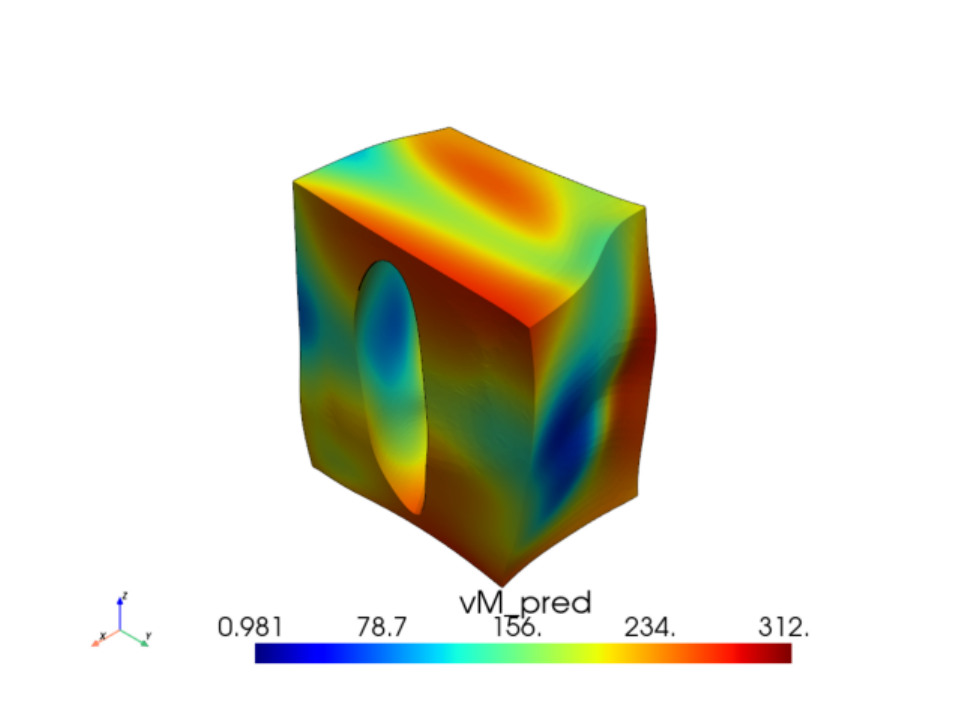}
        }
    \end{minipage} \\

    \begin{minipage}[c]{\x\textwidth}
       \centering 
        \subfloat[MAE, best]{\includegraphics[trim={1.4cm 0cm 2.5cm 1.8cm},clip,width=\textwidth]{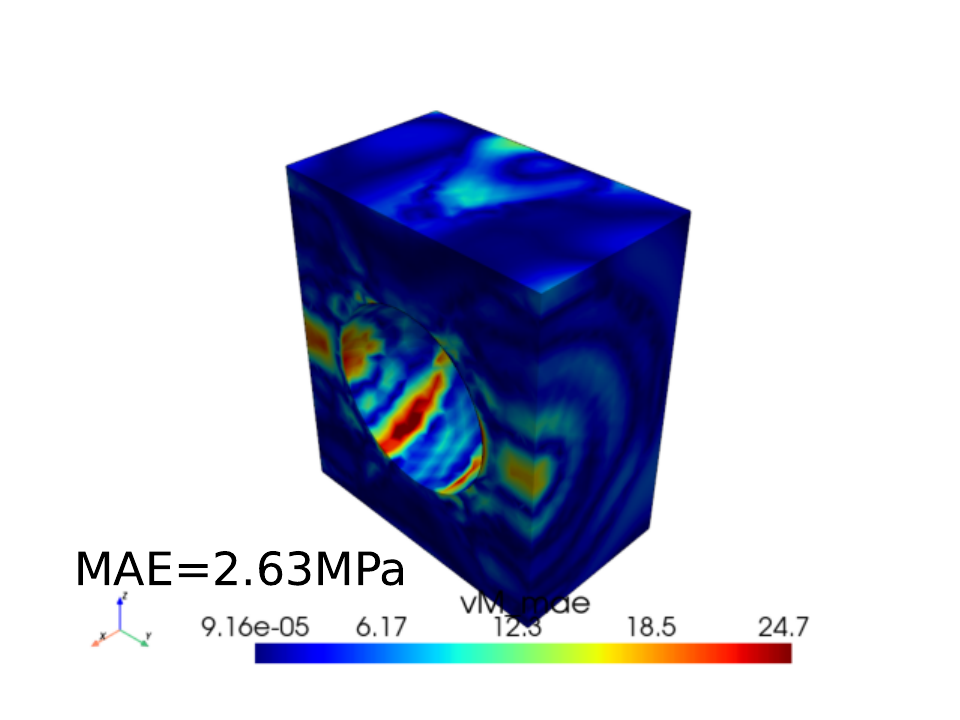}
        }
    \end{minipage} &
    \begin{minipage}[c]{\x\textwidth}
       \centering 
        \subfloat[MAE, 70$^{th}$ pct.]{\includegraphics[trim={1.3cm 0cm 2.5cm 1.8cm},clip,width=\textwidth]{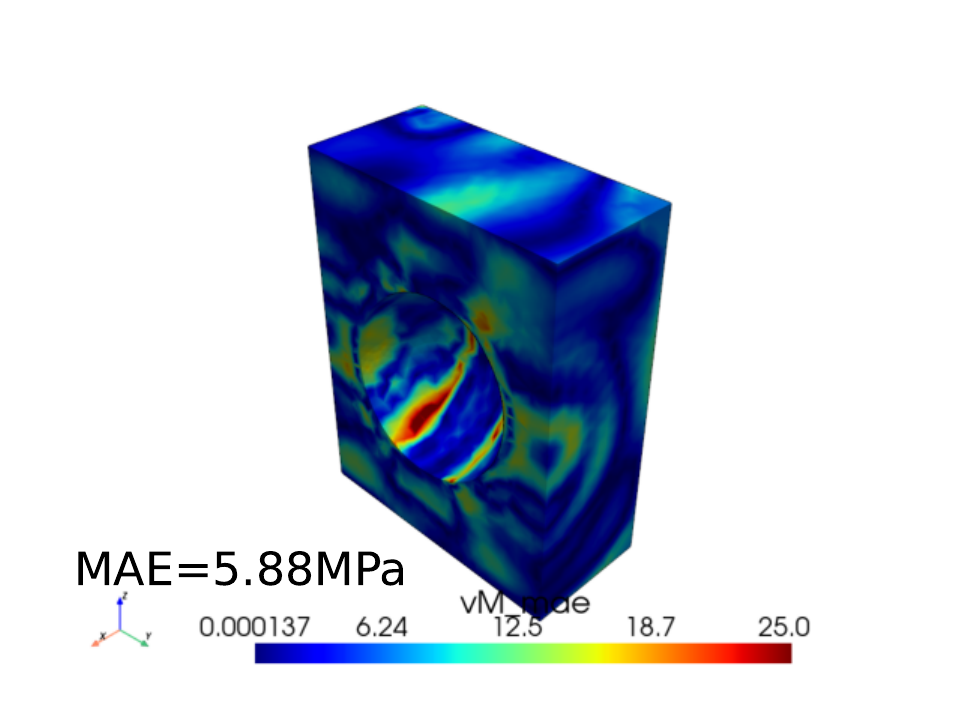}
        }
    \end{minipage} &
    \begin{minipage}[c]{\x\textwidth}
       \centering 
        \subfloat[MAE, 80$^{th}$ pct.]{\includegraphics[trim={1.3cm 0cm 2.5cm 1.8cm},clip,width=\textwidth]{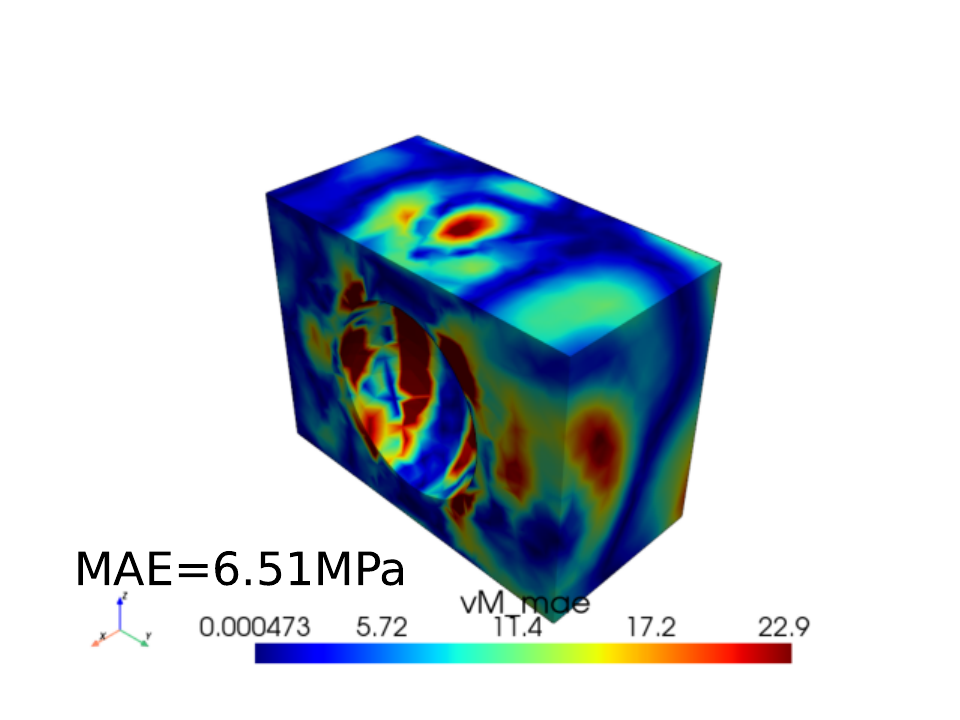}
        }
    \end{minipage} &
    \begin{minipage}[c]{\x\textwidth}
       \centering 
        \subfloat[MAE, 90$^{th}$ pct.]{\includegraphics[trim={1.3cm 0cm 2.5cm 1.8cm},clip,width=\textwidth]{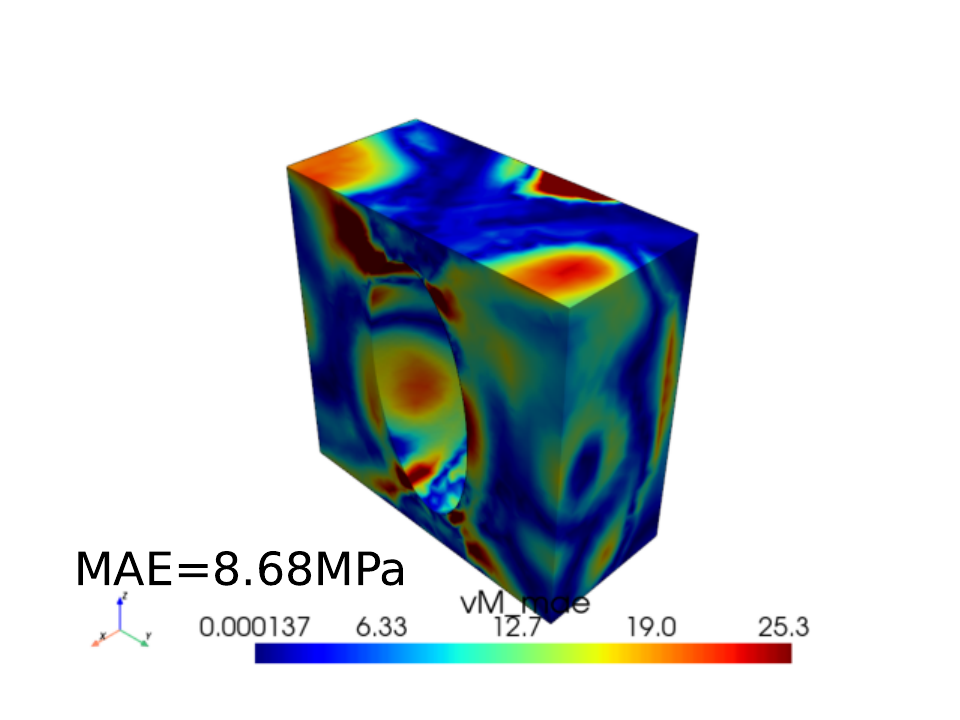}
        }
    \end{minipage} &
    \begin{minipage}[c]{\x\textwidth}
       \centering 
        \subfloat[MAE, worst]{\includegraphics[trim={1.3cm 0cm 2.5cm 1.8cm},clip,width=\textwidth]{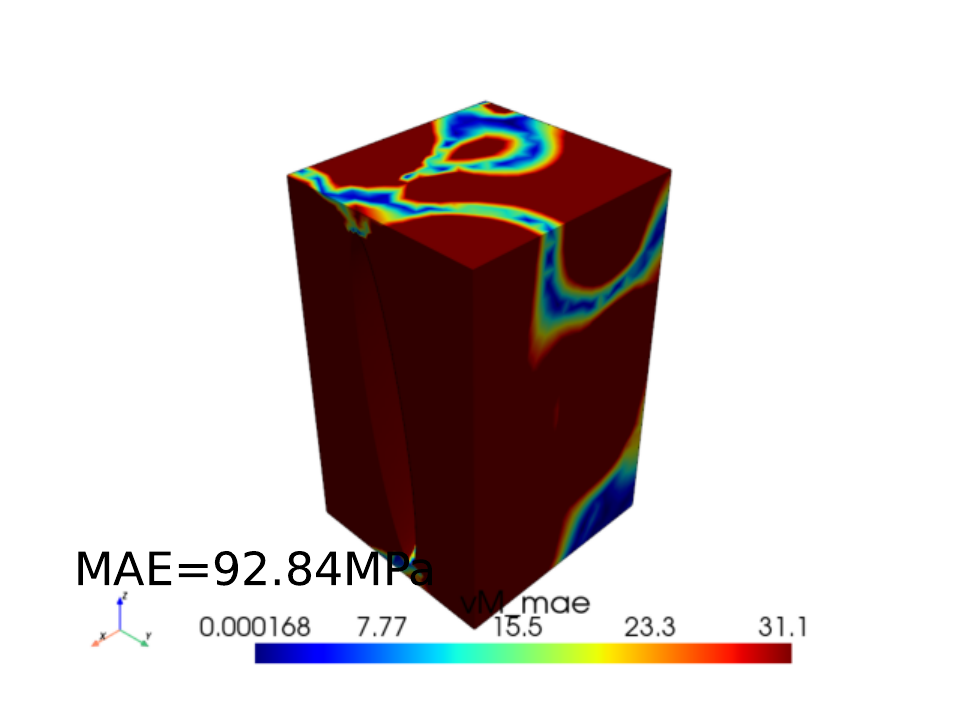}
        }
    \end{minipage} \\

    \end{tabular}
    \caption{Volume predictions by the baseline Geom-DeepONet, ranked by stress MAE. Section view applied to show internal details. The first and second rows show FE results and model predictions with identical color scales. Deformed shapes are rendered with the FE and predicted displacements with a scale factor of 250. The third row shows the stress MAE and the color range is set to $1/10$ of that in the FE ground truth to highlight where the errors are concentrated.}
    \label{c_gdon_pred_small}
\end{figure}

\begin{figure}[h!]
\newcommand\x{0.16}
    \centering
    \begin{tabular}{ c c c c c }
    \begin{minipage}[c]{\x\textwidth}
       \centering 
        \subfloat[FE, best]{\includegraphics[trim={2.6cm 0cm 2.5cm 2.2cm},clip,width=\textwidth]{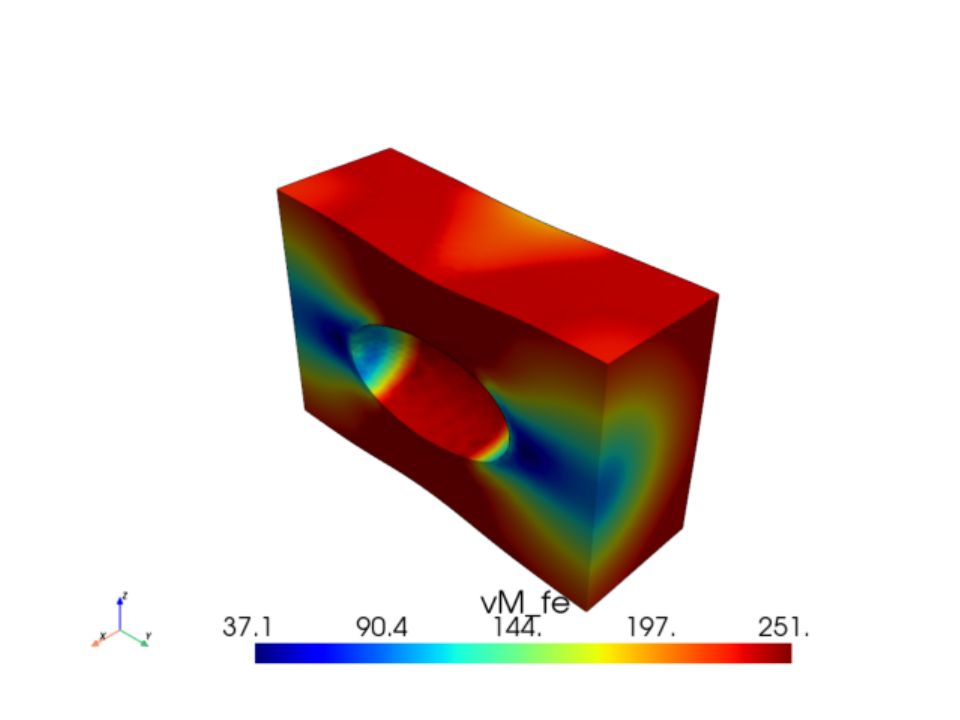}
        }
    \end{minipage} &
    \begin{minipage}[c]{\x\textwidth}
       \centering 
        \subfloat[FE, 70$^{th}$ pct.]{\includegraphics[trim={2.6cm 0cm 2.5cm 2.2cm},clip,width=\textwidth]{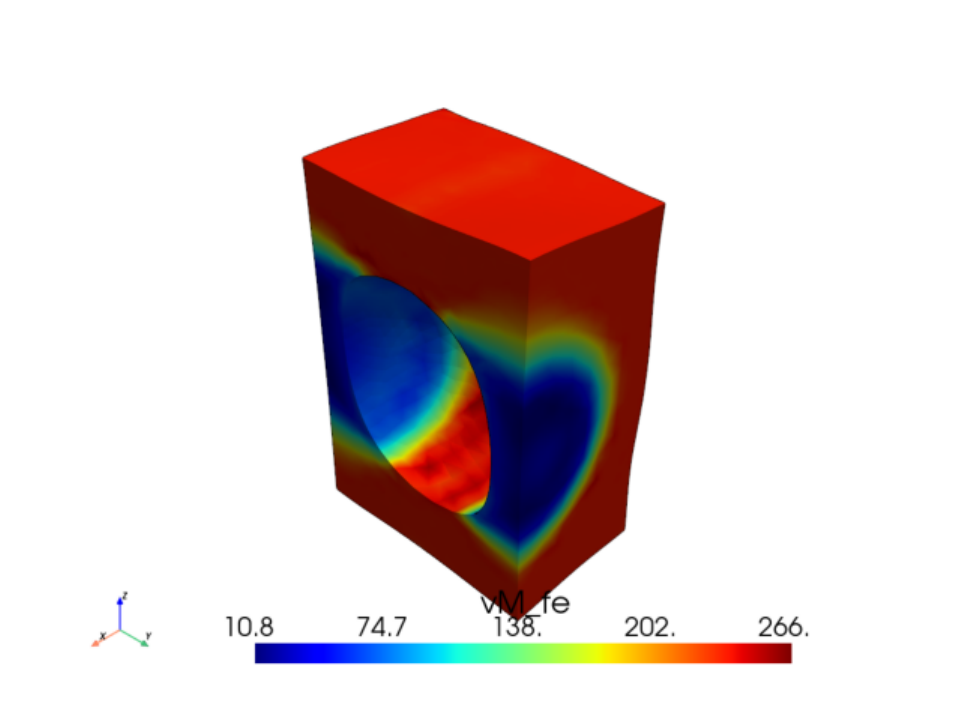}
        }
    \end{minipage} &
    \begin{minipage}[c]{\x\textwidth}
       \centering 
        \subfloat[FE, 80$^{th}$ pct.]{\includegraphics[trim={2.6cm 0cm 2.5cm 2.2cm},clip,width=\textwidth]{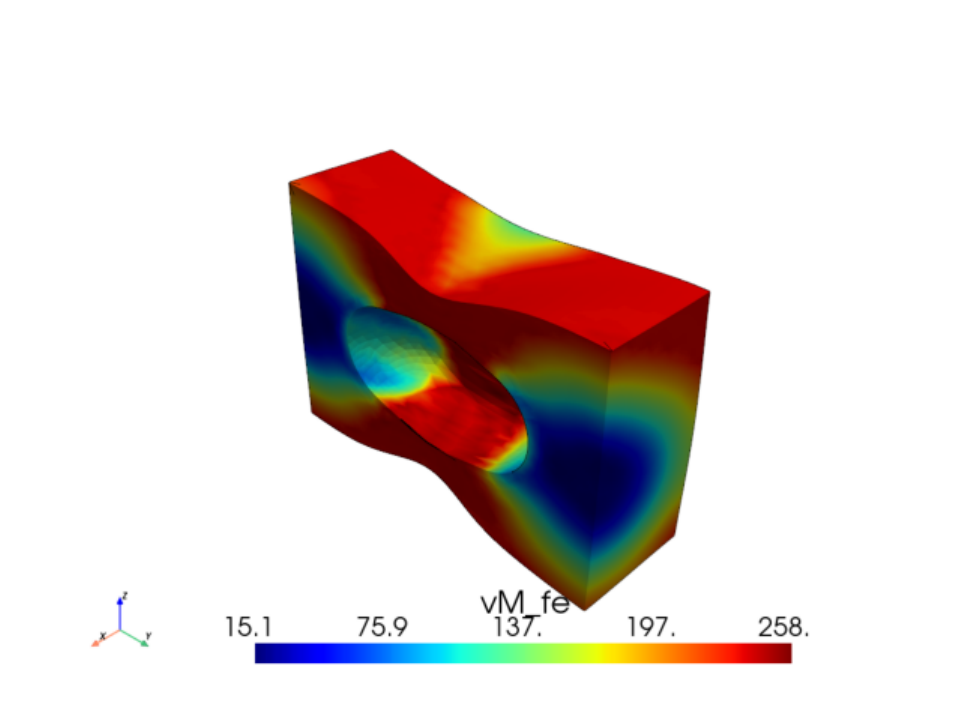}
        }
    \end{minipage} &
    \begin{minipage}[c]{\x\textwidth}
       \centering 
        \subfloat[FE, 90$^{th}$ pct.]{\includegraphics[trim={2.6cm 0cm 2.5cm 2.2cm},clip,width=\textwidth]{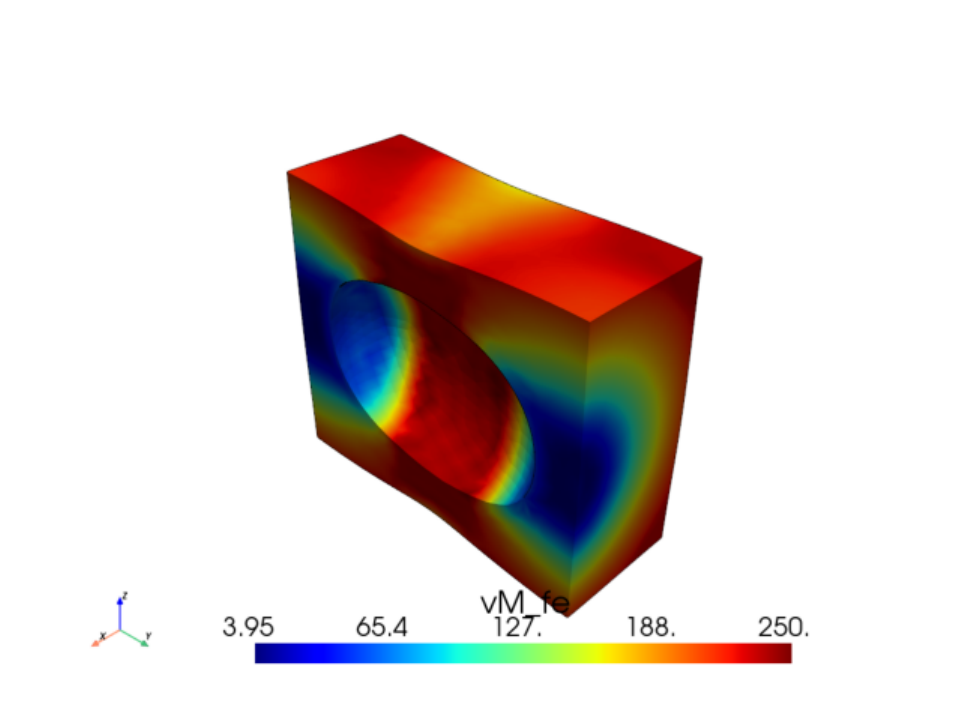}
        }
    \end{minipage} &
    \begin{minipage}[c]{\x\textwidth}
       \centering 
        \subfloat[FE, worst]{\includegraphics[trim={2.6cm 0cm 2.5cm 2.2cm},clip,width=\textwidth]{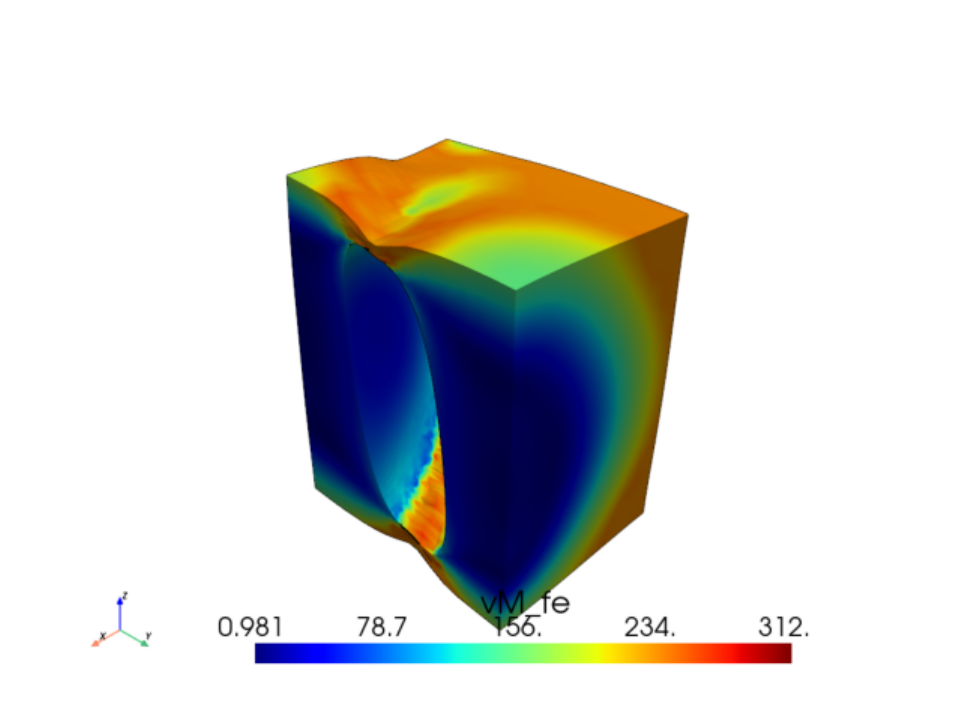}
        }
    \end{minipage} \\

    \begin{minipage}[c]{\x\textwidth}
       \centering 
        \subfloat[Pred., best]{\includegraphics[trim={2.6cm 0cm 2.5cm 2.2cm},clip,width=\textwidth]{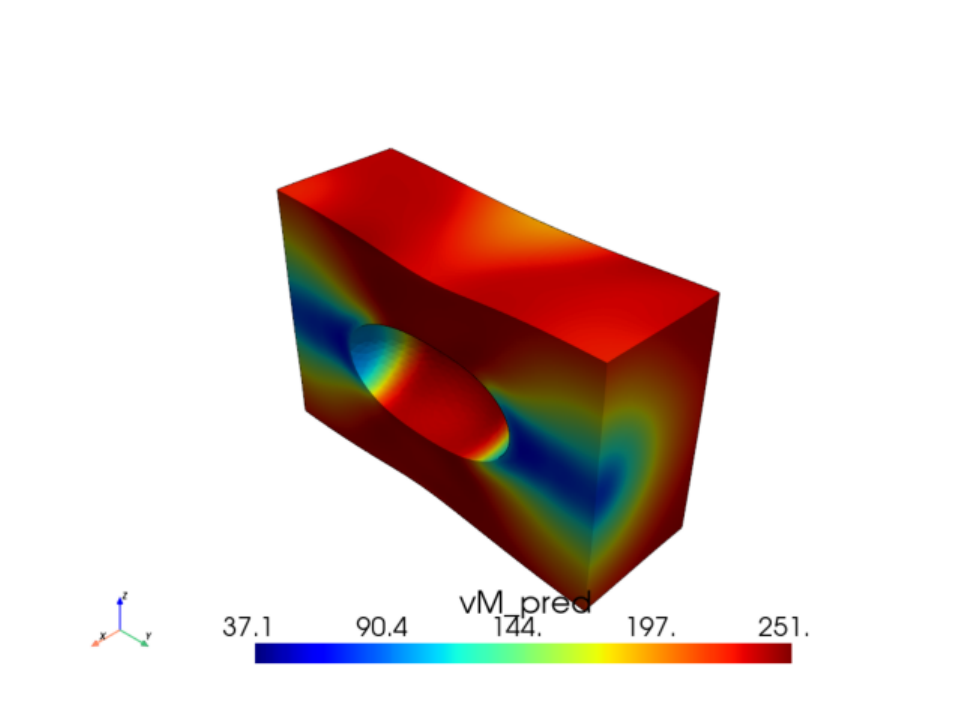}
        }
    \end{minipage} &
    \begin{minipage}[c]{\x\textwidth}
       \centering 
        \subfloat[Pred., 70$^{th}$ pct.]{\includegraphics[trim={2.6cm 0cm 2.5cm 2.2cm},clip,width=\textwidth]{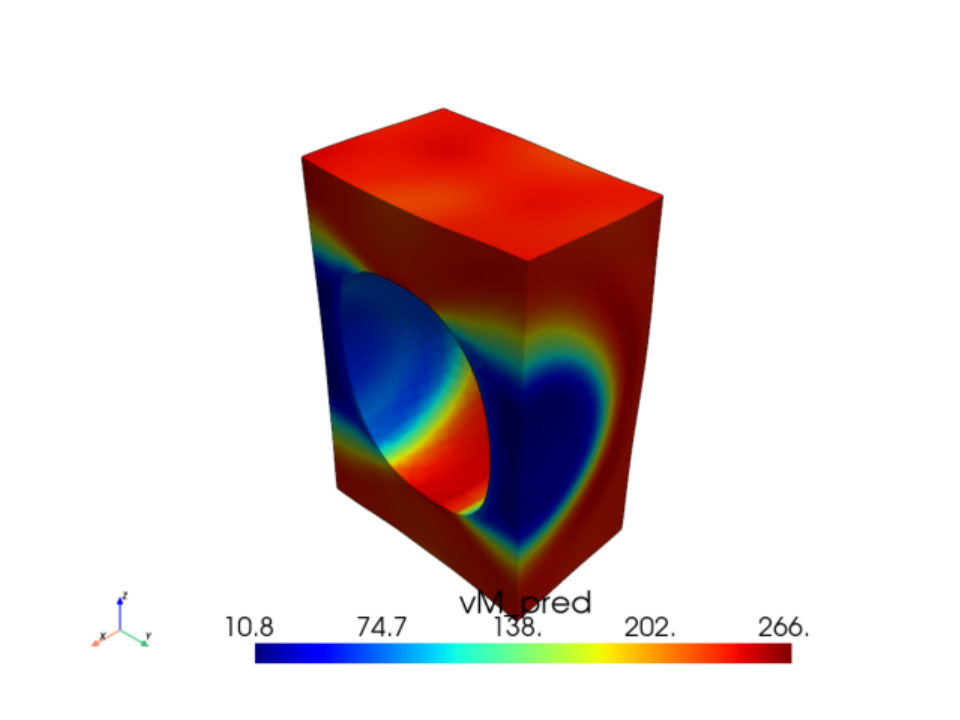}
        }
    \end{minipage} &
    \begin{minipage}[c]{\x\textwidth}
       \centering 
        \subfloat[Pred., 80$^{th}$ pct.]{\includegraphics[trim={2.6cm 0cm 2.5cm 2.2cm},clip,width=\textwidth]{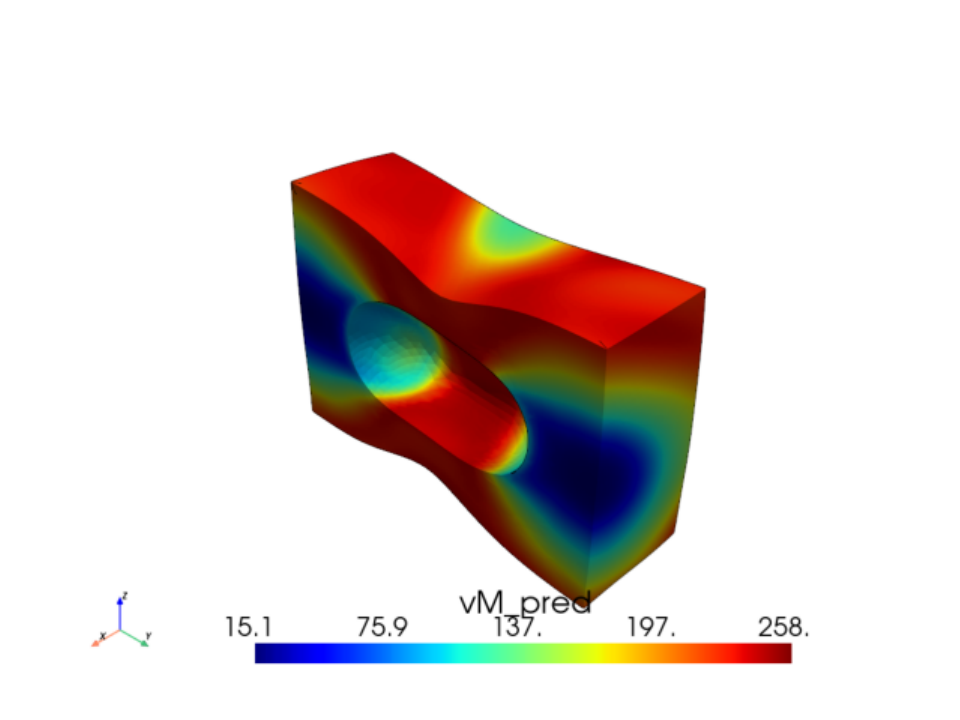}
        }
    \end{minipage} &
    \begin{minipage}[c]{\x\textwidth}
       \centering 
        \subfloat[Pred., 90$^{th}$ pct.]{\includegraphics[trim={2.6cm 0cm 2.5cm 2.2cm},clip,width=\textwidth]{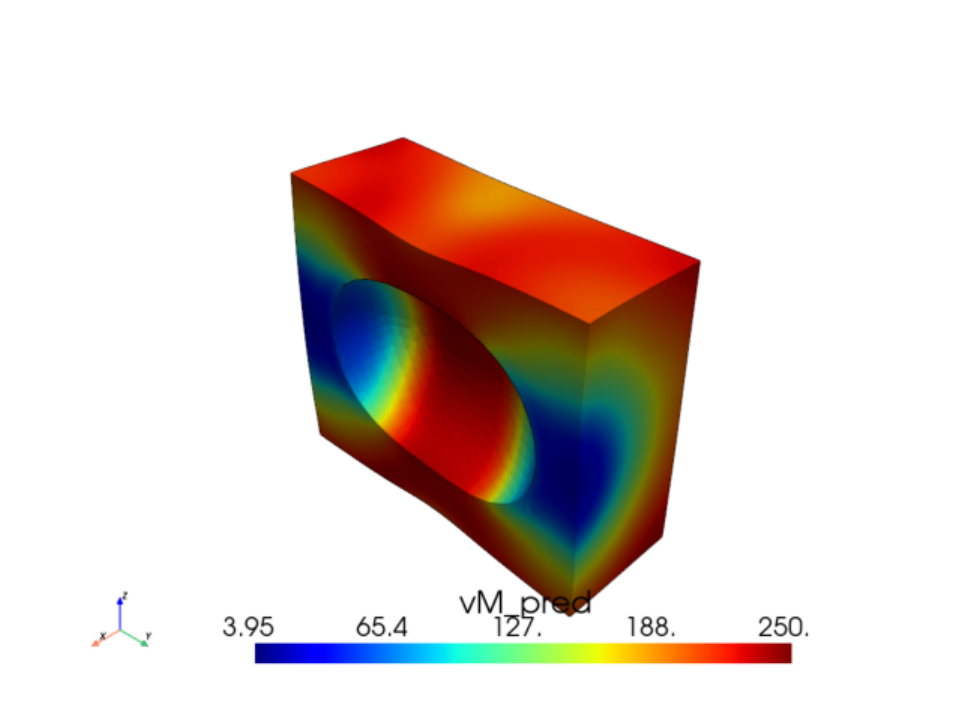}
        }
    \end{minipage} &
    \begin{minipage}[c]{\x\textwidth}
       \centering 
        \subfloat[Pred., worst]{\includegraphics[trim={2.6cm 0cm 2.5cm 2.2cm},clip,width=\textwidth]{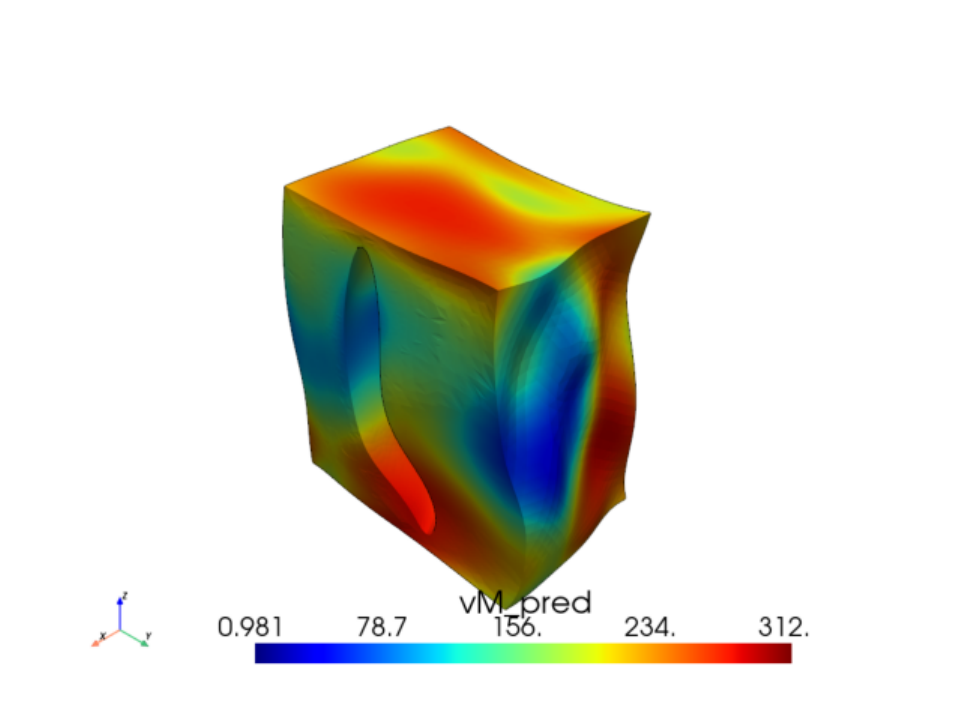}
        }
    \end{minipage} \\

    \begin{minipage}[c]{\x\textwidth}
       \centering 
        \subfloat[MAE, best]{\includegraphics[trim={1.4cm 0cm 2.5cm 1.8cm},clip,width=\textwidth]{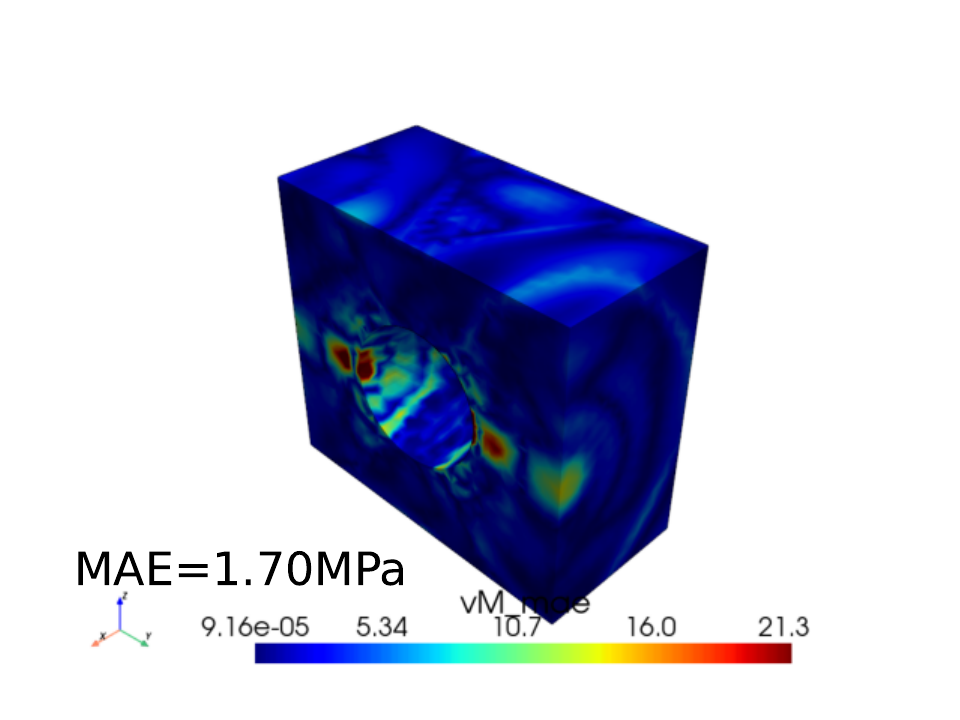}
        }
    \end{minipage} &
    \begin{minipage}[c]{\x\textwidth}
       \centering 
        \subfloat[MAE, 70$^{th}$ pct.]{\includegraphics[trim={1.3cm 0cm 2.5cm 1.8cm},clip,width=\textwidth]{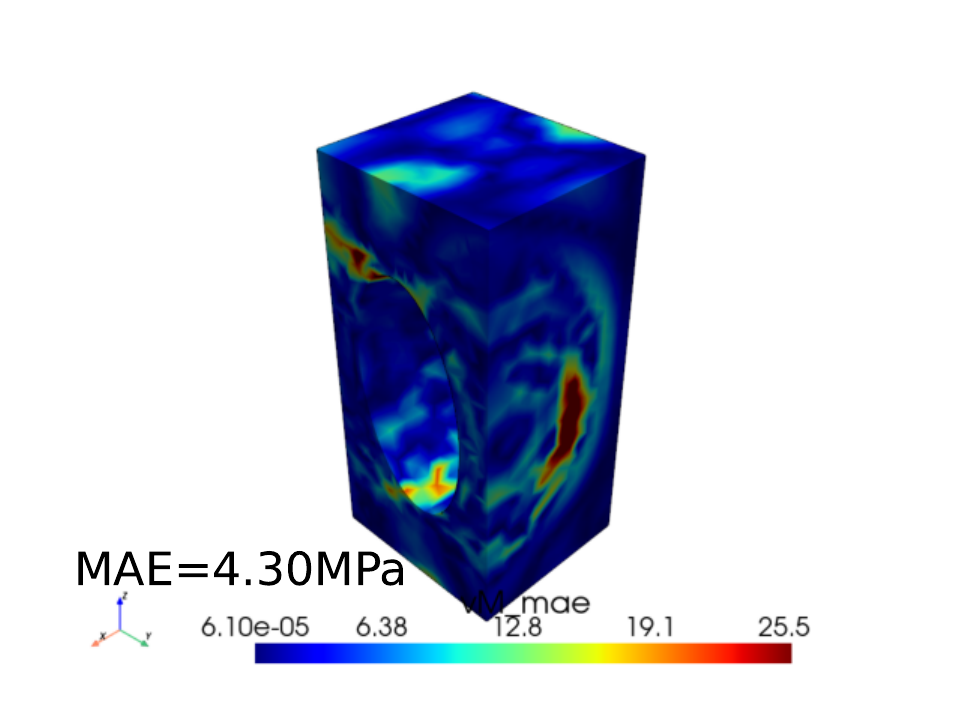}
        }
    \end{minipage} &
    \begin{minipage}[c]{\x\textwidth}
       \centering 
        \subfloat[MAE, 80$^{th}$ pct.]{\includegraphics[trim={1.3cm 0cm 2.5cm 1.8cm},clip,width=\textwidth]{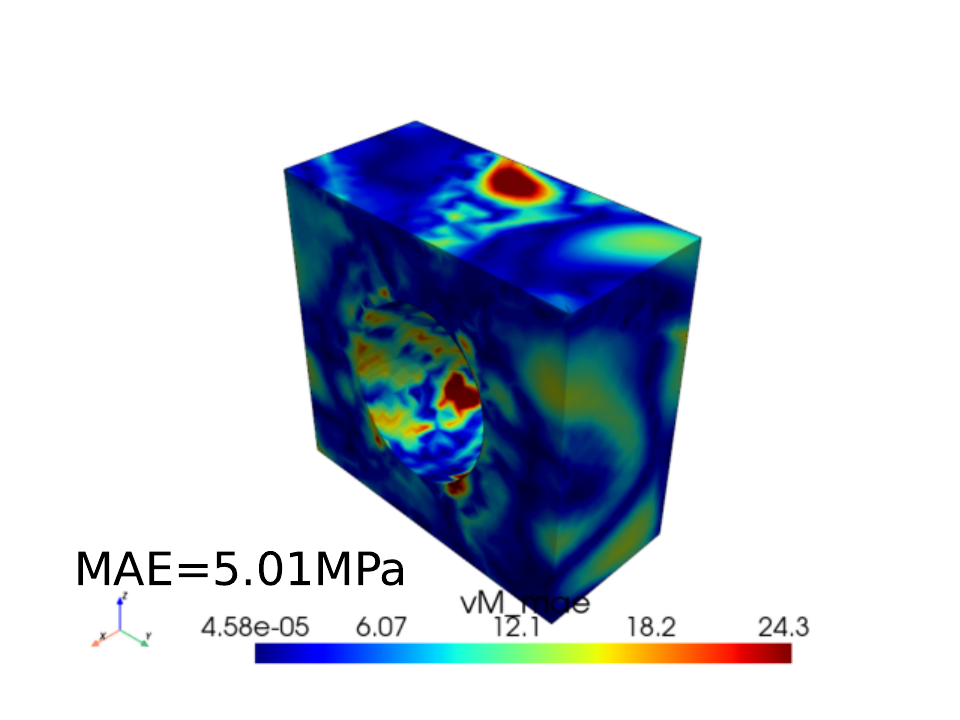}
        }
    \end{minipage} &
    \begin{minipage}[c]{\x\textwidth}
       \centering 
        \subfloat[MAE, 90$^{th}$ pct.]{\includegraphics[trim={1.3cm 0cm 2.5cm 1.8cm},clip,width=\textwidth]{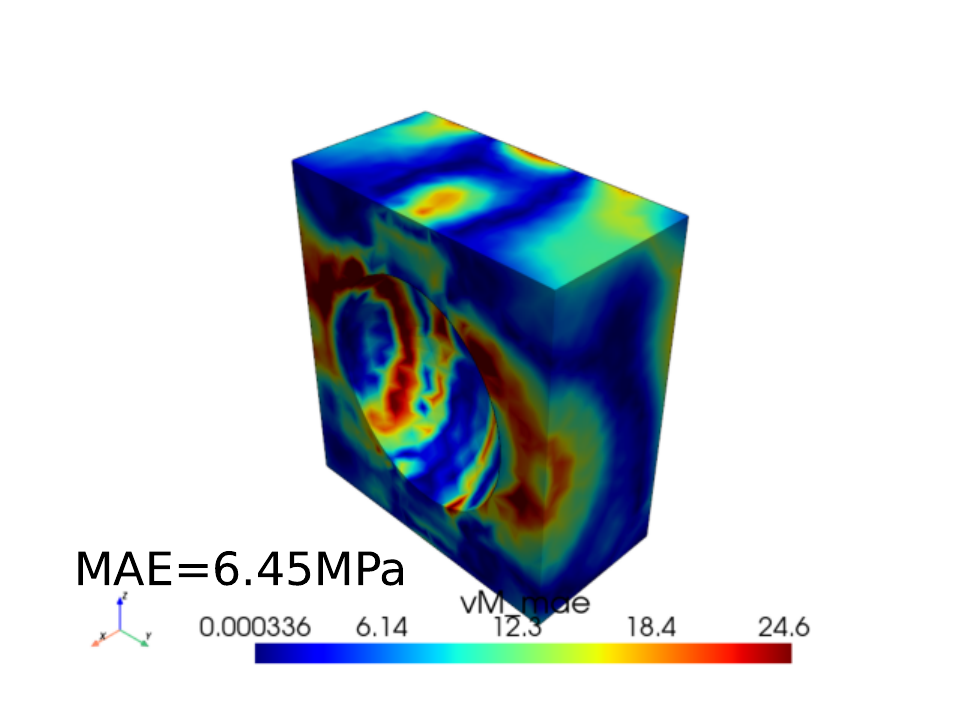}
        }
    \end{minipage} &
    \begin{minipage}[c]{\x\textwidth}
       \centering 
        \subfloat[MAE, worst]{\includegraphics[trim={1.3cm 0cm 2.5cm 1.8cm},clip,width=\textwidth]{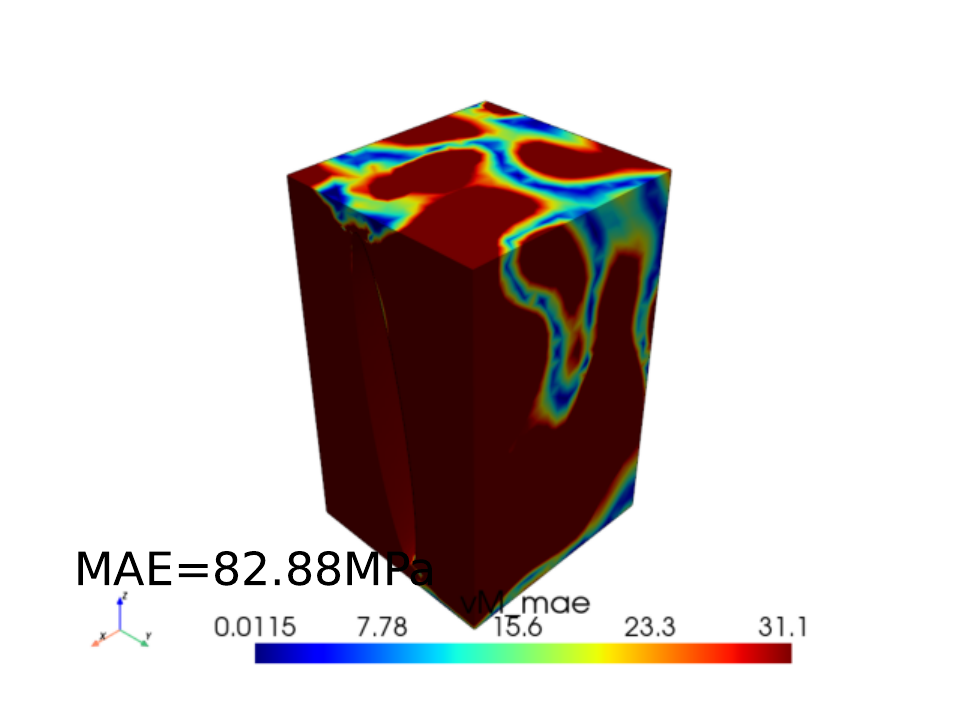}
        }
    \end{minipage} \\

    \end{tabular}
    \caption{Volume predictions by the larger Geom-DeepONet, ranked by different percentiles in stress MAE. The first and second rows show the FE ground truth and the model predictions, and they share identical color scale. The third row shows the MAE and the color scale range is set to $1/10$ of that in the FE ground truth to highlight where the errors are concentrated.}
    \label{c_gdon_pred_large}
\end{figure}

With 4 output vector components, the baseline Geom-DeepONet only has 38298 trainable parameters, and a sheer 95.24\% of the meshes in the cuboid dataset has more degrees of freedom (DoFs) than the Geom-DeepONet trainable parameters. When the geometric variation is large, and the baseline Geom-DeepONet model does not provide sufficient prediction accuracy, it is of practical interest to investigate how the model performance is affected by simply increasing the number of trainable parameters of the model without any further hyper-parameter tuning. To this end, a larger Geom-DeepONet model can be created by simply doubling the number of neurons in each Dense and SIREN layers in \fref{gdon_schematic}. Doing so increases the number of trainable parameters to 150340, about 3 times more than the baseline model. The smaller baseline and larger models were trained independently for 3 times with random 80/20 data split for 600000 iterations. The average (over three runs) training time and performance of both models are shown in \tref{small_v_large_stats}. The FE simulation and prediction times vary due to different numbers of DoFs in each geometry, and a scatter plot comparing the simulation and NN prediction times for different DoF counts is shown in \fref{pred_time}. The amount of speed up comparing the NN prediction times to simulation times in the test geometries is shown in \fref{pred_speedup}. The volume rendering of the predicted fields by the baseline and larger Geom-DeepONets are shown in \fref{c_gdon_pred_small} and \fref{c_gdon_pred_large}, respectively. Contour plots are shown for the stress field, while the FE-simulated and NN-predicted displacement vectors are used to deform the geometries to render the deformed shapes at a scale factor of 250 to highlight any differences.
\begin{table}[h]
\caption{Performance of Geom-DeepONet models of different sizes}
\centering
\begin{tabular}{ccccccccc}
\hline
Model  &  \vline &  Training & MAE, stress & Rel. err, stress  & MAE, $\bm{u}$ & Rel. err, $\bm{u}$  \\
\hline
Baseline &  \vline & 4636s & 7.25 MPa & 5.05\% & 1.68$\times 10^{-4}$ mm & 9.14\% \\
Larger model &  \vline & 5604s & 6.16 MPa & 4.28\% & 1.39$\times 10^{-4}$ mm & 7.87\% \\
\hline
\end{tabular}
\label{small_v_large_stats}
\end{table}

\begin{figure}[h!] 
    \centering
     \subfloat[]{
         \includegraphics[trim={0cm 0cm 0cm 0cm},clip,width=0.23\textwidth]{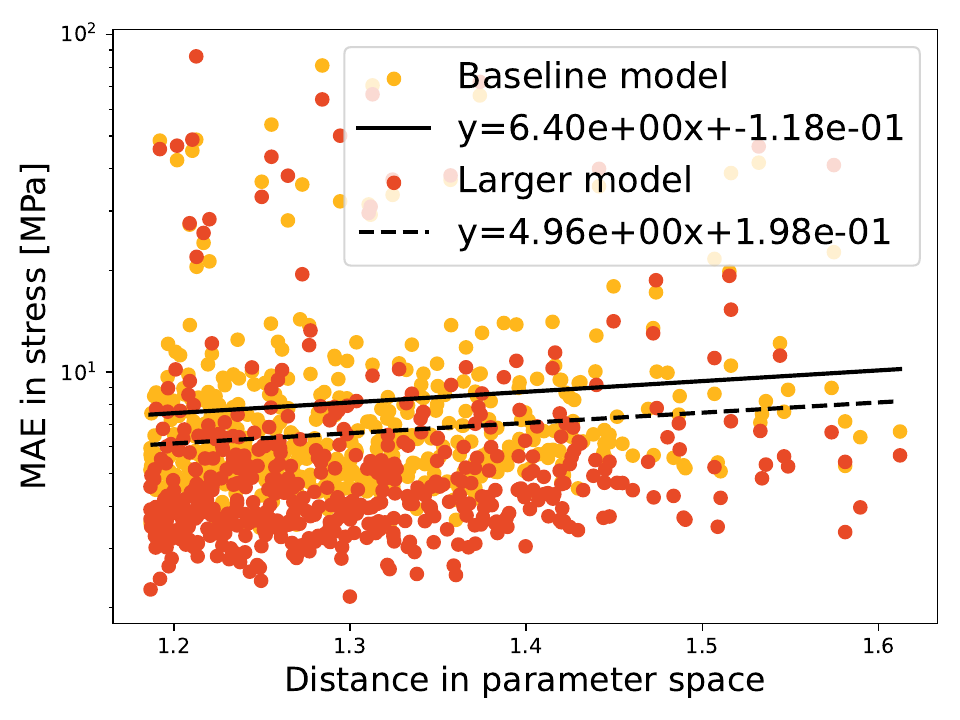}
         \label{s}
     }
     \subfloat[]{
         \includegraphics[trim={0cm 0cm 0cm 0cm},clip,width=0.23\textwidth]{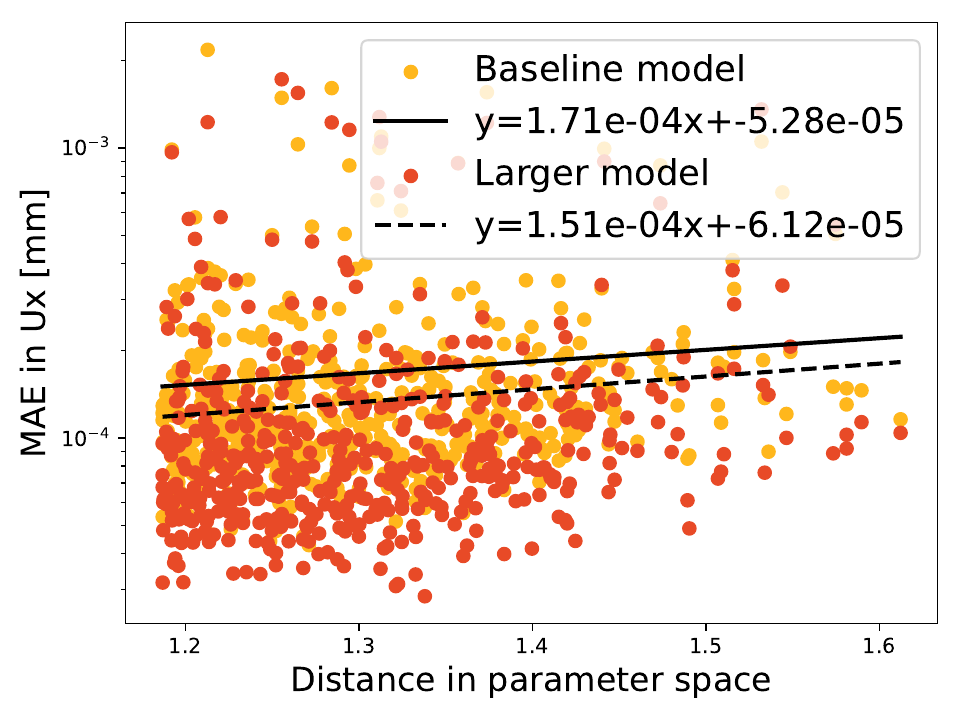}
         \label{ux}
     }
     \subfloat[]{
         \includegraphics[trim={0cm 0cm 0cm 0cm},clip,width=0.23\textwidth]{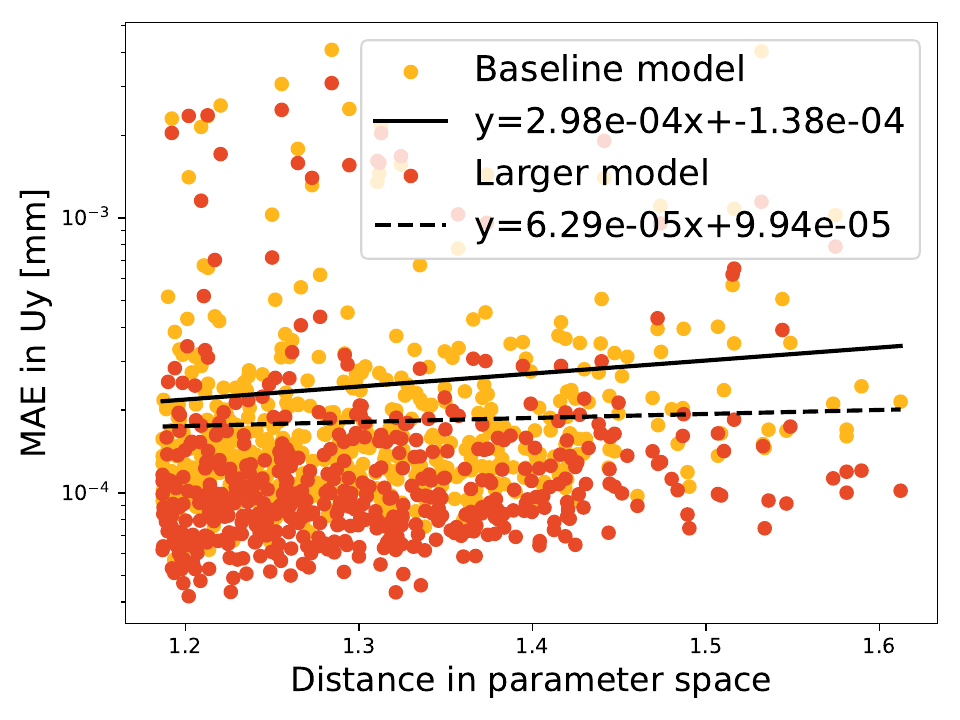}
         \label{uy}
     }
     \subfloat[]{
         \includegraphics[trim={0cm 0cm 0cm 0cm},clip,width=0.23\textwidth]{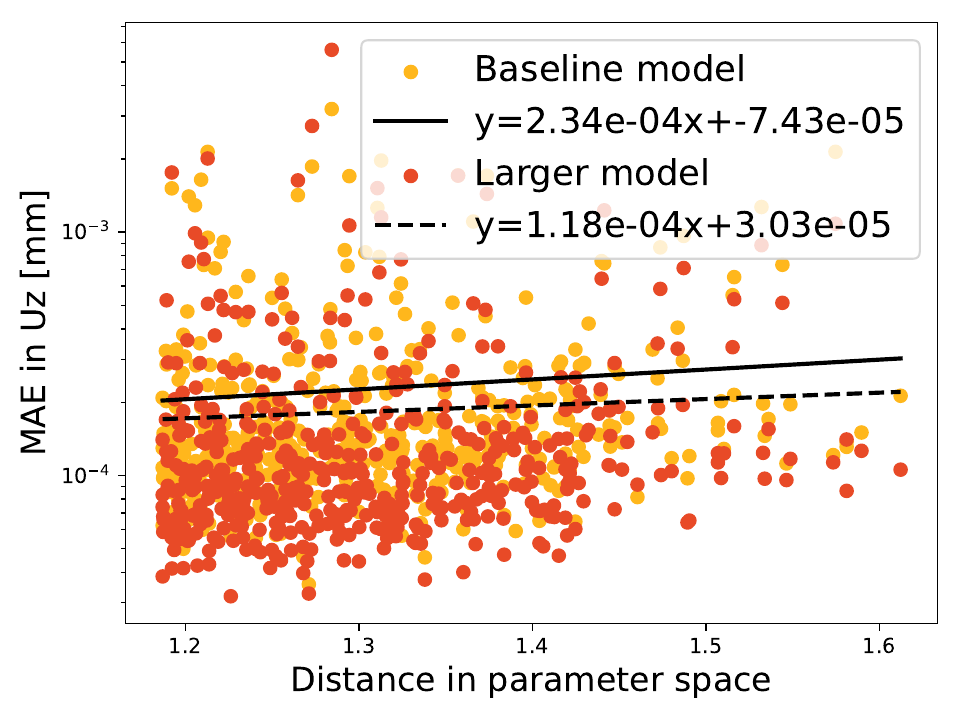}
         \label{uz}
     }
    \caption{Generalization error in parameter space for different components: \psubref{s} von Mises stress, \psubref{ux} $U_x$, \psubref{ux} $U_y$, \psubref{uz} $U_z$. }
    \label{gen_ex2}
\end{figure}

Similar to \sref{generalization}, we trained an instance of the baseline and larger Geom-DeepONets using a similarity-based data split using a similarity score computed from the 8 geometric parameters in the cuboid family of geometries. The comparison of generalization errors for the two models and different output components are shown in \fref{gen_ex2}.

The results in \tref{small_v_large_stats} show that the prediction accuracy can increase by simply increasing the model size without changing any other model hyper-parameters. The MAE in stress decreased by 1.1 MPa, and the relative error in the displacement vector decreased by 1.2\%. It is also interesting to note that although the model size increased by almost 3 folds, the training time only increased by a modest 21\%, indicating that the training of the Geom-DeepONet is efficient thanks to the DeepXDE platform. \fref{speedup} provides insights about the model prediction times on meshes of different node counts and how that compares to FE simulation times. As expected, both show an increasing trend with increasing number of nodes. However, the simulation time increases at a much faster rate as the number of DoFs increases compared to both Geom-DeepONet predictions. As for the speedup in time compared to FE simulation, it again varies with the number of DoFs, ranging from about 1000 times faster for smaller models to over $10^5$ times faster for larger models. These results indicate that, although the predictions cannot be done in batches for maximum efficiency (due to different numbers of nodes in different geometries), the prediction speed of the trained Geom-DeepONet models is still much faster than direct FE simulation.  

\fref{c_gdon_pred_small} and \fref{c_gdon_pred_large} compare the volume predictions generated from the Geom-DeepONets of different sizes. Note that besides the best and worst cases, we showed the 70$^{th}$, 80$^{th}$, and 90$^{th}$ percentiles, aiming to showcase the model performance towards the bad end of the performance spectrum. We see that even the smaller baseline model can capture the general stress contours up to the 90$^{th}$ percentile, and it captures the lateral contractions near the center of the cuboid created due to the internal void (e.g., \fref{c_gdon_pred_small}(d,i)), indicating decent accuracy in all four output components. The stress MAE remains lower than 9 MPa up to the 90$^{th}$ percentile and eventually spikes to over 90 MPa in the worst case, indicating most cases have mean stress errors less than 10 MPa and are quite accurate. The distributions of the prediction errors (last row in \fref{c_gdon_pred_small} and \fref{c_gdon_pred_large}) show no particular concentration in the geometries and appear to be randomly distributed inside the volume. For the larger model, the observation is similar. The model offers accurate stress contour predictions up to 90$^{th}$ percentile, further lowering the error at that point to less than 7 MPa. The lateral contractions due to internal voids are again well captured by the model. In the worst case, neither model is capable of predicting the approximate stress contour and corresponding deformed shape. However, in general, it can be concluded that the Geom-DeepONet predicted solution fields are accurate even for cases with varying geometries. The generalizability of the Geom-DeepONet models is studied in \fref{gen_ex2}. While both models show increased generalization error as the design becomes less and less similar to the reference design (e.g., first design in the family), the rate of error increase is reduced when the model size is increased, indicating that increasing the model size can effectively increase the generalizability of the model.

\subsubsection{Prediction on different meshes}
\label{predictions}
\begin{figure}[h!] 
    \centering
     \subfloat[Mesh M1]{
         \includegraphics[trim={18cm 3cm 18cm 2cm},clip,width=0.23\textwidth]{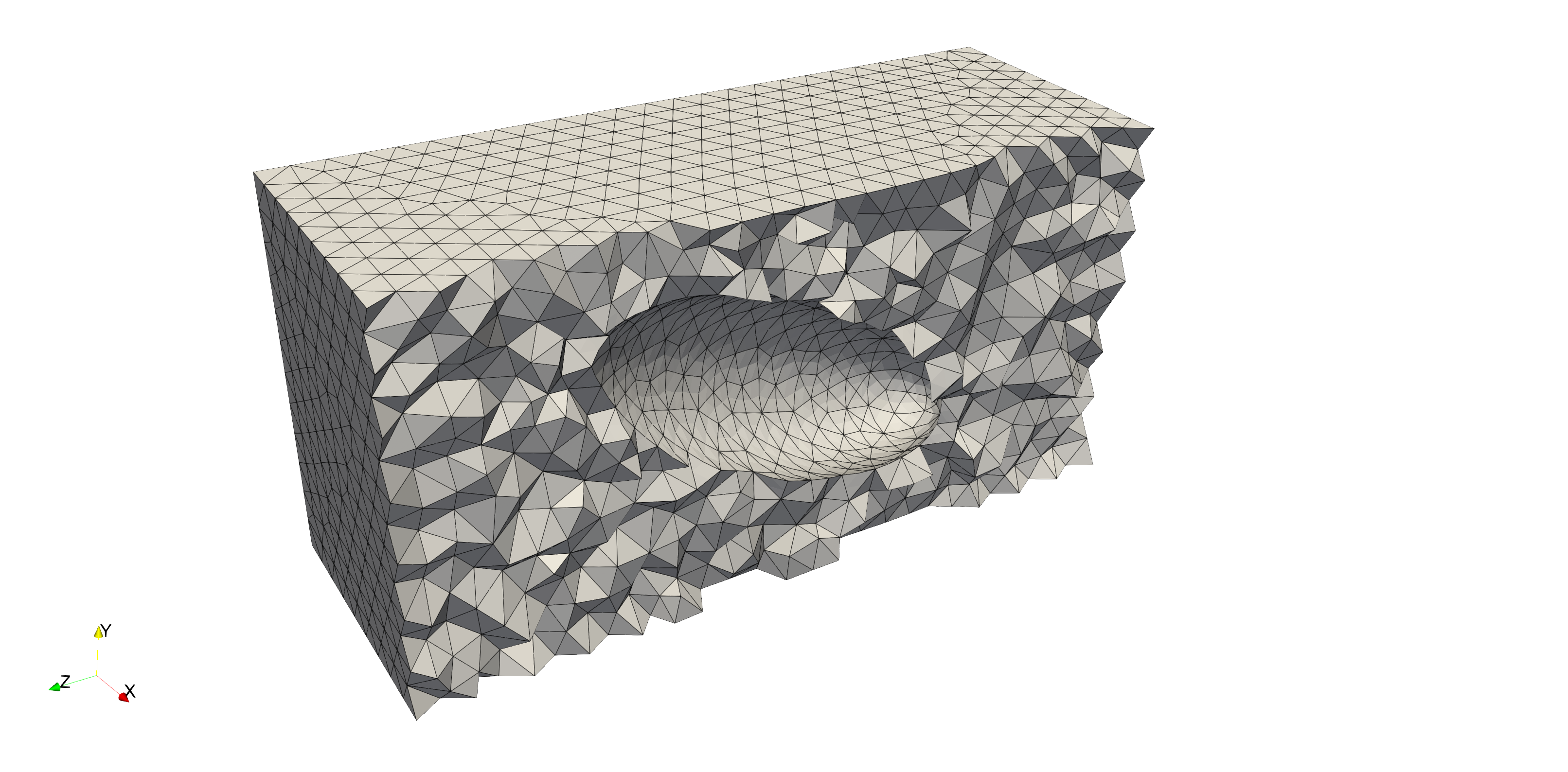}
         \label{m1}
     }
     \subfloat[Mesh M1-1]{
         \includegraphics[trim={18cm 3cm 18cm 2cm},clip,width=0.23\textwidth]{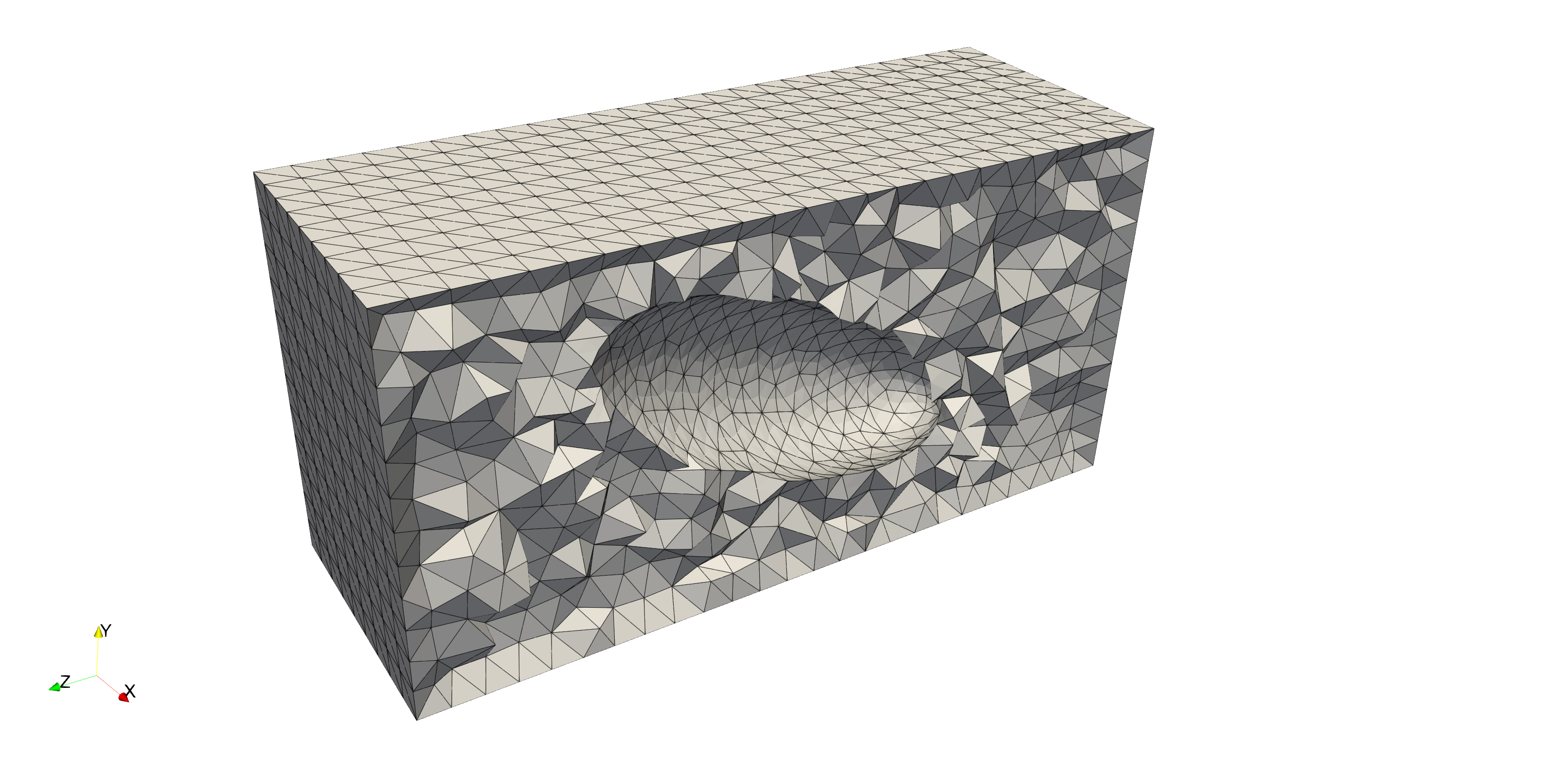}
         \label{m11}
     }
     \subfloat[Mesh M2]{
         \includegraphics[trim={18cm 3cm 18cm 2cm},clip,width=0.23\textwidth]{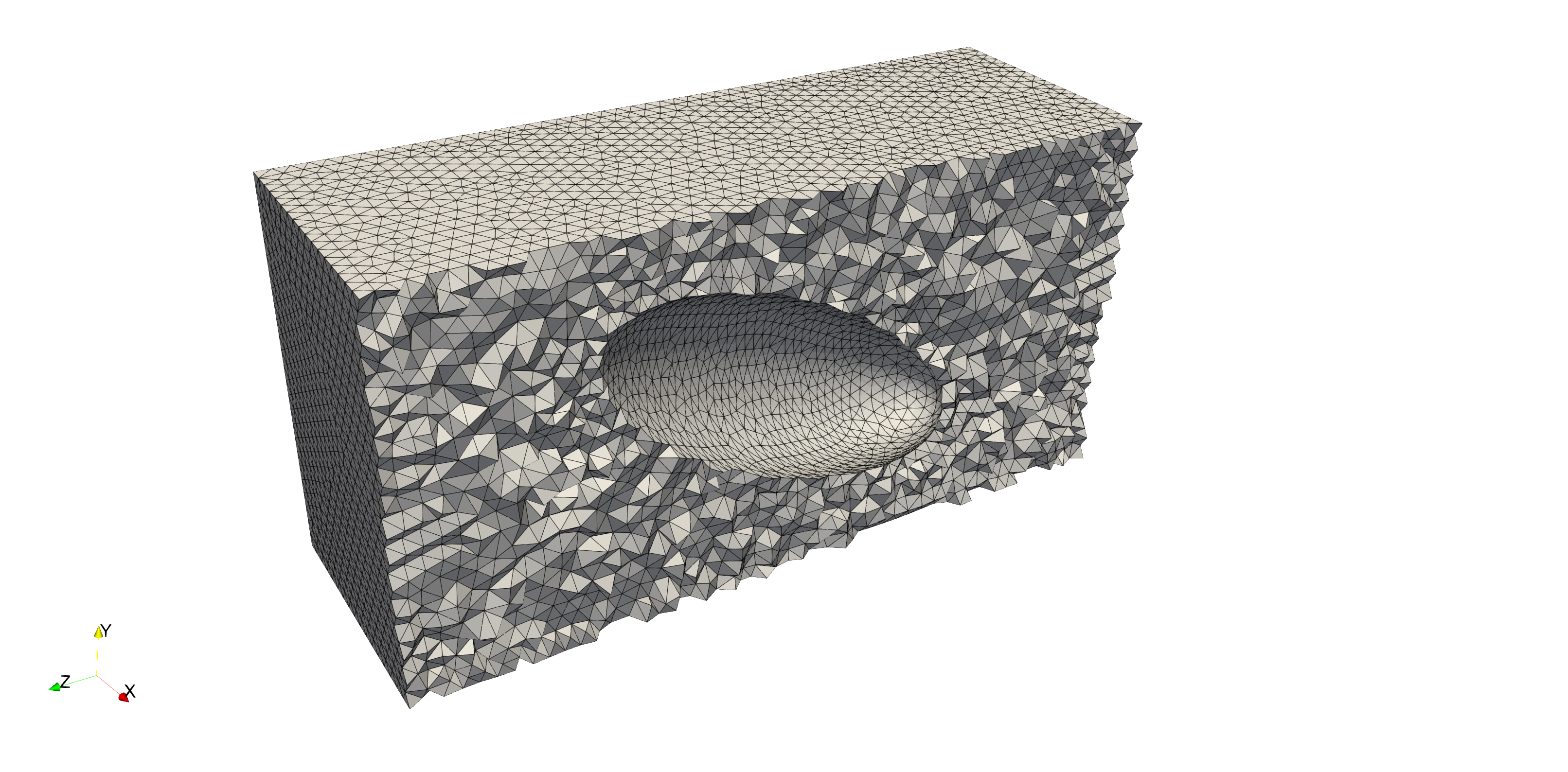}
         \label{m2}
     }
     \subfloat[Mesh M3]{
         \includegraphics[trim={18cm 3cm 18cm 2cm},clip,width=0.23\textwidth]{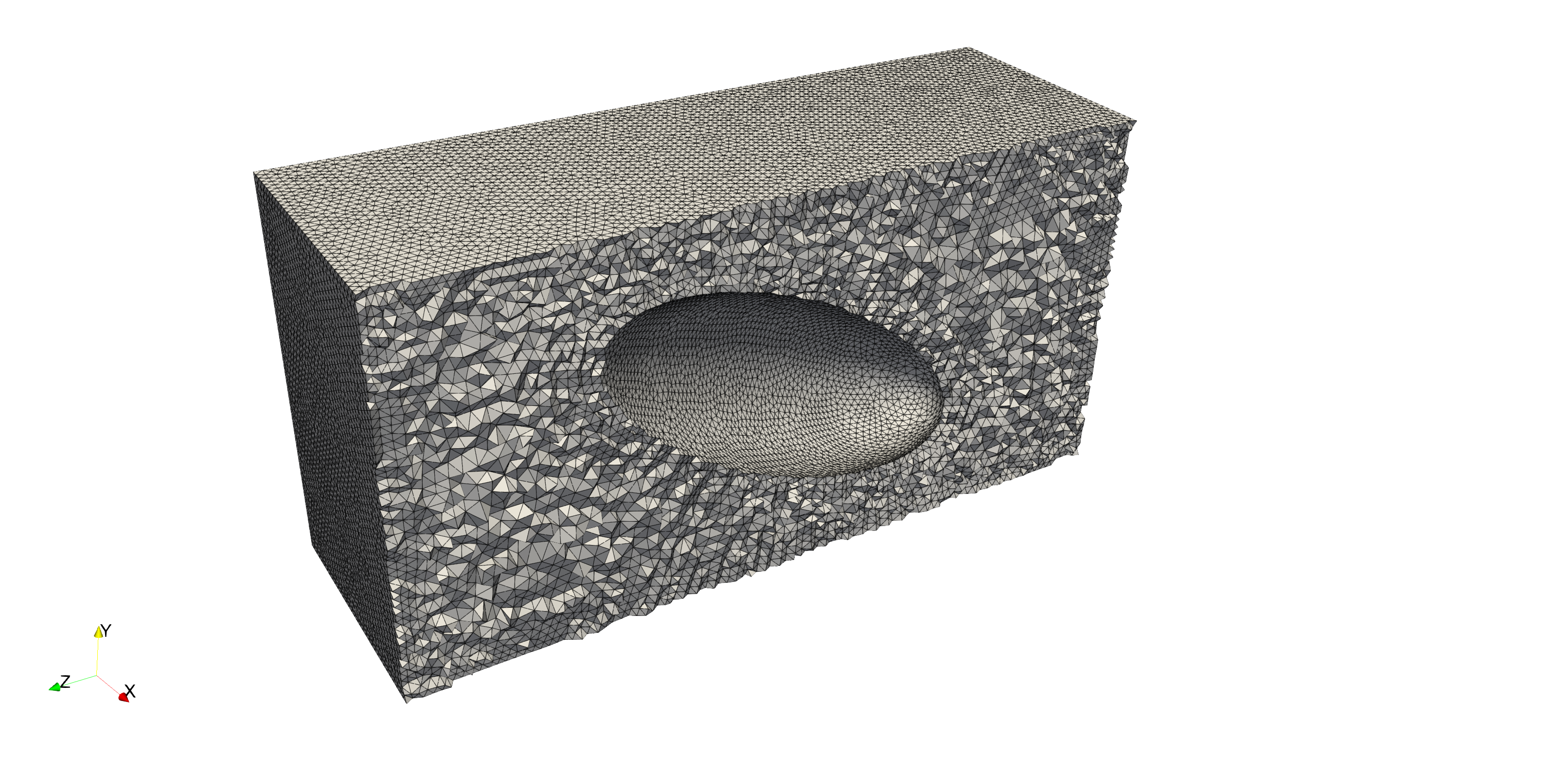}
         \label{m3}
     }
    \caption{Four different meshes of the same geometry used in Geom-DeepONet prediction.}
    \label{meshes}
\end{figure}

\begin{figure}[h!] 
    \centering
     \subfloat[]{
         \includegraphics[trim={0cm 0cm 0cm 0cm},clip,width=0.32\textwidth]{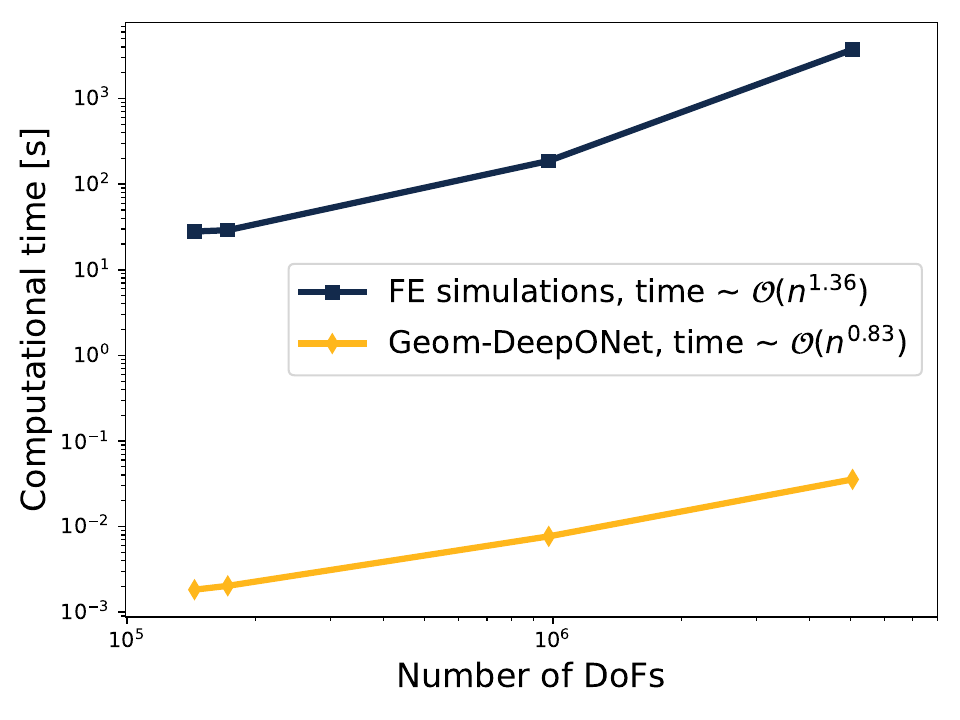}
         \label{pred_time_v_fe}
     }
     \subfloat[]{
         \includegraphics[trim={0cm 0cm 0cm 0cm},clip,width=0.32\textwidth]{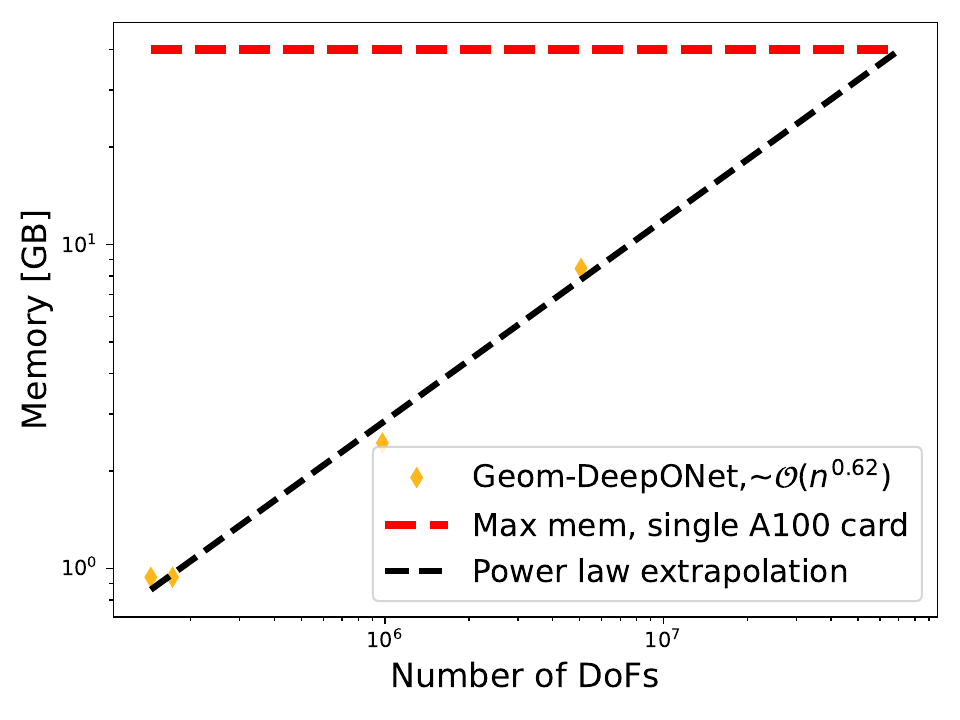}
         \label{mem}
     }
     \subfloat[]{
         \includegraphics[trim={0cm 0cm 0cm 0cm},clip,width=0.32\textwidth]{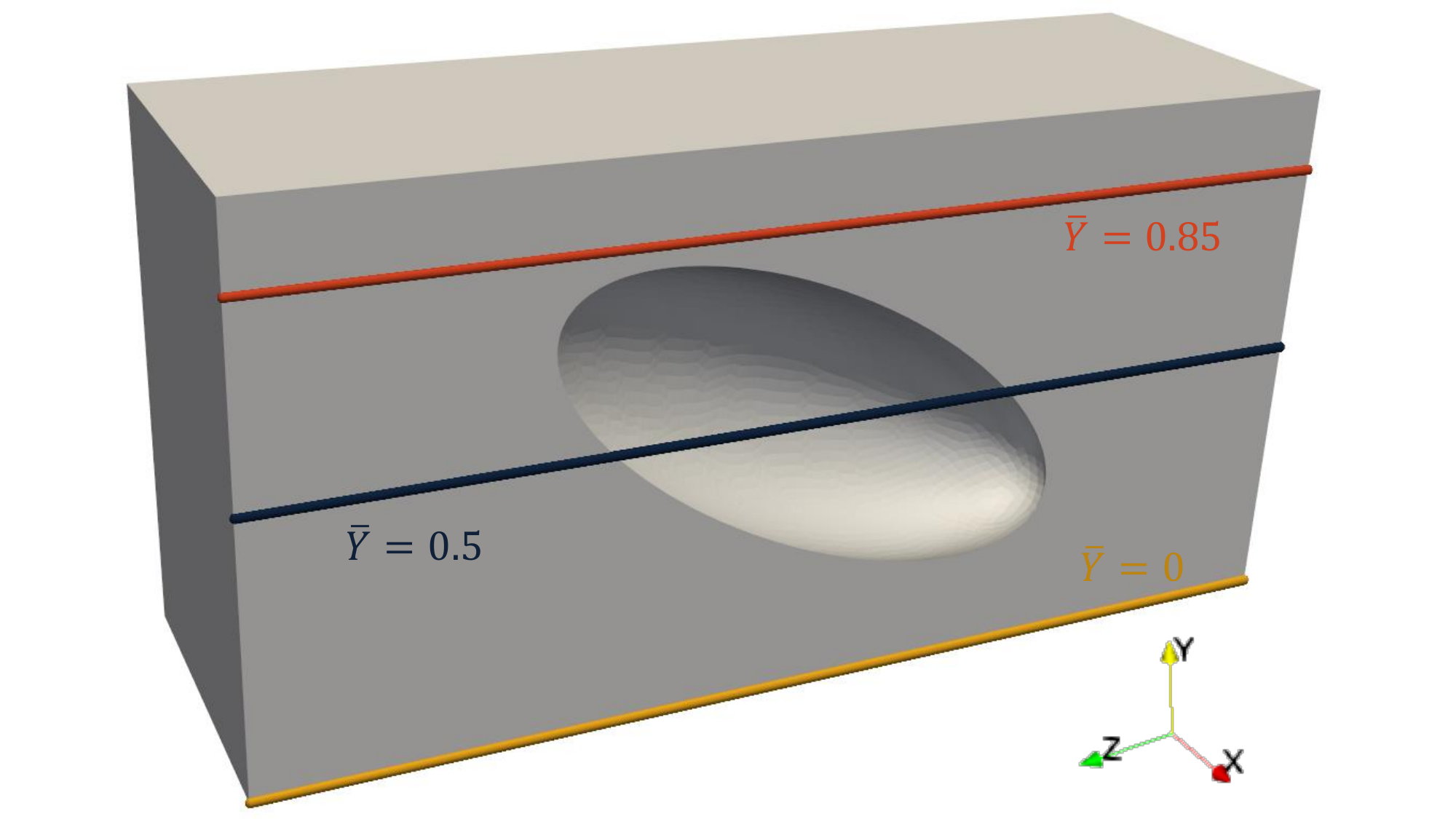}
         \label{3line}
     }
    \caption{\psubref{pred_time_v_fe} Comparison of finite element simulation and neural network prediction times on four meshes. \psubref{mem} GPU memory usage for predictions on different meshes. \psubref{3line} Three lines used to sample the volumetric solution field. Each line contains 1000 evaluation points.}
    \label{time_and_line}
\end{figure}

\begin{figure}[h!] 
    \centering
     \subfloat[$\Bar{Y}=0$]{
         \includegraphics[trim={0cm 0cm 0cm 0cm},clip,width=0.33\textwidth]{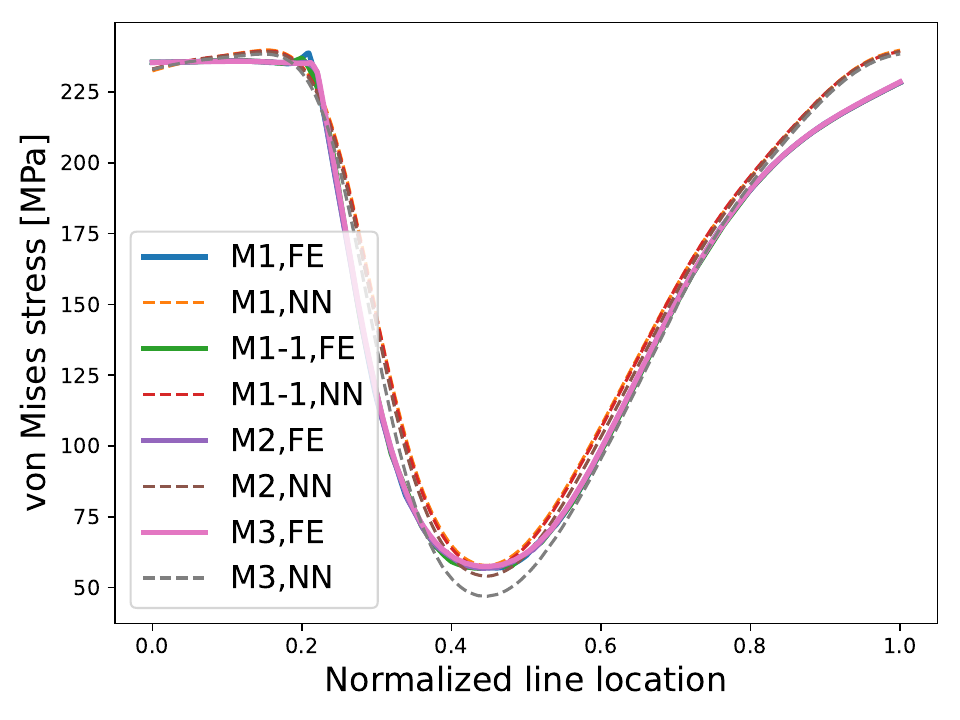}
         \label{l1}
     }
     \subfloat[$\Bar{Y}=0.5$]{
         \includegraphics[trim={0cm 0cm 0cm 0cm},clip,width=0.33\textwidth]{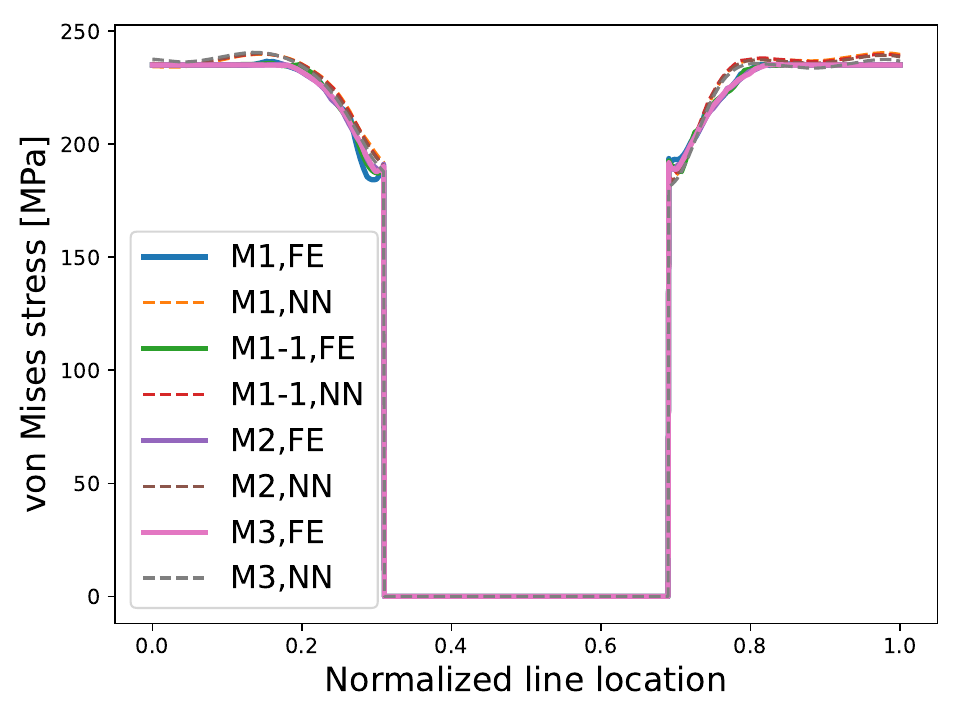}
         \label{l2}
     }
     \subfloat[$\Bar{Y}=0.85$]{
         \includegraphics[trim={0cm 0cm 0cm 0cm},clip,width=0.33\textwidth]{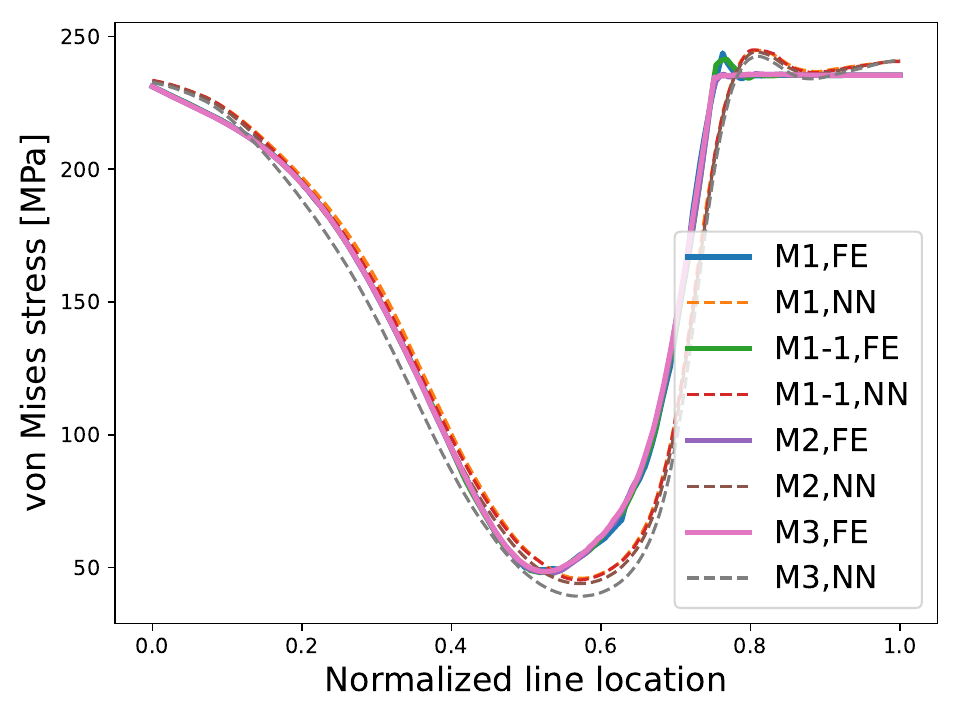}
         \label{l3}
     }
    \caption{Comparison of ground-truth and predictions von Mises stress at the three sample lines. Sampled stress value is 0 at the interior of the void.}
    \label{line_plots_}
\end{figure}

The current Geom-DeepONet is a coordinate-based framework; namely, only point clouds (but not element connectivity) are fed into the network to help describe the geometry. Therefore, it is also of interest to investigate how the model predictions vary as the point cloud used to describe certain geometry changes. To this end, a geometry was randomly chosen from the test set of the larger Geom-DeepONet trained using a similarity-based data splitting. Besides the original mesh (here denoted as M1), three additional meshes were generated. The first one (denoted as M1-1) has identical characteristic element size but was meshed with a mapped surface mesh, while the two other meshes have $\frac{1}{2}$ and $\frac{1}{4}$ for the mesh sizes, respectively (denoted as M2 and M3). The four meshes are shown in \fref{meshes}. FE simulations were performed on all four meshes to obtain the ground truth values. The trained Geom-DeepONet model was loaded and used to make field predictions on the four meshes. The mesh node count, computational time, and MAE in stress and displacement predictions are shown in \tref{new_mesh_data}. The FE simulation and NN prediction times are compared in \fref{pred_time_v_fe}. To further compare the FE simulation stress results and the Geom-DeepONet predictions, we sampled the 3D volumetric data along three lines in the middle cut-plane of the geometry at three different normalized $\Bar{Y}$ values of 0, 0.5, and 0.85, respectively. The three sample lines are shown in \fref{3line}, and the extracted results are compared in \fref{line_plots_}.
\begin{table}[h]
\caption{Model performance comparison on 4 different meshes}
\centering
\begin{tabular}{ccccccccc}
\hline
Mesh  &  \vline & Node \# & FE time & Prediction time & MAE, stress  & MAE, $\bm{u}$  \\
\hline
M1 &  \vline & 57408 & 29 s & 2.03 ms & 5.52 MPa & 1.46$\times 10^{-4}$ mm \\
M1-1 &  \vline & 47962 & 28 s & 1.83 ms & 5.52 MPa & 1.42$\times 10^{-4}$ mm \\
M2 &  \vline & 325726 & 187 s & 7.68 ms & 5.14 MPa & 1.42$\times 10^{-4}$ mm \\
M3 &  \vline & 1684604 & 3742 s & 35.58 ms & 5.20 MPa & 1.66$\times 10^{-4}$ mm \\
\hline
\end{tabular}
\label{new_mesh_data}
\end{table}

The comparison shown in \tref{new_mesh_data} and \fref{pred_time_v_fe} shows that the Geom-DeepONet prediction on sequentially refined meshes is far more efficient than finite element simulations, more than $10^4$ times faster. A power-law curve fit revealed that the FE simulation times scale as $\mathcal{O}(n^{1.36})$, where $n$ denotes the number of DoFs in the problem, while that from the NN predictions is $\mathcal{O}(n^{0.83})$. This indicates that the NN prediction also scales better than the FE simulation as the number of DoFs increases. \fref{mem} shows GPU memory consumption during prediction time for different mesh sizes. Sub-linear scaling was observed for the required memory, and a power-law curve fit reveals that a single A100 GPU card with 40GB memory can handle predictions with approximately $7\times10^7$ DoFs before running out of memory. Comparing the MAEs in stress and displacement, we see similar levels of errors for the two meshes with identical mesh sizes (M1 and M1-1) as well as meshes of different densities (M1, M2, and M3), indicating that the performance of the Geom-DeepONet is relatively stable with respect to changes in the point cloud (nodes) used to describe the geometry. It is insensitive to different element connectivity used to form the volume meshes. 

When comparing the line plot results in \fref{line_plots_}, it is evident that the FE results from different mesh densities are almost identical, indicating that the FE results converge even at the coarsest mesh level. The NN predictions remain generally close to the FE ground truth values. However, for lines (a) and (c), the minimum predicted value near the mid-point of the sample line decreases with finer mesh size, thus increasing the prediction error on the sample line. Further comparing the shape of the FE stress contours with the NN-predicted contours, it is evident that a spectral bias \citep{rahaman2019spectral} exists in the predictions. The sharp stress transitions in the FE ground truth (high-frequency components) are approximated by Geom-DeepONet with smooth transitions (low-frequency components). This is a limitation of the current model and can be detrimental when the geometric features in the part, such as cracks, defects, and sharp corners, induce regions of sharp stress gradients.

\section{Conclusions, limitations, and future work}
\label{sec:conc}
Solution field prediction via neural networks for 3D objects with varying geometries has significant practical importance as all engineering components are 3D. Repeated evaluation over many design iterations might be necessary during the initial design process. PointNet \citep{kashefi2021point}, a neural network architecture that can handle geometry variations, suffers from the limitation of a fixed point cloud size used for input and output. The DeepONet \citep{lu2021learning}, although having low generalization errors, is not designed to handle variable geometries. Therefore, the current work proposed a novel DeepONet architecture termed the Geom-DeepONet, specifically designed to handle parameterized geometries (i.e., can be fully defined via a set of geometric parameters like length, width, height, and radius) in 3D discretized by different numbers of nodes and finite elements. This is the first attempt in the literature to the best of the authors' knowledge. In addition to the traditional nodal coordinate information used in the trunk network of vanilla DeepONet, Geom-DeepONet augments the trunk input with signed distance functions, thus providing spatial information on the current node in terms of its nearest distance to the exterior surface of the geometry. Geom-DeepONet also leverages intermediate data fusion between the branch and trunk networks, a technique shown to improve model performance. We also employed SIREN in the trunk network to encode the 3D point clouds instead of simple dense layers, another improvement from the vanilla DeepONet.

Furthermore, by introducing an additional flexible dimension, the current architecture can seamlessly generate predictions on meshes of an arbitrary number of nodes, making the network naturally suitable for varying geometries. Using a parameterized beam dataset, a comprehensive benchmark was conducted between PointNet, vanilla DeepONet, and Geom-DeepONet. The results indicate that Geom-DeepONet is more accurate than PointNet and vanilla DeepONet. Geom-DeepONet is also much more computationally efficient (in terms of training time and required GPU memory) to train than PointNet running on identical hardware. Using the same dataset, we also demonstrate that Geom-DeepONet can generalize much better in the design parameter space for unseen and dissimilar designs than the vanilla DeepONet, proving its strong capability in capturing geometry changes and their influence on the underlying fields. It is further demonstrated that the proposed model can accurately predict vector fields (e.g., stress and three displacement components) using a cuboid with the random elliptical void dataset. When the geometry variation is considerable, simply increasing the number of neurons in each layer proves to be a straightforward way to obtain improved performance without the need for intricate hyperparameter optimization, and increasing the model size also lowers the generalization errors of the model for unseen and dissimilar designs. It is shown that the predictions of the proposed model are relatively insensitive to the input point cloud used to describe the geometry, and the time for neural network prediction can be over $10^5$ times faster than finite element simulations for models with a large number of degrees of freedom. Finally, the proposed network has a relatively small memory footprint in training and prediction, indicating that it is possible to employ DeepONets in variable 3D real-world geometries today with the current GPU hardware capabilities.

The current work provides a novel tool for generating field predictions on variable geometries. The relatively short training time, small memory footprint, high prediction accuracy, and much faster prediction speed compared to finite element simulations make it a powerful asset for design engineers, where repeated preliminary design evaluations are required. The model can provide valuable directional insights to the design engineer and quickly filter out undesirable designs. Compared to previous neural network models that aim to serve a similar purpose, the ability of the current model to handle truly three-dimensional variable geometries makes it one step closer to real-world engineering applications. 

Although the current methodology yielded high accuracy, the proposed methods have certain limitations. First, the current model relies on a design parameter-based description of the input geometries, which may only sometimes be feasible or available for complex engineering components and assemblies. Therefore, in future works, we will seek to explore implicit and parameter-free means to encode varying input geometries. Secondly, from the comparison of the ground truth and predicted stress curves, it is seen that the current model suffers from spectral bias, and the high-frequency components of the solution field are not learned effectively. Since previous works \citep{jiang2023fourier,li2022fourier} have leveraged the Fourier transform to capture high-frequency components of the solution field, future works will also include improving the current Geom-DeepONet to minimize spectral bias.

\section*{Replication of results}
The data and source code that support the findings of this study can be  made available upon reasonable request to the corresponding author.

\section*{Conflict of interest}
The authors declare that they have no conflict of interest.

\section*{Acknowledgements}
The authors would like to thank the National Center for Supercomputing Applications (NCSA) at the University of Illinois, particularly its Research Consulting Directorate, the Industry Program, and the Center for Artificial Intelligence Innovation (CAII) for their support and hardware resources. This research is a part of the Delta research computing project, which is supported by the National Science Foundation (award OCI 2005572) and the State of Illinois, as well as the Illinois Computes program supported by the University of Illinois Urbana-Champaign and the University of Illinois System.

\section*{CRediT author contributions}
\textbf{Junyan He}: Methodology, Formal analysis, Investigation, Writing - Original Draft. 
\textbf{Seid Koric}: Supervision, Resources, Writing - Original Draft, Writing - Review \& Editing, Funding Acquisition.
\textbf{Diab Abueidda}: Supervision, Writing - Review \& Editing.
\textbf{Ali Najafi}: Supervision, Writing - Review \& Editing.
\textbf{Iwona Jasiuk}: Supervision, Writing - Review \& Editing.

\bibliographystyle{unsrtnat}
\setlength{\bibsep}{0.0pt}
{\scriptsize \bibliography{References.bib} }
\end{document}